\documentclass[a4paper,12pt]{article}
\usepackage{jcappub} 
\usepackage{lineno}
\usepackage[utf8]{inputenc}
\usepackage{amsmath,amssymb}
\usepackage{graphicx}
\usepackage{hyperref}
\usepackage{natbib}
\usepackage[normalem]{ulem}
\usepackage{wrapfig}
\usepackage{subcaption}
\usepackage{relsize}
\newcommand{\om}{\Omega_{m}}
\newcommand{\ok}{\Omega_{k}}
\newcommand{\ol}{\Omega_{r}}
\newcommand{\ode}{\Omega_{de}}
\newcommand{\paren}[1]{\left(#1\right)}
\usepackage[top=1in, bottom=2.5cm, left=2cm, right=2cm]{geometry}
\usepackage{multibib}
\usepackage[misc]{ifsym}
\usepackage{adjustbox}
\usepackage{multirow}
\usepackage{makecell}
\usepackage{natbib}

\allowdisplaybreaks

\title{Barrow holographic dark energy interacting model in the presence of radiation and matter}
\author[a,1]{Gopinath Guin,\note{Corresponding author.}}
\author[a]{Souvik Paul,}
\author[a]{Sunandan Gangopadhyay}

\affiliation[a]{Department of Astrophysics and High Energy Physics,\\
	S.N.~Bose National Centre for Basic Sciences,\\
	Salt Lake, Kolkata 700106, India}

\emailAdd{gopinath.guin@bose.res.in}
\emailAdd{souvik.paul@bose.res.in}
\emailAdd{sunandan.gangopadhyay@bose.res.in}

\abstract{We have studied the effect of dynamical radiation in the interacting barrow holographic dark energy model for a non-flat universe. For both open and closed universes, we have obtained the evolution equation for the energy density parameters for dark energy, dark matter and radiation for four different kinds of interaction among the seven possible linear phenomenological interactions. We have then numerically solved those coupled differential equations to show their behaviour with the redshift parameter. Also, the dynamics of the dark energy equation of state parameter with redshift for different interaction models are shown. For all four interaction models, it is also found that for higher values of the Barrow exponent, the dark energy equation of state parameter shows a transition into the phantom region from the quintessence region in the early time, that is, for lower redshift values. We have also found different epochs corresponding to dark energy-dark matter, dark energy-radiation and dark matter-radiation crossings. These crossing points are also consistent with the thermal history of the universe. We have also obtained various observational constraints for different cosmological parameters for our interacting Barrow holographic dark matter model using the Cosmic chronometer, Baryon Acoustic Oscillator and Pantheon+ data sets. The constraint values of the Hubble parameter in our cosmological shows higher values compared to the $\Lambda$CDM model, therefore indicating towards a possible resolution to the Hubble tension problem.
}

\begin{document}
\maketitle
\flushbottom
\section{Introduction}
In cosmology, one of the most challenging tasks is to explain the late-time accelerated expansion of the universe. The accelerated expansion is supported by several observations, supernova SNIa data \cite{SupernovaSearchTeam:1998fmf,SupernovaCosmologyProject:1998vns,BOSS:2012bus,Planck:2014loa}, CMB data \cite{WMAP:2010sfg}. In order to explain the late time acceleration untill now  one of the most successful model is the $\Lambda$CDM model\cite{DES:2018paw,SupernovaSearchTeam:1998fmf,Perivolaropoulos:2021jda}. Although the $\Lambda$CDM model successfully explains late time acceleration and several other things but it can't successfully explain problems like Hubble tension\cite{Bernal:2016gxb,Riess:2016jrr}, cosmological problem\cite{Weinberg:1988cp,Straumann:2002tv} etc. That is why people tried with some other models to remove the subtlity with Holographic dark energy models \cite{Bousso:2002ju,Fischler:1998st,Li:2004rb,Li:2009bn,Wang:2016och,Nojiri:2021iko,Jawad:2018juh,Huang:2025tqr,Das:2020rmg,Shaikh:2019ppk,Samaddar:2024web,Kaur:2022lgk,Nojiri:2023nop, Li:2024qus,Nojiri:2005pu,
Nojiri:2017opc}, agegraphic dark energy model\cite{Wei:2007ty,Wei:2007ut,Sheykhi:2009rk}, Ricci dark energy\cite{delCampo:2013hka,Pan:2014afa} Chaplygin gas \cite{Bento:2002ps} etc. Although all of these models have their own successes and drawbacks.\\  
Recently, the the origin of dark energy through the holographic principle has been studied in several articles.  The application of the holographic principle in a cosmological context is very interesting, as it relates the ultraviolet length scale with the largest length scale like the horizon\cite{Saridakis:2020zol}. Also, several observationion contraints support the theory directly or indirectly.\cite{Micheletti:2010cm,Micheletti:2009jy,Feng:2007wn,Lu:2009iv}. 
On the other hand, there have also been several attempts to develop a model of dark energy within the framework of quantum gravity using string theory techniques. In this context, the Kachru, Kallosh, Linde, and Trivedi (KKLT) \cite{Kachru:2003aw} scenario and the Large Volume Scenario (LVS) \cite{Balasubramanian:2005zx} indicate that a metastable de Sitter vacuum can be produced using flux compactifications with moduli stabilization \cite{Giddings:2001yu}. However, this model suffers from several theoretical challenges, such as instability, backreaction, and the Swampland conjectures \cite{Obied:2018sgi,Ooguri:2018wrx,Vafa:2005ui,Palti:2019pca,Agrawal:2018own}. This issue raises questions about the existence of de-Sitter vacua in string backgrounds. For example, the Swampland conjecture states that ds vacua should not exist in a UV-complete theory, which makes the validity of the KKLT theory doubtful.\\
Without considering these complications of string theory and taking inspiration from the COVID-19 virus structure, Barrow \cite{Barrow:2020tzx} questioned whether at the Planck scale the horizon surface has any intricate structure that leads to a higher area in a constant volume. In fact, at the higher intricacy the horizon surface and correspondingly the entropy (scales according to the area law) even go to infinity. However, there is hardly any physical meaning to the infinite surface area of a black hole. That is why one needs a cut off length scale to make the horizon area finite. But whatever the cut off is, the area with intricacy and fractal structure is always greater than the classical horizon area\cite{Barrow:2020tzx}. 
Considering the geometry that has fractal structure and intricacy, the entropy associated with it called the Barrow entropy, and it has the following form \cite{Barrow:2020tzx}
\begin{equation}
    S_{B}=\Bigg(\frac{A}{A_{p}}\Bigg)^{1+\frac{\Delta}{2}}
\end{equation}
where $\Delta$ is called the Barrow exponent and $A_p$ is the plank area. It can take values in the range $0\leq \Delta\leq 1$. The value of $\Delta$ is $1$ for the most complex fractal structure, and  is $0$ for the smoothest structure. In the limit $\Delta\to 0$, the Barrow entropy reduces to the Bekenstein-Hawking entropy for the cosmological horizon. The horizon in the  literature has been chosen as the future event horizon\cite{Li:2004rb}.
Using the Holographic principle, one can say that physical quantities of the universe inside a horizon length $L$ can be described by some parameters in the boundary of the universe. Using dimension analysis, one gets the form of the dark energy density following \cite{Wang:2016och},
\begin{equation}
    \rho_{de}=BM_p^4+\Tilde{C}M_P^2L^{-2}+DL^{-4}+\cdots~.
\end{equation}
Following the  Cohen-Kaplan-Nelson bound\cite{Cohen:1998zx}, one gets energy inside a black hole with horizon radius $L$ cannot be greater than the mass of the black hole; the B term should not be present in the expression. Also, compared to the second term, the latter terms are negligible. So effectively, one can write 
\begin{equation}
    \rho_{de}=CL^{-2}~.
\end{equation}
Taking the holographic dual of the black hole horizon to the cosmological horizon, and with the fractal structure present in the horizon, one can write the form of the Barrow holographic dark energy as 
\begin{equation}\label{dark energy density}
    \rho_{de}=CL^{\Delta-2}
\end{equation}
where $C=3c^2 M^2_{p}$ and $L$ is some cosmological length scale, as said earlier, we will take it as the future event horizon. This dark energy density takes care of the complex fractal structure of the universe through the Barrow exponent. \\
There have been studies in which a universe filled with dark matter and holographic dark energy has been considered with zero spatial curvature \cite{Basilakos:2023seo,Sheykhi:2022fus,Saridakis:2020zol}. However, several observations suggest that the spatial curvature cannot be constrained to zero \cite{Handley:2019tkm,Planck:2018vyg,Anselmi:2022uvj,DiValentino:2019qzk}. So, without knowing whether we live in an exactly flat universe or not, it is not a wise choice to proceed theoretically with taking the curvature density parameter ($\Omega_k$) equal to zero. That is why in the literature there has been extended work on a universe filled with dark matter and dark energy in a non-flat universe \cite{Luciano:2022hhy,Saridakis:2020lrg,Adhikary:2021xym,Adhikary:2024sax,Motaghi:2024rag,Kotal:2025bof,Dixit:2021phd,Mamon:2020spa,Luciano:2025elo,Luciano:2025hjn}.
On the other hand, the exact behaviour of dark energy and dark matter is unclear, and studies have argued that the possibility of interaction between the dark components of the universe may be possible \cite{Bolotin:2013jpa,Wetterich:1994bg} and may resolve problems like the cosmological constant \cite{amendola2010dark,Bolotin:2013jpa,Weinberg:1988cp,Sahni:2002kh}, Hubble tension \cite{DiValentino:2017iww,Kumar:2019wfs,Clifton:2024mdy,Bernal:2016gxb}, coincidence problem, etc.
Even with all this development, the reason to consider interaction only between matter and dark energy is not sufficiently justified because in the real universe where we live, it is a clear truth that although radiation is less in amount \cite{Planck:2018vyg}, light is all around. Also, it is well known in literature that at the very beginning, from the time of reheating the universe had radiation domination followed by dark matter domination and dark energy domination \cite{Ryden:1970vsj,Mukhanov:2005sc}. In \cite{Luciano:2022hhy}, a model where interaction between dark matter and dark energy have been considered,  where the radiation field was untouched by any interaction. There are also other studies that generalise the analysis of Barrow holographic dark energy beyond Einstein gravity, like the reconstructed $f(R)$, $f(R,T)$, $f(Q,T)$ and $f(Q,C)$ gravity etc. \cite{Ens:2024zzs,Devi:2024gcr,Myrzakulov:2024jvg,Sharma:2022dzc,Samaddar:2024uwx,Yang:2024tkw,Wang:2020xwn}. also holographic inflation was studied in \cite{Nojiri:2019kkp,
Nojiri:2020wmh}. In this article, we generalise the interaction where radiation and dark matter continuously convert to dark energy. The rate of conversion for matter and radiation can be the same or different depending on the choice of the parameter $\alpha$  (defined in the following sections) whether it is equal to one or not. Besides doing the analytical calculations for deriving the evolution equation for different energy density parameters, we have also ran Markov-Chain Monte Carlo (MCMC) on several datasets like Cosmic Chronometer data, Baryon Acoustic Oscillator (BAO) data, Pantheon+ data, etc., to constrain different free parameters like the present Hubble constant, different density parameters at the current time, the interaction strength, etc., which appears in our theoretical model. \\
This article is organised as follows. At first, we have given a brief review of the FLRW universe in the presence of spatial curvature and different energy content in Section (\ref{review}). In section (\ref{3}), we have shown that the dynamical relation emerges due to the consideration of interacting barrow holographic dark energy with matter and radiation in a closed universe. Here, we have considered four different phenomenological interaction models for both the closed and open universe.  In section (\ref{close universe}), we have done a similar kind of analysis as (\ref{3}) for an open universe with a constant negative curvature. Section \eqref{sec:5} discusses about the numerical solutions and their graphical representations of different evolution equations from section \eqref{3} and section \eqref{close universe}. We have also shown the numerical solutions of the dark energy equation of state (EOS) parameter corresponding to different interaction models and for both open and closed universe scenario. In section (\ref{Observational Constraints}), we have obtained various observational constraints in our model using different cosmological datasets like the CC \& BAO data and Pantheon+ \& SH0ES data sets. We finally conclude our findings in the section (\ref{Conclusion}). An Appendix section \eqref{appendix}, is also added to provide the evolution equations of dark energy, dark matter and radiation density parameters along with dark energy EOS parameter for two component liner interaction terms.

\section{Interacting Barrow holographic dark energy for non-zero spatial curvature}\label{review}
In this section, we will give a short review of the IBHDE model for a non-flat universe (the spatial curvature is non-zero).\\
We shall start with the Friedmann-Lemaitre-Robertson-Walker (FLRW)\cite{Friedman:1922kd,Friedmann:1924bb,Lemaitre:1931zza,Lemaitre:1933gd,Robertson:1935jpx,Robertson:1935zz,Walker:1937qxv} line element for the non-flat case in $(3+1)$-dimensions, which is given in the radial coordinate system as 
\begin{equation}
    ds^2 =-dt^2 +a(t)^2 \Bigg[\frac{dr^2}{1-kr^2}+r^2 d\Omega^2_{2}\Bigg]
\end{equation}
where $k$ is the spatial curvature, $\Omega_{2}$ is the volume of a two sphere and $a(t)$ is the scale factor. $k$ can have three values, +1,0,-1, depending upon which one can determine whether the spacetime is closed, flat or open. Using the metric, one can calculate the Einstein metric $G_{\mu\nu}$ in a straightforward way. If we consider the flow of the universe as a flow of perfect fluid, then the stress energy tensor associated with it can be written as
\begin{equation}
    T_{\mu\nu}=\left(\rho+p\right)u_\mu u_\nu+pg_{\mu\nu}
\end{equation}
where $\rho$ and $p$ are the total energy density and total pressure of the fluid. $u_\mu$ is the four velocity of the universe, in the comoving frame, which has the components as $\{1,0,0,0\}$
Using the Einstein equation
\begin{equation}
    G_{\mu\nu}=8\pi G T_{\mu\nu}~,
\end{equation}
one arrived at the Friedmann equations
\begin{align}
    3H^2+\frac{3k}{a^2}=\frac{\rho}{M_P^2} \nonumber \\
    \frac{\ddot a}{a}=-\frac{1}{6M_P^2}\left(3p+\rho\right)
\end{align}
where $H=\frac{\dot a}{a}$ is the Hubble parameter and $M_P$ is the Planck mass.
If we consider the energy density that comes from different components, then the vanishing divergence of the total  stress energy tensor reads
\begin{align}\label{traceless}
    \nabla_\mu T^{\mu\nu}_{Total}=0 \nonumber \\
    \implies\nabla_\mu\sum_iT^{\mu\nu}_{i}=0~.
\end{align}
If there is no coupling between the components of the total stress tensor, then each divergence of the stress energy tensor independently becomes zero. However, the true behaviour is still unclear. Transformation from radiation to matter and vice versa has experimentally been observed (for example, creation and annihilation of particle and antiparticle). So, considering the presence of interaction between dark energy, matter, and radiation is a well-motivated and mathematically more generalised version. Some studies have considered such an interaction.\cite{Saridakis:2020zol,Luciano:2022hhy,Nojiri:2021jxf,Anagnostopoulos:2020ctz,Adhikary:2021xym,Adhikary:2024sax}\\
From eq.(\ref{traceless}) one can directly writes considering the interaction 
\begin{equation}\label{interaction}
    \dot \rho_i+3H\left(\rho_i +p_i\right)=\mathcal{Q}_i~,
\end{equation}
with 
\begin{equation}\label{Q Sum zero}
    \sum_iQ_i=0~.
\end{equation}
The sign of the $\mathcal{Q}_i$ in the $i^{th}$ field determines whether energy is transferring from another field to that field or the opposite. 
Now finding the form of the $\mathcal{Q}$'s from the action level is very difficult (although there has been such attempts\cite{Wang:2016lxa})and suffers from several problems like fine tunning.\cite{Costa:2016tpb}. That is why in most of the studies have been done choosing the form of $\mathcal{Q}_i$ in a phenomenological way. From eq.(\ref{interaction}) we can clearly see the dimension of  $\mathcal{Q}$ is same as $\dot\rho $ that is [energy density][time]$^{-1}$. So the simplest form of $\mathcal{Q}$ should be $\mathcal{Q}\propto \frac{\rho}{t}$. Again in cosmology we have better time measure as $\frac{1}{H}$. So we will choose $\mathcal{Q}_i=-\Gamma H \rho_i$ \footnote{In this current manuscript we have only taken $\mathcal{Q}\sim \rho_{de}$,$\rho_m$, $\rho_r$,$\paren{\rho_{de}+\rho_m+\rho_r}$ the other possible linear forms of $\mathcal{Q}$ can be proportional to  $\paren{\rho_{de}+\rho_m}$, $\paren{\rho_{r}+\rho_m}$, $\paren{\rho_{r}+\rho_{de}}$ which we have shown only the calculation in the appendix also non linear terms can be considered like $\mathcal{Q}\sim \frac{\rho_{de}^2}{H^2}$ etc.}, where $\Gamma$ determines the interaction strength and also higher power nonlinear interaction can be taken  $\rho_i$ can be a single component as well as a multi-component. The exact composition of the total energy density is still unclear, although observationally dark matter and radiation have been seen, there have been theoretical predictions about the existence of dark energy and stiff matter too. In previous studies, mostly the interaction between two energy components has been considered. In this article, we are going to take multi-component interactions because the physical universe is not made by only two component. The possible multi-component interaction is the interaction between radiation, matter and dark energy. As from the observation, it is almost established that the very late time evolution of the universe is driven by the dark energy sector, so we will consider throughout this article the continuous decay of dark matter and radiation to dark energy for both open and close universes.
\section{Closed Universe}\label{3}
In this section, we will derive the evolution equation corresponding to the energy density parameter of dark energy, dark matter and radiation for a closed universe with positive curvature ($k=1$). In order to derive the evolution equations, we have considered four different kinds of interactions, that is, interaction only depending on dark energy, dark matter, radiation and the sum of all of them.
The definition of the future event horizon is 
\begin{equation}\label{horizon1}
    R_h\equiv  a(t)\int_t^\infty\frac{d\tau}{a(\tau)}=a(t)\int_{a(t)}^\infty\frac{d\tilde{a}}{\dot{a}a}=\int_{a(t)}^\infty\frac{d\tilde{a}}{\tilde{a}^2H}
\end{equation}
Again for radial null geodesic we have 
\begin{equation}
    \frac{dt}{a(t)}=\frac{dr}{1-r^2}
\end{equation}
So we get,
\begin{equation}\label{horizon2}
R_h=a(t)\int_0^{r_h}\frac{d\tilde{r}}{1-\tilde{r}^2}=a(t)\sin^{-1}{r_h}
\end{equation}
Using eq.(\ref{horizon1}) and eq.(\ref{horizon2}) we can easily write 
\begin{equation}
    r_h= \sin{\paren{\frac{R_h}{a}}}=\sin{\paren{\int_{a(t)}^\infty\frac{d\tilde{a}}{\tilde{a}^2H}}}=\sin y
\end{equation}
where $y=\int_{a(t)}^\infty\frac{d\tilde{a}}{\tilde{a}^2H}$.\\
Taking the definition of horizon length scale $L\equiv a(t)r_h$,  one gets
\begin{equation}\label{Length Scale}
    L=a\sin{\int_a^\infty\frac{da}{a^2H}}=a\sin y~.
\end{equation}
The Friedmann equation for the closed universe can be written in terms of the energy density parameters as follows
\begin{equation}
    \frac{1}{H^2 a^2}=\ode+\om+\ol-1~.
\end{equation}
Now, taking the ratio of $\frac{1}{H^2 a^2}$ with its present value $\frac{1}{H_0^2 a_0^2}$, one gets
\begin{equation}
    \frac{H_0^2 a_{0}^2}{H^2 a^2}=\frac{\ode+\om+\ol-1}{\Omega_{de,0}+\Omega_{m,0}+\Omega_{r,0}-1}~.
\end{equation}
After a little bit of simplification, we get an expression for $\rho_{de}$, which is given by
\begin{equation}
    \rho_{de}=\frac{(1+z)^2 (H_{0}^2 \Omega_{k0})\ode}{c^2(\ode+\om+\ol-1)}
\end{equation}
where we have used the relation $H^2 =\frac{\rho_{de}}{3M^2_{p}\ode}$. Therefore to get the expression of length scale $L$, we can use the relation $\rho_{de}=3c^2 M^2_{p}L^{\Delta-2}$, this finally gives 
\begin{equation}\label{Length scale k plus}
    L=\Big[\frac{\ode (H^2_{0}\Omega_{k0})(1+z)^2}{c^2 (\ode+\om+\ol-1)}\Big]^{\frac{1}{\Delta-2}}~.
\end{equation}
This expression of $L$ will be later useful while obtaining the numerical solution for various energy density parameters and finding various observational constraints in our model.
\subsection{Interaction Depends on Dark Energy Density Only}
Let us begin with the Friedmann equations in eq.\eqref{interaction} where the interaction solely depends on dark energy density $\rho_{de}$.
Therefore, taking $\mathcal{Q}_m=-\Gamma H \rho_{de}$, $\mathcal{Q}_r=-\Gamma \alpha H \rho_{de}$ in eq.(\ref{interaction}) and eq.(\ref{Q Sum zero}), leads to the following set of equations.
\begin{equation}\label{matter equation}
    \dot \rho_{m}+3H\rho_{m} =\mathcal{Q}_{m}=-\Gamma H\rho_{de}~,
\end{equation}
\begin{equation}\label{radiation equation}
     \dot \rho_{r}+3H \left(\rho_{r} +p_{r}\right)=\mathcal{Q}_{r}=- \Gamma \alpha H \rho_{de}~,
\end{equation}
\begin{equation}\label{dark energy equation}
    \dot \rho_{de}+3H\left(\rho_{de} +p_{de}\right)=\mathcal{Q}_{de}=\Gamma \left(1+\alpha\right)\rho_{de}~,
\end{equation}
We will see in a while that this set of three equations will be useful to derive the evolution equations for dark energy, dark matter and radiation. For convenience, we will do a change of variable as $x=ln(a)$. This change of variable will shift the dynamical equations of different energy densities from time derivative to derivative with respect to $x$ and eventually with respect to the redshift parameter. Performing this transformation the above set of eq.(s)(\ref{matter equation}-\ref{dark energy equation}) can be recast in the following forms
\begin{eqnarray}\label{differential equation of energy density}
        \frac{\rho_{m}'}{\rho_m}=-\paren{3+\frac{\Gamma}{r_1}}~\label{m},\\
       \frac{\rho_{r}'}{\rho_r}=-\paren{4+\frac{\Gamma\alpha}{r_2}}~,\label{r}  \\
        \frac{\rho_{de}'}{\rho_{de}}=-\paren{3\left(1 +\omega_{de}\right)+\Gamma \left(1+\alpha\right)}~,\label{de}
\end{eqnarray}
where we have taken $p_r=\frac{1}{3}\rho_r$, $p_{de}=\omega_{de}\rho_{de}$, $p_{m}=0$ and the prime($'$) denotes derivative taken with respect to $x$. In the above equations $r_1 =\frac{\rho_m}{\rho_{de}}$ and $r_2 =\frac{\rho_r}{\rho_{de}}$. It is to be noted that the dark energy equation of state parameter is no more taken to be $-1$, the main reason is the presence of the phenomenological interaction terms $\mathcal{Q}_{de}$, $\mathcal{Q}_{m}$ and $\mathcal{Q}_{r}$ in the right hand side of the Friedmann equations in eq.(s)(\ref{matter equation}-\ref{dark energy equation}). Although one can choose all the equations of state parameters corresponding to dark energy, matter and radiation to be unknown, in order to make our calculations simpler, we have chosen only the equation of state parameter of dark energy ($\omega_{de}$) to be unknown. We have also followed this assumption throughout the paper.\\
Now we will proceed further to find the evolution equation for the dark energy density parameter $\ode=\frac{\rho_{de}}{3H^2 M^2_{p}}$. The main reason is to shifting from energy density to energy density parameter lies in the fact that different observational data sets provide the energy density parameter instead of the true energy density.
Using the Friedmann equation, one can write the expression of $\ode$ in the following manner 
\begin{equation}
    \ode=\frac{\rho_{de}}{\rho_{r}+\rho_{m}+\rho_{k}+\rho_{de}}
\end{equation}
where we have defined $\rho_k:=-\frac{3}{a^2}$.
Then the above equation implies ,
\begin{equation}\label{Omega_de}
    \frac{\ode}{1-\ode}=\frac{\rho_{de}}{\rho_{m}+\rho_{k}+\rho_{r}}~.
\end{equation}
Using the expression of $\rho_{de}$ from the above equation and inserting this into the eq.(\ref{de}) one gets,
\begin{equation}\label{d log main positive}
    \frac{d}{dx}\ln{\left[\paren{\rho_{m}+\rho_{k}+\rho_{r}}\frac{\ode}{1-\ode}\right]}=-\paren{3\left(1 +\omega_{de}\right)+\Gamma \left(1+\alpha\right)}~.
\end{equation}
Before moving further we need to first evaluate the equation state parameter for the dark energy ($\omega_{de}$).
In order to get the functional form of the equation of state parameter $\omega_{de}$ in terms of observable parameters, we can use eq.\eqref{dark energy equation}. 
From eq.(\ref{dark energy density}), we can get an expression for the time derivative of the dark energy density, which reads
\begin{align}\label{dark energy density time derivative}
    \dot\rho_{de}&=\rho_{de}\paren{\Delta-2}\frac{\dot L}{L}
\end{align}
Now we have to calculate $\frac{\dot L}{L}$, which can be written down using eq.(\ref{Length Scale}) as 
\begin{equation}
    \frac{\dot L}{L}=H\paren{1-\sqrt{\frac{3M_P^2\ode}{C}}L^{-\frac{\Delta}{2}}\cos{y}}
\end{equation}
Hence, combining the above expression for $\frac{\dot L}{L}$ with eq.(\ref{dark energy density time derivative}) gives
\begin{equation}\label{dark matter equation 2}
    \dot\rho_{de}= \rho_{de}\paren{\Delta-2}H\paren{1-\sqrt{\frac{3M_P^2\ode}{C}}L^{-\frac{\Delta}{2}}\cos{y}}
\end{equation}
It is now an easy task to write down the form of $\omega_{de}$. Substituting the expression of $\dot\rho_{de}$ and $\rho_{de}$ from eq.(s)(\eqref{dark matter equation 2},\eqref{dark energy density}) in eq.(\ref{dark energy equation}) one can read of the expression for $\omega_{de}$, which reads
\begin{equation}\label{omega de positive}
    \omega_{de}=\frac{\paren{1+\alpha}\Gamma}{3}-\Big(\frac{1+\Delta}{3}\Big)+\Big(\frac{\Delta-2}{3}\Big)\sqrt{\frac{3M_{p}^2 \Omega_{de}}{C}}L^{-\frac{\Delta}{2}}\cos{y}~.
\end{equation}
Finally, using eq.(s)(\eqref{differential equation of energy density}) along with eq.\eqref{omega de positive} in eq.\eqref{d log main positive} gives the expression for the evolution equation of the energy density parameter $\ode$ for dark energy. This reads
\begin{align}\label{ode final eq}
    \frac{\ode '}{\ode(1-\ode)}&=-\frac{1}{1-\ode}\left[2\paren{\ode+\om+\ol-1}-\ol\paren{\frac{\Gamma\alpha}{r_2}+4}-\om\paren{\frac{\Gamma}{r_1}+3}\right]  \\& \nonumber
    +(\Delta-2)\Big(1-\sqrt{\frac{3M_{p}^2 \Omega_{de}}{C}}L^{-\frac{-\Delta}{2}}\cos{y}\Big)~.
    \end{align} 
    From the above equation, it is clear that the differential equation of $\ode$ depends on $\om$ and $\ol$, therefore, to get any solution we at least need another two differential equations of $\om$ and $\ol$. Hence, we will now proceed further to derive the evaluation equation of $\om$. In order to do that, we will substitute $\rho_m=3H^2\om M_P^2$ in eq.(\ref{differential equation of energy density}), this leads us to the following equation 
    \begin{equation}\label{om eval initial}
        \frac{\om'}{\om}+2\frac{H'}{H}=-\paren{\frac{\Gamma}{r_1}+3}~.
    \end{equation}
    Now the main task is to get an expression of $\frac{H^{\prime}}{H}$ and substitute it back into the above equation to get the required evolution equation for $\om$. In order to calculate the expression for $\frac{H^{\prime}}{H}$, we will start with the relation $H^2 =\frac{\rho_{de}}{3M^2_{p}\ode}$. Now, differentiating this expression with respect to $x$ and deciding it with $H^2$, we get
    \begin{equation}
        2\frac{H^{\prime}}{H}=\frac{\rho^{\prime}_{de}}{\rho_{de}}-\frac{\ode^{\prime}}{\ode}~.
    \end{equation}
    Now substituting the expression of $\frac{\rho^{\prime}_{de}}{\rho_{de}}$ and $\frac{\ode^{\prime}}{\ode}$ from eq.(s)(\eqref{de},\eqref{ode final eq}), the above equation gives the following expression for $\frac{H'}{H}$
\begin{align}\label{H prime by H}
        2\frac{H'}{H}=&\ode\paren{\Delta-2}\paren{1-\sqrt{\frac{3M_P^2\ode}{C}}L^{-\frac{\Delta}{2}}\cos{y}}\\&\nonumber+\left[2\paren{\ode+\om+\ol-1}-\ol\paren{\frac{\Gamma\alpha}{r_2}+4}-\om\paren{\frac{\Gamma}{r_1}+3}\right]~.
\end{align}
Using the above expression of $\frac{H'}{H}$ in eq.\eqref{om eval initial}, we get
\begin{align}
    \frac{\om'}{\om}&=-\paren{\frac{\Gamma}{r_1}+3}-\ode\paren{\Delta-2}\paren{1-\sqrt{\frac{3M_P^2\ode}{C}}L^{-\frac{\Delta}{2}}\cos{y}}\\&\nonumber
    -\left[2\paren{\ode+\om+\ol-1}-\ol\paren{\frac{\Gamma\alpha}{r_2}+4}-\om\paren{\frac{\Gamma}{r_1}+3}\right]   
\end{align}
The above expression takes care of the evaluation of $\om$ with respect to $x=\ln{a}$, therefore with respect to the redshift parameter $z$. Now, in a similar fashion, we can derive the evaluation for $\ol$. At first, we will substitute $\rho_{r}=3H^2 M^2_{p}\ol$ in eq.\eqref{r}, this gives the following relation
\begin{equation}
    \frac{\ol'}{\ol}+2\frac{H'}{H}=-\paren{\frac{\Gamma\alpha}{r_2}+4}~.
\end{equation}
Again, substituting the value of $\frac{H'}{H}$ from eq.\eqref{H prime by H} in the above equation and simplifying, one gets the evaluation equation for $\ol$, which is given as follows
\begin{align}
    \frac{\ol'}{\ol}&=-\paren{\frac{\Gamma\alpha}{r_2}+4}-\ode\paren{\Delta-2}\paren{1-\sqrt{\frac{3M_P^2\ode}{C}}L^{-\frac{\Delta}{2}}\cos{y}}\\&\nonumber
    -\left[2\paren{\ode+\om+\ol-1}-\ol\paren{\frac{\Gamma\alpha}{r_2}+4}-\om\paren{\frac{\Gamma}{r_1}+3}\right]
\end{align}
Now with all three evolution equations for $\ode$, $\om$ and $\ol$ in hand, we can numerically solve the three coupled differential equations for different values of redshift. We would like to say that we have used the $RK4/5$ method of the python scipy library to solve these equations numerically. We have then plotted the evaluation of the energy density parameters with respect to the redshift parameter $z$.\\
\subsection{Interaction Depends on Matter Density Only}
If we consider the interaction terms depend on dark matter density, then we get a set of equations following eq.(\ref{interaction}) and eq.(\ref{Q Sum zero}) ,
\begin{eqnarray}
    \dot \rho_{m}+3H\rho_{m} =-\Gamma H\rho_{m}~,\\
\dot \rho_{r}+3H \left(\rho_{r} +p_{r}\right)=- \Gamma \alpha H \rho_{m}~,\\
\dot \rho_{de}+3H\left(\rho_{de} +p_{de}\right)=\Gamma \left(1+\alpha\right)\rho_{m}\label{diff rho-de Q rho-m closed}~,
\end{eqnarray}
Now, defining a new variable $x=\ln{a}$ as before, we can recast the above equations in the following form
\begin{eqnarray}
    \frac{d}{dx}\paren{\ln{\rho_m}}=-\paren{\Gamma+3}\label{ddx rho-m Q rho-m closed}\\
    \frac{d}{dx}\paren{\ln{\rho_r}}=-\paren{\Gamma\alpha\frac{r_1}{r_2}+4}\label{ddx rho-r Q rho-m closed}\\
    \frac{d}{dx}\paren{\ln{\rho_{de}}}=\Gamma r_1\paren{\alpha+1}-3\paren{1+\omega_{de}}~.\label{ddx rho-de Q rho-m closed}
\end{eqnarray}
Using eq.(\ref{Omega_de}) the above equation can be transformed in to a differential equation of dark energy energy density parameter, which reads 
\begin{equation}\label{frac ode Q prop to rho-m}    
    \frac{\ode'}{1-\ode}+\frac{1}{\rho_k+\rho_m+\rho_r}\frac{d}{dx}\paren{\rho_k+\rho_m+\rho_r}=\Gamma r_1\paren{\alpha+1}-3\paren{1+\omega_{de}}~.
\end{equation}
In the above equation, the dark energy EOS parameter, $\omega_{de}$, can be calculated by substituting the expression of $\dot \rho_{de}$ and $\rho_{de}$ in eq.\eqref{diff rho-de Q rho-m closed}. The expression of $\dot \rho_{de}$ can be calculated using eq.(s)(\eqref{dark energy density},\eqref{Length Scale}), then putting back this value in eq.\eqref{diff rho-de Q rho-m closed} gives the following relation for $\omega_{de}$
\begin{equation}
    \omega_{de}=\frac{\Gamma r_1}{3}\paren{1+\alpha}-\frac{\Delta+1}{3}+\frac{\Delta-2}{3}\sqrt{\frac{3M_P^2\ode}{C}}L^{-\frac{\Delta}{2}}\cos{y}
\end{equation}
Therefore, using this relation in eq.\eqref{frac ode Q prop to rho-m} along with eq.(s)(\eqref{ddx rho-m Q rho-m closed},\eqref{ddx rho-r Q rho-m closed},\eqref{ddx rho-de Q rho-m closed}) gives the coupled differential equation for $\ode$, which reads 
    \begin{align}
    \frac{\Omega^{\prime}_{de}}{\Omega_{de}(1-\Omega_{de})}&=(\Delta-2)\Big(1-\sqrt{\frac{3M_{p}^2 \Omega_{de}}{C}}L^{-\frac{\Delta}{2}}\cos{y}\Big)\nonumber\\&-\frac{1}{(1-\Omega_{de})}\Bigg[2(\Omega_{de}+\Omega_{m}+\Omega_{r}-1)-\Omega_{m}(\Gamma+3)-\Omega_{r}\paren{\Gamma\alpha\frac{r_1}{r_2} +4}\Bigg]
\end{align}
In a similar fashion to subsection (3.1), the above equation can be used to obtain differential equations corresponding to $\om$ and $\ol$ respectively, which reads
\begin{align}
    \frac{\om^{\prime}}{\om}&=-\Big(\Gamma+3\Big)-(\Delta-2)\Big(1-\sqrt{\frac{3M_{p}^2 \Omega_{de}}{C}}L^{\frac{-\Delta}{2}}\cos{y}\Big)\ode\nonumber\\&-\Bigg[2(\Omega_{de}+\Omega_{m}+\Omega_{r}-1)-\Omega_{m}\Big(\Gamma+3\Big)-\Omega_{r}(\Gamma\alpha\frac{r_1}{r_2} +4)\Bigg]
\end{align}
\begin{align}
    \frac{\ol^{\prime}}{\ol}&=-\Big(\Gamma\alpha\frac{r_1}{r_2}+4\Big)-(\Delta-2)\Big(1-\sqrt{\frac{3M_{p}^2 \Omega_{de}}{C}}L^{-\frac{\Delta}{2}}\cos{y}\Big)\ode\nonumber\\&-\Bigg[2(\Omega_{de}+\Omega_{m}+\Omega_{r}-1)-\Omega_{m}\Big(\Gamma\frac{r_2}{r_1}+3\Big)-\Omega_{r}(\Gamma\alpha\frac{r_1}{r_2} +4)\Bigg]
\end{align}
With all these coupled differential equations in hand, we have performed numerical solution to find the evolution profile of $\ode$, $\om$ and $\ol$ with respect to redshift parameter ($z$), this is shown in Fig.\eqref{fig:evolution matter}. The numerical solution for the EOS parameter with respect to the redshift are shown in Fig.\eqref{fig:eos_m}.
\subsection{Interaction Depends on Radiation Density}
The Freedman equations corresponding to the interaction $\mathcal{Q}=-\Gamma H \rho_r$ are given by
\begin{eqnarray}
    \dot \rho_{de} + 3 (1+\omega_{de})H \rho_{de}=\Gamma (1+\alpha)r_2 H \rho_{r}=\Gamma (1+\alpha)r_2 H \rho_{de}\label{diff rho-de Q rho-r closed}\\
    \dot \rho_{m} + 3 H\rho_{m}=-\Gamma H \rho_{r}=-\Gamma r_2 H \rho_{de}\label{diff rho-m Q rho-r closed}\\
    \dot \rho_{r} + 4 H\rho_{r}=-\Gamma \alpha H \rho_{r}=-\Gamma \alpha r_2 H \rho_{de}\label{diff rho-r Q rho-r closed}
\end{eqnarray}
The equation of state parameter of dark energy is given by
\begin{equation}
    \omega_{de}=\frac{\Gamma(1+\alpha)r_2}{3}-\Big(\frac{\Delta+1}{3}\Big)+\Big(\frac{\Delta-2}{3}\Big)cos y \sqrt{\frac{3M_{p}^2 \Omega_{de}}{C}}L^{\frac{-\Delta}{2}}
\end{equation}
The equation describing the dynamics of the dark energy density parameter is governed by the following expression
\begin{align}
    \frac{\Omega^{\prime}_{de}}{\Omega_{de}(1-\Omega_{de})}&=(\Delta-2)\Big(1-\sqrt{\frac{3M_{p}^2 \Omega_{de}}{C}}L^{\frac{-\Delta}{2}}\cos{y}\Big)\nonumber\\&-\frac{1}{(1-\Omega_{de})}\Bigg[2(\Omega_{de}+\Omega_{m}+\Omega_{r}-1)-\Omega_{m}(\Gamma\frac{r_2}{r_1}+3)-\Omega_{r}(\Gamma\alpha +4)\Bigg]
\end{align}
Now we will proceed to derive the evaluation equation for the dark matter density parameter ($\om$). 
\begin{align}
    &\frac{d}{dx}\ln(3M^2_{p}H^2 \om)=-\Big(\Gamma\frac{r_2}{r_1}+3\Big)\nonumber\\
    & \implies \frac{2H^{\prime}}{H}+\frac{\om^{\prime}}{\om}=-\Big(\Gamma\frac{r_2}{r_1}+3\Big)
\end{align}
We will then use the following relation to obtain an expression for $\frac{2H^{\prime}}{H}$
\begin{equation}
    H^2=\frac{\rho_{de}}{3M^2_{p}\ode}~.
\end{equation}
Taking the derivative of the above equation with respect to $x=\ln{a}$ implies
\begin{align}
    \frac{2H^{\prime}}{H}&=\frac{\rho^{\prime}_{de}}{\rho_{de}}-\frac{\ode^{\prime}}{\ode}\nonumber\\&=\Gamma(1+\alpha)-3(1+\omega_{de})-\frac{\ode^{\prime}}{\ode}
\end{align}
The above equation can be written in the following form
\begin{align}
    \frac{2H^{\prime}}{H}&=(\Delta-2)\Big(1-\sqrt{\frac{3M_{p}^2 \Omega_{de}}{C}}L^{\frac{-\Delta}{2}}\cos{y}\Big)\ode\nonumber\\&+\Bigg[2(\Omega_{de}+\Omega_{m}+\Omega_{r}-1)-\Omega_{m}(\Gamma\frac{r_2}{r_1}+3)-\Omega_{r}(\Gamma\alpha +4)\Bigg]
\end{align}
The expression for the evolution of the dark matter density parameter is given by
\begin{align}
    \frac{\om^{\prime}}{\om}&=-\Big(\Gamma\frac{r_2}{r_1}+3\Big)-(\Delta-2)\Big(1-\sqrt{\frac{3M_{p}^2 \Omega_{de}}{C}}L^{\frac{-\Delta}{2}}\cos{y}\Big)\ode\nonumber\\&-\Bigg[2(\Omega_{de}+\Omega_{m}+\Omega_{r}-1)-\Omega_{m}\Big(\Gamma\frac{r_2}{r_1}+3\Big)-\Omega_{r}(\Gamma\alpha +4)\Bigg]
\end{align}
The evolution equation for the radiation density parameter is given by 
\begin{align}
    \frac{\ol^{\prime}}{\ol}&=-\Big(\Gamma\alpha+4\Big)-(\Delta-2)\Big(1-\sqrt{\frac{3M_{p}^2 \Omega_{de}}{C}}L^{\frac{-\Delta}{2}}\cos{y}\Big)\ode\nonumber\\&-\Bigg[2(\Omega_{de}+\Omega_{m}+\Omega_{r}-1)-\Omega_{m}\Big(\Gamma\frac{r_2}{r_1}+3\Big)-\Omega_{r}(\Gamma\alpha +4)\Bigg]
\end{align}
The numerical solutions for these differential equations are displayed in Fig. \eqref{fig:radiation}. Additionally, Fig. \eqref{fig:eos_r} depicts how the dark energy equation of state (EOS) parameter varies with redshift for various values of the Barrow exponent ($\Delta$).
\subsection{Interaction depends on Dark Energy, Matter and Radiation density}
In this subsection, we will derive the evolution equation corresponding to the case when the phenomenological interaction term $\mathcal{Q}$ only depends on Dark Energy, Matter and Radiation density. In this scenario, the set of the Friedmann equations is as follows
\begin{eqnarray}
  \dot \rho_{m}+3H\rho_{m} =\mathcal{Q}_{m}=-\Gamma H\paren{\rho_{de}+\rho_{r}+\rho_{m}}=-\Gamma H \rho_{de}\paren{1+r_1+r_2}\label{diff rho-m Q rho-all closed}~\\
  \dot \rho_{r}+3H \left(\rho_{r} +p_{r}\right)=\mathcal{Q}_{r}=- \Gamma \alpha H \paren{\rho_{de}+\rho_{r}+\rho_{m}}=- \Gamma \alpha H\rho_{de}\paren{1+r_1+r_2}\label{diff rho-r Q rho-all closed}~\\
      \dot \rho_{de}+3H\left(\rho_{de} +p_{de}\right)=\mathcal{Q}_{de}=\Gamma \left(1+\alpha\right)\paren{\rho_{de}+\rho_{r}+\rho_{m}}=\Gamma \left(1+\alpha\right)H\rho_{de}\paren{1+r_1+r_2}\label{diff rho-de Q rho-all closed}~.
\end{eqnarray}
Now, following the same procedure as the previous subsections, we get a set of three coupled differential equations corresponding to dark energy, dark matter and radiation energy density parameters, which reads
\begin{align}
    \frac{\ode '}{\ode(1-\ode)}&=-\frac{1}{1-\ode}\left[2\paren{\ode+\om+\ol-1}-\ol\paren{\frac{\Gamma\alpha\paren{1+r_1+r_2}}{r_2}+4}-\right.\\& \nonumber \left.\om\paren{\frac{\Gamma\paren{1+r_1+r_2}}{r_1}+3}\right] 
    +(\Delta-2)\Big(1-\sqrt{\frac{3M_{p}^2 \Omega_{de}}{C}}L^{-\frac{-\Delta}{2}}\cos{y}\Big)
    \end{align}
\begin{align}
    \frac{\om'}{\om}&=-\paren{\frac{\Gamma\paren{1+r_1+r_2}}{r_1}+3}
    -\left[2\paren{\ode+\om+\ol-1}-\ol\paren{\frac{\alpha\Gamma\paren{1+r_1+r_2}}{r_2}+4}\right.\\&\nonumber \left. -\om\paren{\frac{\Gamma\paren{1+r_1+r_2}}{r_1}+3}\right]-\ode\paren{\Delta-2}\paren{1-\sqrt{\frac{3M_P^2\ode}{C}}L^{-\frac{\Delta}{2}}\cos{y}}
\end{align}
\begin{align}
    \frac{\ol'}{\ol}&=-\paren{\frac{\alpha\Gamma\paren{1+r_1+r_2}}{r_2}+4} -\left[2\paren{\ode+\om+\ol-1}-\ol\paren{\frac{\alpha\Gamma\paren{1+r_1+r_2}}{r_2}+4}\right.\\&\nonumber\left.-\om\paren{\frac{\Gamma\paren{1+r_1+r_2}}{r_1}+3}\right]-\ode\paren{\Delta-2}\paren{1-\sqrt{\frac{3M_P^2\ode}{C}}L^{-\frac{\Delta}{2}}\cos{y}}
\end{align}
Also the equation of state parameter for dark energy reads,
\begin{equation}
    \omega_{de}=\frac{\Gamma\paren{1+\alpha}\paren{1+r_1+r_2}}{3}-\frac{\Delta+1}{3}+\frac{\Delta-2}{3}\sqrt{\frac{3M_P^2\ode}{C}}L^{-\frac{\Delta}{2}}\cos{y}
\end{equation}
The numerical solution for the evolution of all these energy density parameters along with the EOS parameter are discussed later in section 5.
\section{Open Universe}\label{close universe}
In an open universe, the spatial curvature constant (k) is negative one. Taking this into consideration, we have performed a similar calculation for the cosmic length scale like sec.(\ref{3}), which reads 
\begin{equation}
    L=ar_H=a\sinh{y}=a\sinh{\int_a^\infty\frac{da}{a^2H}}~.
\end{equation}
For $k=-1$, the Friedmann equation gives the following relation
\begin{equation}
    \frac{1}{H^2 a^2}=1-\ode-\om-\ol~.
\end{equation}
Now, taking the ratio of $\frac{1}{H^2 a^2}$ with its present value $\frac{1}{H_0^2 a_0^2}$, we obtain an expression for $\rho_{de}$ and using $\rho_{de}=CL^{\Delta-2}$, we get
\begin{equation}\label{Length scale k minus}
    L=\Big[\frac{\ode (H^2_{0}\Omega_{k0})(1+z)^2}{c^2 (\ode+\om+\ol-1)}\Big]^{\frac{1}{\Delta-2}}~.
\end{equation}
One thing to be noted is that although the mathematical expressions of $L$ in eq.\eqref{Length scale k plus} and eq.\eqref{Length scale k minus} are same, the value of $\Omega_{k0}$ is negative in the above equation.
\subsection{Interaction Depends on Dark Energy Density Only}
For a closed universe, the three energy evolution eq.(s)(\ref{matter equation},\ref{radiation equation},\ref{dark energy equation},\ref{m},\ref{r},\ref{de}) does not have any explicit change. However, the definition of $\rho_k$ changes to
\begin{equation}
    \rho_k:=\frac{3}{a^2}~.
\end{equation}
With this definition of $\rho_k$ in the above equation and using the expression of cosmic length scale in eq.\eqref{Length scale k minus}, the expression of $\dot \rho_{de}$ reads 
\begin{equation}
\dot\rho_{de}=\rho_{de}\paren{\Delta-2}H\paren{1-\sqrt{\frac{3M_P^2\ode}{C}}L^{-\frac{\Delta}{2}}\cosh{y}}~.
\end{equation}
Now, using the above expression of $\dot \rho_{de}$ in eq.\eqref{dark energy equation}, one can easily write down the equation of state parameter for dark energy as 
\begin{equation}\label{omegade negative}
    \omega_{de}=\frac{\paren{1+\alpha}\Gamma}{3}-\Big(\frac{1+\Delta}{3}\Big)+\Big(\frac{\Delta-2}{3}\Big)\sqrt{\frac{3M_{p}^2 \Omega_{de}}{C}}L^{-\frac{\Delta}{2}}\cosh{y}~.
\end{equation}
This expression of $\omega_{de}$, along with eq.(s)(\eqref{m},\eqref{r},\eqref{de},\eqref{Omega_de}) gives the differential equation corresponding to dark energy density, which is given as follows
\begin{align}\label{ode final eq open}
    \frac{\ode'}{\ode(1-\ode)}&=-\frac{1}{1-\ode}\left[2\paren{\ode+\om+\ol-1}-\ol\paren{\frac{\Gamma\alpha}{r_2}+4}-\om\paren{\frac{\Gamma}{r_1}+3}\right]  \\& \nonumber
    +(\Delta-2)\Big(1-\sqrt{\frac{3M_{p}^2 \Omega_{de}}{C}}L^{-\frac{-\Delta}{2}}\cosh{y}\Big)~.
    \end{align} 
Now, just like the previous section, we will derive the evolution equation for $\om$. Using equations \eqref{ode final eq open}, we finally get
\begin{align}
    \frac{\om'}{\om}&=-\paren{\frac{\Gamma}{r_1}+3}-\ode\paren{\Delta-2}\paren{1-\sqrt{\frac{3M_P^2\ode}{C}}L^{-\frac{\Delta}{2}}\cosh{y}}\\&\nonumber
    -\left[2\paren{\ode+\om+\ol-1}-\ol\paren{\frac{\Gamma\alpha}{r_2}+4}-\om\paren{\frac{\Gamma}{r_1}+3}\right]   ~.
\end{align}
Similarly, eq.(s)(\eqref{ode final eq open}) can be used to derive the evolution equation for the  radiation energy density parameter, which reads 
\begin{align}
    \frac{\ol'}{\ol}&=-\paren{\frac{\Gamma\alpha}{r_2}+4}-\ode\paren{\Delta-2}\paren{1-\sqrt{\frac{3M_P^2\ode}{C}}L^{-\frac{\Delta}{2}}\cosh{y}}\\&\nonumber
    -\left[2\paren{\ode+\om+\ol-1}-\ol\paren{\frac{\Gamma\alpha}{r_2}+4}-\om\paren{\frac{\Gamma}{r_1}+3}\right]~.
\end{align}
With all of these evaluation equations corresponding to $\ode$, $\om$ and $\ol$ in hand, we have now solved them numerically for different values of the redshift parameters. We have also graphically shown our results in Fig. and Fig..\\
\subsection{Interaction depends on Matter density only}
We will now proceed to derive the evolution equation for energy density parameters in the presence of a phenomenological interaction term depending upon dark matter density ($\mathcal{Q}=\Gamma H \rho_m$) for an open universe. \\
Similar calculation of $\dot \rho_{de}$ to the previous subsections and using its value in eq.\eqref{diff rho-de Q rho-m closed} leads to the following expression of $\omega_{de}$
\begin{equation}
    \omega_{de}=\frac{\Gamma r_1}{3}\paren{1+\alpha}-\frac{\Delta+1}{3}+\frac{\Delta-2}{3}\sqrt{\frac{3M_P^2\ode}{C}}L^{-\frac{\Delta}{2}}\cosh{y}~.
\end{equation}
Now following the same manner as subsection (4.1), we can get the coupled differential equations corresponding to $\ode$, $\om$ and $\ol$, which reads
\begin{align}
    \frac{\Omega^{\prime}_{de}}{\Omega_{de}(1-\Omega_{de})}&=(\Delta-2)\Big(1-\sqrt{\frac{3M_{p}^2 \Omega_{de}}{C}}L^{-\frac{\Delta}{2}}\cosh{y}\Big)\nonumber\\&-\frac{1}{(1-\Omega_{de})}\Bigg[2(\Omega_{de}+\Omega_{m}+\Omega_{r}-1)-\Omega_{m}(\Gamma+3)-\Omega_{r}\paren{\Gamma\alpha\frac{r_1}{r_2} +4}\Bigg]
\end{align}
\begin{align}
    \frac{\om^{\prime}}{\om}&=-\Big(\Gamma+3\Big)-(\Delta-2)\Big(1-\sqrt{\frac{3M_{p}^2 \Omega_{de}}{C}}L^{\frac{-\Delta}{2}}\cosh{y}\Big)\ode\nonumber\\&-\Bigg[2(\Omega_{de}+\Omega_{m}+\Omega_{r}-1)-\Omega_{m}\Big(\Gamma+3\Big)-\Omega_{r}(\Gamma\alpha\frac{r_1}{r_2} +4)\Bigg]
\end{align}
\begin{align}
    \frac{\ol^{\prime}}{\ol}&=-\Big(\Gamma\alpha\frac{r_1}{r_2}+4\Big)-(\Delta-2)\Big(1-\sqrt{\frac{3M_{p}^2 \Omega_{de}}{C}}L^{-\frac{\Delta}{2}}\cosh{y}\Big)\ode\nonumber\\&-\Bigg[2(\Omega_{de}+\Omega_{m}+\Omega_{r}-1)-\Omega_{m}\Big(\Gamma\frac{r_2}{r_1}+3\Big)-\Omega_{r}(\Gamma\alpha\frac{r_1}{r_2} +4)\Bigg]
\end{align}
The numerical solutions for this differential equations are plotted in Fig.\eqref{fig:evolution matter}. Also variation of the dark energy EOS parameter with respect to redshift is graphically shown in Fig.\eqref{fig:eos_m} for different values of the Barrow exponent $\Delta$.
\subsection{Interaction depends on Radiation density only}
For an open universe when the phenomenological interaction term is only proportional to radiation density, we can compute the expression of $\omega_{de}$ using eq.(s)(\eqref{dark energy density},\eqref{Length scale k minus},\eqref{diff rho-de Q rho-r closed}), which reads
\begin{equation}
    \omega_{de}=\frac{\Gamma(1+\alpha)r_2}{3}-\Big(\frac{\Delta+1}{3}\Big)+\Big(\frac{\Delta-2}{3}\Big) \sqrt{\frac{3M_{p}^2 \Omega_{de}}{C}}L^{\frac{-\Delta}{2}}cosh y~.
\end{equation}
Now the above expression of $\omega_{de}$ along with the Friedmann equations in eq.(s)(\eqref{diff rho-de Q rho-r closed},\eqref{diff rho-m Q rho-r closed},\eqref{diff rho-r Q rho-r closed}) gives the evolution equations corresponding to the energy density parameters $\ode$, $\om$ and $\ol$ respectively, and these equations are given as follows
\begin{align}
    \frac{\Omega^{\prime}_{de}}{\Omega_{de}(1-\Omega_{de})}&=(\Delta-2)\Big(1-\sqrt{\frac{3M_{p}^2 \Omega_{de}}{C}}L^{\frac{-\Delta}{2}}\cosh{y}\Big)\nonumber\\&-\frac{1}{(1-\Omega_{de})}\Bigg[2(\Omega_{de}+\Omega_{m}+\Omega_{r}-1)-\Omega_{m}(\Gamma\frac{r_2}{r_1}+3)-\Omega_{r}(\Gamma\alpha +4)\Bigg]~,
\end{align}
\begin{align}
    \frac{\om^{\prime}}{\om}&=-\Big(\Gamma\frac{r_2}{r_1}+3\Big)-(\Delta-2)\Big(1-\sqrt{\frac{3M_{p}^2 \Omega_{de}}{C}}L^{\frac{-\Delta}{2}}\cosh{y}\Big)\ode\nonumber\\&-\Bigg[2(\Omega_{de}+\Omega_{m}+\Omega_{r}-1)-\Omega_{m}\Big(\Gamma\frac{r_2}{r_1}+3\Big)-\Omega_{r}(\Gamma\alpha +4)\Bigg]~,
\end{align}
\begin{align}
    \frac{\ol^{\prime}}{\ol}&=-\Big(\Gamma\alpha+4\Big)-(\Delta-2)\Big(1-\sqrt{\frac{3M_{p}^2 \Omega_{de}}{C}}L^{\frac{-\Delta}{2}}\cosh{y}\Big)\ode\nonumber\\&-\Bigg[2(\Omega_{de}+\Omega_{m}+\Omega_{r}-1)-\Omega_{m}\Big(\Gamma\frac{r_2}{r_1}+3\Big)-\Omega_{r}(\Gamma\alpha +4)\Bigg]~.
\end{align}
We have plotted the numerical solutions for these differential equations in Fig. \eqref{fig:radiation}. Figure \eqref{fig:eos_r} also illustrates the dark energy EOS parameter's variation with redshift, considering different Barrow exponent ($\Delta$) values.
\subsection{Interaction depends on Dark Energy, Matter and Radiation density}
Here, for an open universe when the interaction term is proportional to sum of $\rho_{de}$, $\rho_{m}$ and $\rho_{r}$, the dark energy EOS parameter is given by 
\begin{equation}
    \omega_{de}=\frac{\Gamma\paren{1+\alpha}\paren{1+r_1+r_2}}{3}-\frac{\Delta+1}{3}+\frac{\Delta-2}{3}\sqrt{\frac{3M_P^2\ode}{C}}L^{-\frac{\Delta}{2}}\cosh{y}~.
\end{equation}
Like in the previous subsections, the above expression can be derived using eq.(s)(\eqref{dark energy density},\eqref{Length scale k minus},\eqref{diff rho-de Q rho-all closed}). This expression of $\omega_{de}$, further helps to derive the evolution equations for various energy density parameters. To do so, we use the Friedmann equations in eq.(s)(\eqref{diff rho-de Q rho-all closed},\eqref{diff rho-m Q rho-all closed},\eqref{diff rho-r Q rho-all closed}), this gives the following set of coupled differential equations corresponding to $\ode$, $\om$ and $\ol$, which reads
\begin{align}
    \frac{\ode '}{\ode(1-\ode)}&=-\frac{1}{1-\ode}\left[2\paren{\ode+\om+\ol-1}-\ol\paren{\frac{\Gamma\alpha\paren{1+r_1+r_2}}{r_2}+4}-\right.\\& \nonumber \left.\om\paren{\frac{\Gamma\paren{1+r_1+r_2}}{r_1}+3}\right] 
    +(\Delta-2)\Big(1-\sqrt{\frac{3M_{p}^2 \Omega_{de}}{C}}L^{-\frac{-\Delta}{2}}\cosh{y}\Big)~.
    \end{align}
\begin{align}
    \frac{\om'}{\om}&=-\paren{\frac{\Gamma\paren{1+r_1+r_2}}{r_1}+3}
    -\left[2\paren{\ode+\om+\ol-1}-\ol\paren{\frac{\alpha\Gamma\paren{1+r_1+r_2}}{r_2}+4}\right.\\&\nonumber \left. -\om\paren{\frac{\Gamma\paren{1+r_1+r_2}}{r_1}+3}\right]-\ode\paren{\Delta-2}\paren{1-\sqrt{\frac{3M_P^2\ode}{C}}L^{-\frac{\Delta}{2}}\cosh{y}}~.
\end{align}
\begin{align}
    \frac{\ol'}{\ol}&=-\paren{\frac{\alpha\Gamma\paren{1+r_1+r_2}}{r_2}+4} -\left[2\paren{\ode+\om+\ol-1}-\ol\paren{\frac{\alpha\Gamma\paren{1+r_1+r_2}}{r_2}+4}\right.\\&\nonumber\left.-\om\paren{\frac{\Gamma\paren{1+r_1+r_2}}{r_1}+3}\right]-\ode\paren{\Delta-2}\paren{1-\sqrt{\frac{3M_P^2\ode}{C}}L^{-\frac{\Delta}{2}}\cosh{y}}~.
\end{align}
The numerical solutions to these differential equations are depicted in Fig.\eqref{for all}. Additionally, the variation of the dark energy equation of state (EOS) parameter with respect to redshift is illustrated in Fig.\eqref{fig:eos_all} for various values of the Barrow exponent $\Delta$.
\section{Cosmological Behaviour}\label{sec:5}
Cosmological evolution is greatly influenced by it's constituent matters, hence by the parameters associated with it. Thus, constraining them becomes very important. However, before grasping the understanding of the dynamics of different variables, it is important. In this section we will try to understand the dynamics of the equation of state(EOS) for the dark energy($\omega_{de}(z)$) with redshift($z$) for different values of the barrow exponent $\Delta$.

\subsection{Case when the interaction is proportional to dark energy density}
\begin{figure}[ht]
    \centering
    \includegraphics[width=0.7\linewidth]{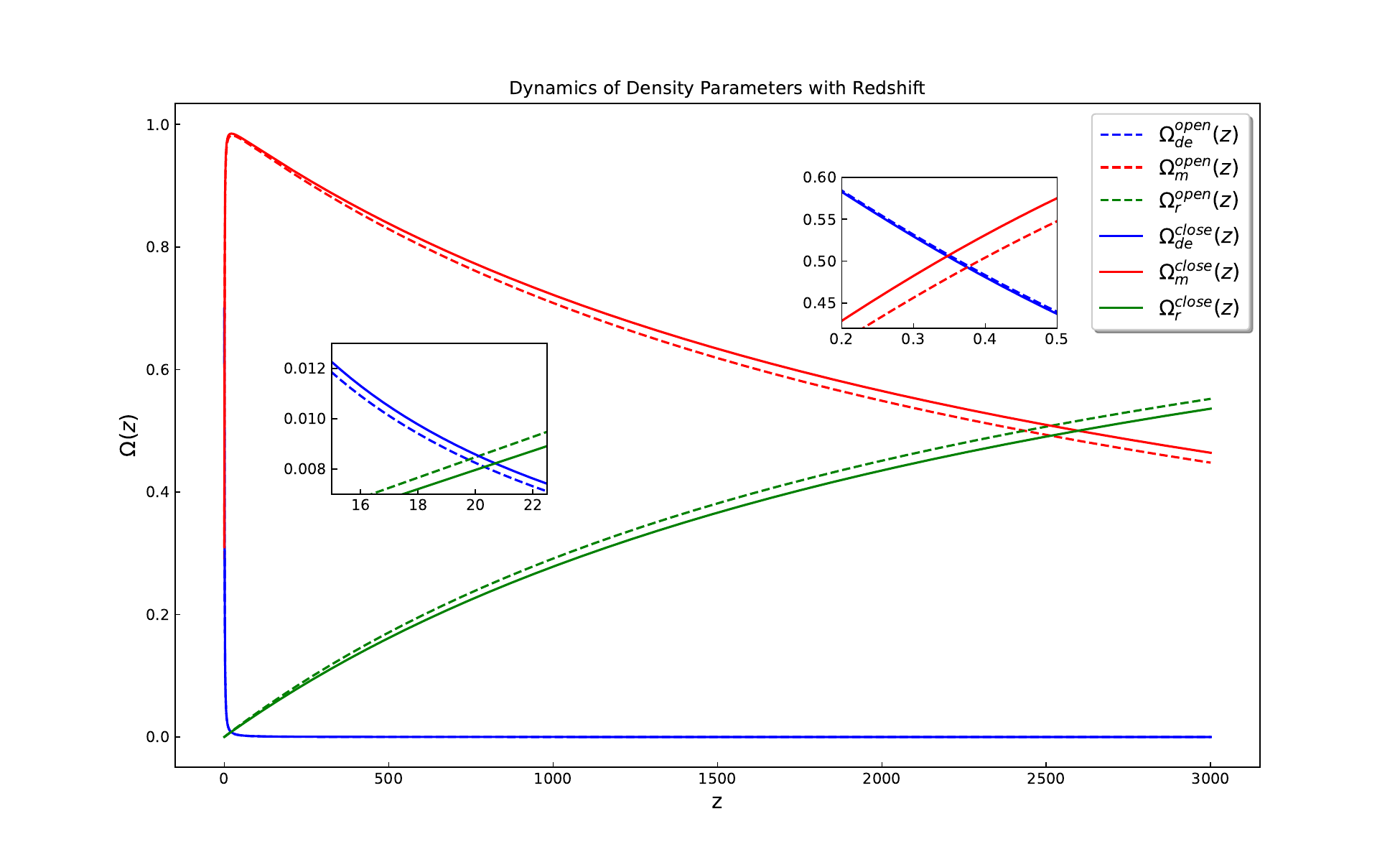}
    \caption{Energy density parameters vs redshift graphs when the phenomenological interaction term in the model $\mathcal{Q}$ is proportional to $\rho_{de}$. The solid lines denotes evolution of energy density parameters for closed universe and the dashed one are for open universe. Two inset plots are also provided to show different epoch points.}
    \label{fig: evolution de}
\end{figure}

In order to see the evolution of all the energy density parameters corresponding to dark energy, dark matter and radiation for both closed and open universes,
we have plotted the numerical solution for different energy density parameters and the dark energy equation of state parameter $\omega_{de}$, with respect to the redshift parameter.\\
In Fig.\eqref{fig: evolution de}, we have graphically represented the evaluation of the energy density parameters with respect to the redshift parameter. To do so, we have numerically solved the coupled evolution equation corresponding to $\ode$, $\om$ and $\ol$ in subsections (3.1) and (4.1). While obtaining the plots, we have taken the present values of the evaluation parameters as ${\ode}_{,0}=0.75$, ${\om}_{,0}=0.26$, ${\ol}_{,0}=10^{-4}$ for closed universe and ${\ode}_{,0}=0.73$, ${\om}_{,0}=0.26$, ${\ol}_{,0}=10^{-4}$ for open universe. Also the interaction parameter and the Barrow exponent are taken to be $\Gamma=0.05$ and $\Delta=0.1$ respectively. The reason for not choosing the same parameter value is because for closed universe, one has to be careful about the condition ${\ode}_{,0}+{\om}_{,0}+{\ol}_{,0}>1$ never violets and for open universe ${\ode}_{,0}+{\om}_{,0}+{\ol}_{,0}<1$ always satisfy strictly, otherwise analysis will be wrong. The blue, red and green curves shows the evolution of energy density parameters $\ode$, $\om$ and $\ol$ respectively. We have also added two inset plots to show the epochs of dark energy-dark matter equality and dark energy-radiation crossovers. It is evident from Fig.\eqref{fig: evolution de} that, for a closed universe, the dark energy and dark matter equality epoch took place at $z\sim 0.3470$, the dark energy-radiation equality epoch happened at $z\sim 20.686 $ and the dark matter-radiation equality epoch happened somewhere close to $z\sim 2596.9114$. For an open universe, these epochs took place at redshift values $z\sim 0.3772$, $z\sim 19.755$ and $z\sim 2433.395$ respectively. We have also included the redshift values corresponding to various epoch points in Table \eqref{tab:dark energy density}.\\

\begin{figure}[ht!]
    \centering
    \includegraphics[width=0.7\linewidth]{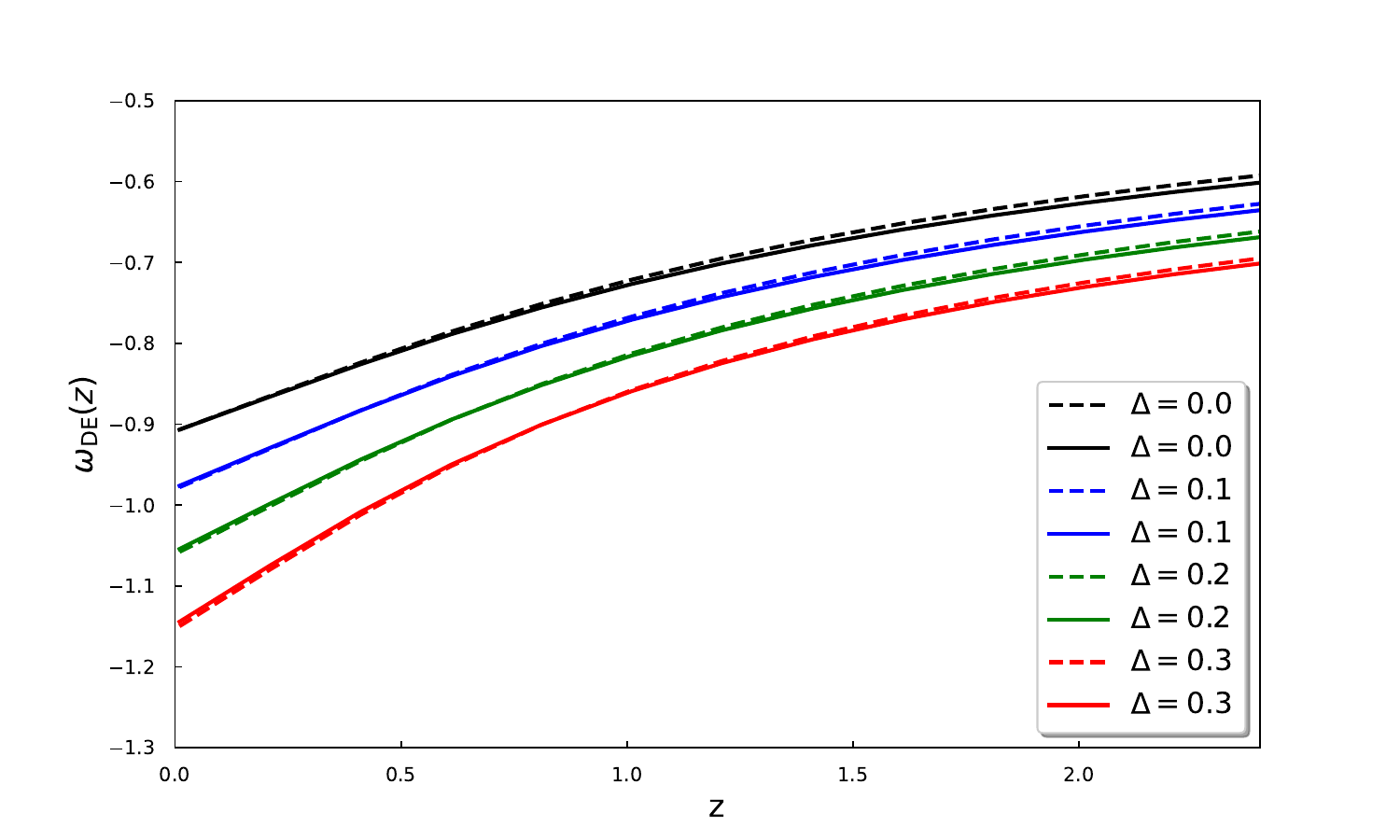}
    \caption{Growth of Equation of state with red shift for $\mathcal{Q}=-\Gamma H \rho_{de}$ for both open and closed universe. Solid lines correspond to the closed universe, and the dashed line is for the open case. The blue, green, red and purple
curves are for barrow exponent $\Delta= 0, 0.1, 0.2 \textit{ and } 0.3$ respectively. While choosing the current values of density parameters we have taken $\ode=0.75$, $\om=0.26$, $\ol=10^{-4}$, $\Omega_{k,0}=-0.0101$ for closed universe and for open universe we choose $\ode=0.73$, $\om=0.26$, $\ol=10^{-4}$,$\Omega_{k,0}=0.0099$. $\Gamma=10^{-4}$ for both the cases}  
    \label{fig:eos_de}
\end{figure}

\noindent In Fig.\eqref{fig:eos_de}, the numerical solution corresponding to the dark energy EOS parameter with respect to the redshift parameter has been shown graphically. We have used the evolution equations for $\ode$, $\om$ and $\ol$ from subsections (3,4) along with the expressions of $\omega_{de}$ from eq.(s)(\eqref{omega de positive},\eqref{omegade negative}) to obtain the plots in Fig.\eqref{fig:eos_de}. The curves in blue, green, red and purple are for barrow exponent $\Delta$= $0$, $0.1$, $0.2$ and $0.3$ respectively. It is evident from the plot that $\omega_{de}$ always has values that are less than zero. Another important observation from the graph is that for $\Delta=0.0,0.1$ the EOS parameter always lies in the quintessence regime, although the EOS correspond.ing to $\Delta=0.2,0.3$ shows a transition into the phantom region for lower redshift values. For high redshift, $\omega_{de}$ asymptotically approaches a constant value. One can clearly see that more the value of the barrow exponent, earlier the EOS parameter enters into the phantom regime from the quintessence regime. However, in contrast to \cite{Adhikary:2024sax} inclusion of the interaction of radiation make the $\omega_{de}$ grow for open and closed universes in a similar way for a broad values of the Barrow exponent.   
\begin{table}[h!]
    \centering
    \begin{tabular}{|c||c|c|}
    \hline
         \textbf{Epochs} & \textbf{Closed universe} & \textbf{Open universe}\\
    \hline
         Dark energy-Dark matter &  0.3470&0.3772\\
         Dark energy-Radiation & 20.686&19.755\\
         Dark matter-Radiation & 2596.9114&2433.395\\
    \hline
    \end{tabular}
    \caption{Redshift values at different epochs for $\mathcal{Q}=-\Gamma H \rho_{de}$}
    \label{tab:dark energy density}
\end{table}
\subsection{Case when the interaction is proportional to dark matter density only}
\begin{figure}[h!]
    \centering
    \includegraphics[width=0.7\linewidth]{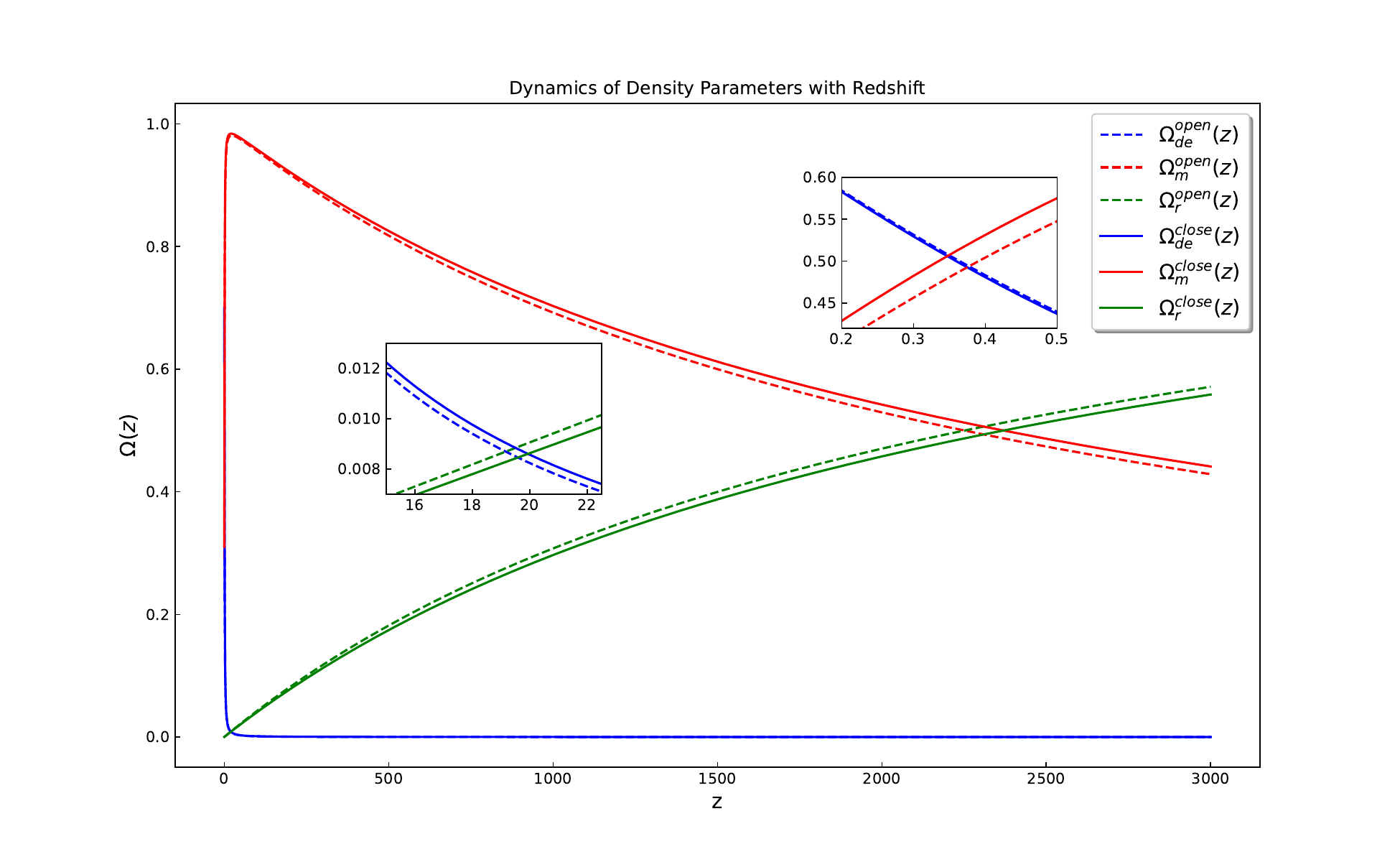}
    \caption{Plot for energy density parameters vs redshift graphs when the phenomenological interaction term in the model $\mathcal{Q}$ is proportional to $\rho_{m}$. The solid lines denotes evolution of energy density parameters for closed universe and the dashed one are for open universe. Two inset plots are also provided to show different epoch points. Interaction prop to matter}
    \label{fig:evolution matter}
\end{figure}

\begin{figure}[h!]
    \centering
    \includegraphics[width=0.7\linewidth]{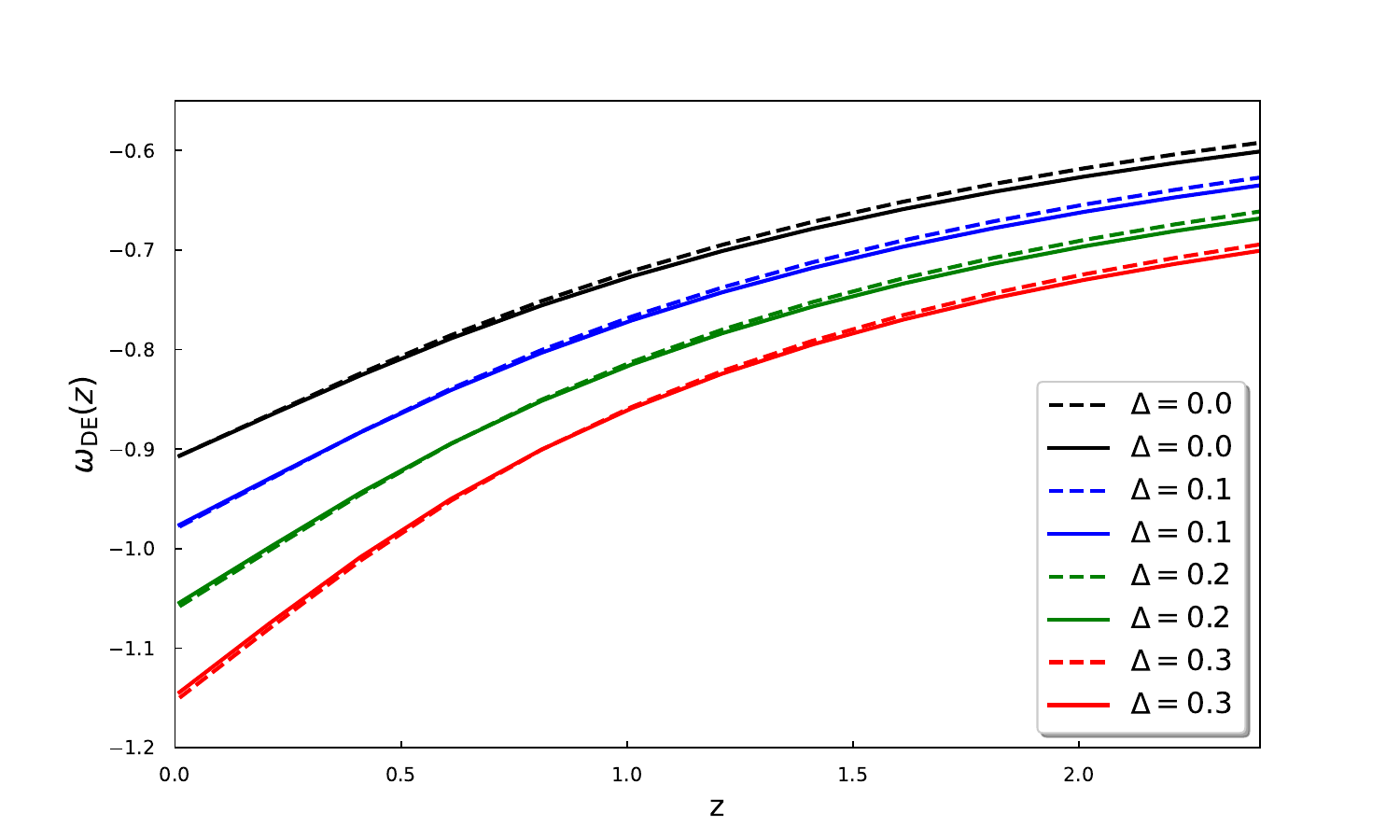}
    \caption{Growth of Equation of state with red shift for $\mathcal{Q}=-\Gamma H \rho_{m}$ for both open and closed universe. Solid lines correspond to the closed universe, and the dashed lines are for the open universe. The blue, green, red and purple curves corresponds to the barrow exponent $\Delta= 0, 0.1, 0.2 \textit{ and } 0.3$ respectively. While choosing the current values of density parameters we have taken $\ode=0.75$, $\om=0.26$, $\ol=10^{-4}$, $\Omega_{k,0}=-0.0101$ for closed universe and for open universe we choose $\ode=0.73$, $\om=0.26$, $\ol=10^{-4}$,$\Omega_{k,0}=0.0099$. $\Gamma=10^{-4}$ for both the cases. Evolution of equation of state for $\mathcal{Q}=-\Gamma H \rho_{m}$}
    \label{fig:eos_m}
\end{figure}

In this interaction model when the phenomenological interaction term $\mathcal{Q}$ is proportional to dark matter density, the evolution of different energy density parameters corresponding to dark energy, dark matter and radiation respectively are shown in Fig.\eqref{fig:evolution matter}. To showcase this evolution, we utilized the coupled differential equations from subsections (3.2) and (4.2), maintaining the initial conditions specified in the preceding subsections of this section. The blue, red and green curves in Fig.\eqref{fig:evolution matter}, shows the evolution of energy density parameters $\ode$, $\om$ and $\ol$ respectively. The solid curves corresponds to closed universe and the dashed curves corresponds to an open universe scenario. We have also added two inset plots to show the epochs of dark energy-dark matter equality and dark energy-radiation crossovers. In this scenario, as illustrated in Fig.\eqref{fig:evolution matter}, the epoch when dark energy equals dark matter occurred at approximately \( z \sim 0.346 \). The epoch of dark energy-radiation equality took place around \( z \sim 19.936  \), while the dark matter-radiation equality occurred near \( z \sim 2369.488 \) for a closed universe. For an open universe, these respective epochs occurred at different redshift values, denoted as \( z \sim 0.37718\), \( z \sim 19.164\), and \( z \sim 2251.251\). In summary, we have also included the redshift values at various epoch points in Table \eqref{tab:mat density}.\\
For the case where interaction is proportional to dark-matter density, the dark energy equation of state parameter is plotted against redshift in Fig.\eqref{fig:eos_m}. We have shown the variation of equation of state parameter for a set of barrow exponents \{$0,0.1,0.2,0.3$\}, for both open and closed universes. During plotting, we have chosen current values of the density parameters ${\ode}_{,0}=0.75,{\om}_{,0}=0.26,{\Omega}_{r,0}=10^{-4}$ for the closed universe and for open universe ${\ode}_{,0}=0.73,{\om}_{,0}=0.26,{\Omega}_{r,0}=10^{-4}$. The black, blue, green, and red curves show the evolution of $\omega_{de}$ with respect to redshift $z$ for $\Delta$ values of 0, 0.1, 0.2, and 0.3, respectively. For each color, solid lines represent the dark energy EOS parameter in a closed universe, while dashed lines illustrate it for an open universe. We can see from the plot that the equation of state for dark energy density is continuously growing with redshift. At a late time (that is for low redshift) $\omega_{de}$ enters the phantom region.
\begin{table}[h!]
    \centering
    \begin{tabular}{|c||c|c|}
    \hline
         \textbf{Epochs} & \textbf{Closed universe} & \textbf{Open universe}\\
    \hline
         Dark energy-Dark matter & 0.346 & 0.37718\\
         Dark energy-Radiation & 19.936&19.164\\
         Dark matter-Radiation & 2369.488&2251.251\\
    \hline
    \end{tabular}
    \caption{Redshift values at different epochs for $\mathcal{Q}=-\Gamma H \rho_{m}$}
    \label{tab:mat density}
\end{table}
\subsection{Case when the interaction is proportional to radiation density only}
\begin{figure}[ht]
    \centering
    \includegraphics[width=0.7\linewidth]{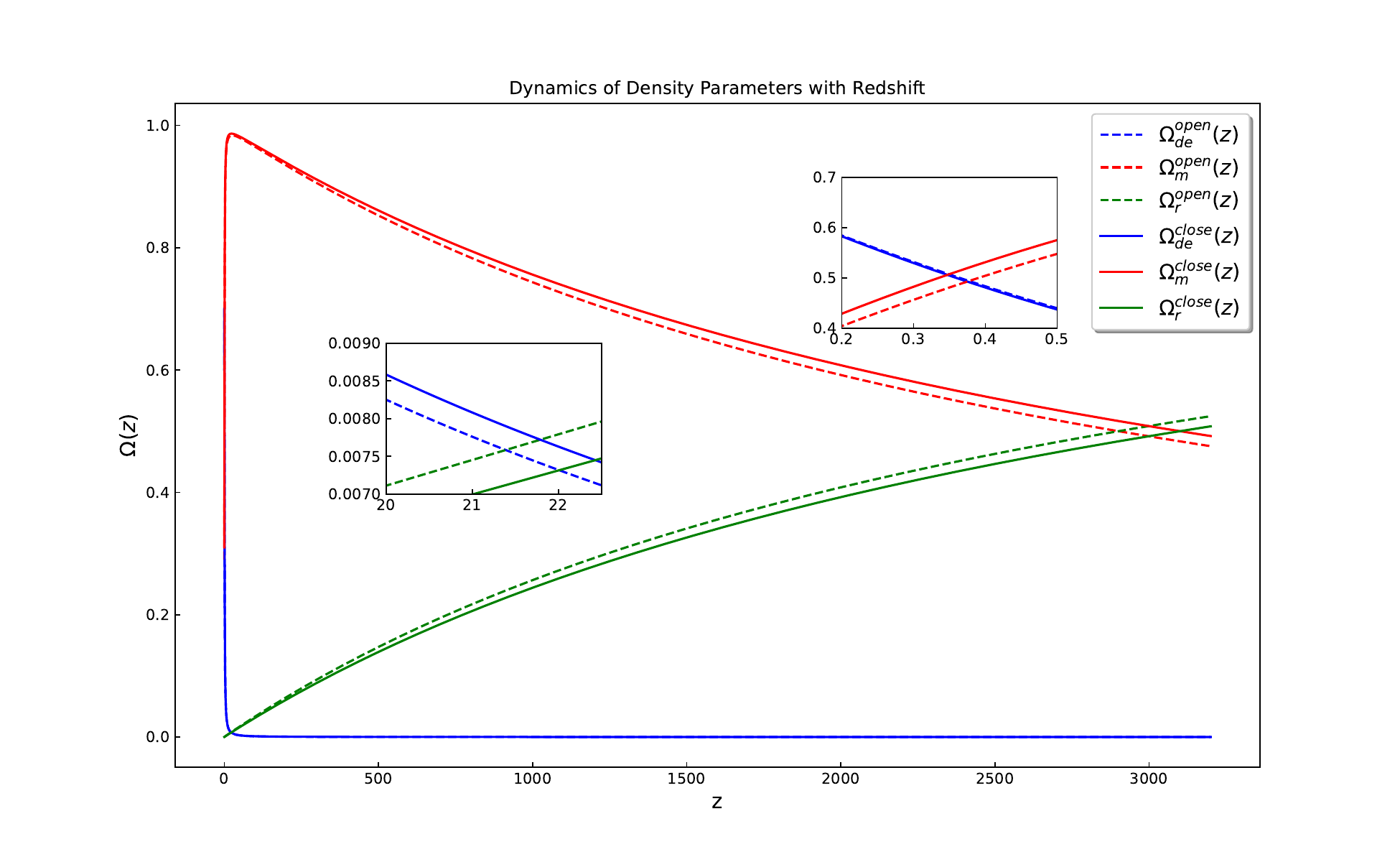}
    \caption{Graphical representation of energy density parameters vs redshift when the phenomenological interaction term in the model $\mathcal{Q}$ is proportional to $\rho_{r}$. The solid lines denotes evolution of energy density parameters for closed universe and the dashed one are for open universe. Two inset plots are also provided to show different epoch points. Interaction prop to radiation}
    \label{fig:radiation}
\end{figure}

\begin{figure}[ht!]
    \centering
    \includegraphics[width=0.7\linewidth]{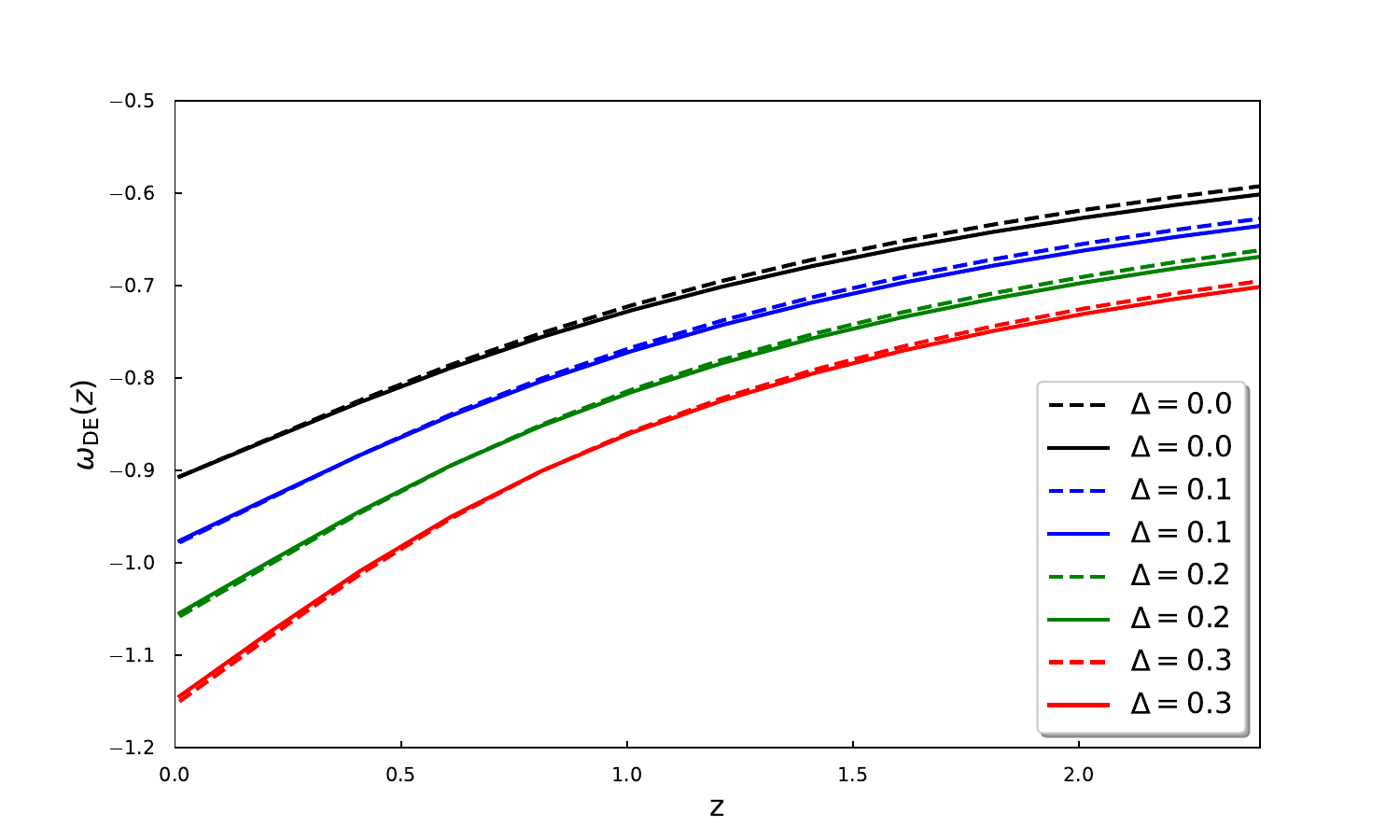}
    \caption{Dynamics of equation of state with red shift for $\mathcal{Q}=-\Gamma H \rho_{r}$ for both open and closed universe shown with respect to redshift $z$. Solid lines correspond to the closed universe, and the dashed lines denotes equation of state parameter for the open universe. The blue, green, red and purple curves are for barrow exponent $\Delta= 0, 0.1, 0.2 \textit{ and } 0.3$ respectively. While choosing the current values of density parameters we have taken $\ode=0.75$, $\om=0.26$, $\ol=10^{-4}$, $\Omega_{k,0}=-0.0101$ for closed universe and for open universe we choose $\ode=0.73$, $\om=0.26$, $\ol=10^{-4}$,$\Omega_{k,0}=0.0099$. $\Gamma=10^{-4}$ for both the cases. Dynamics of equation of state for dark energy with redshift for $\mathcal{Q}=-\Gamma H \rho_r$}
    \label{fig:eos_r}
\end{figure}

For this scenario, in order to see the evolution of different energy density parameters (that is $\ode$, $\om$ and $\ol$), we have numerically solved the set of coupled differential equations in subsections (3.3) and (4.3). The initial values of these various energy density parameters are set to be the same as those used in the previous subsection when deriving the solutions. The blue, red, and green curves in Fig.\eqref{fig:radiation} depict the evolution of the energy density parameters $\ode$, $\om$, and $\ol$, respectively. The solid lines represent a closed universe, while the dashed lines correspond to an open universe scenario.
As illustrated in Fig. \eqref{fig:radiation}, in this scenario, dark energy and dark matter had equal densities at approximately $z \sim 0.347$. Dark energy and radiation were equal around $z \sim 22.428 $, and dark matter and radiation reached equality near $z \sim 3100.028$ in a closed universe. For an open universe, these equalities happened at different redshifts, specifically $z \sim 0.3772$, $z \sim 21.388$, and $z \sim 2900.583$ respectively. To summarize we have also provided the value of redshift at different epoch points in Table \eqref{tab:rad density}.\\
The dynamics of the equation of state parameter for dark energy has been shown in Fig\eqref{fig:eos_r} for four values of $\Delta\in\{0.1,0.2,0.3,0.4\}$. From the graph, we can clearly see that $\omega_{de}$ increases for increasing value of redshift. The black, blue, green, and red curves illustrate the evolution of \(\omega_{de}\) as a function of redshift \(z\) for \(\Delta\) values of 0, 0.1, 0.2, and 0.3, respectively. In each color, the solid lines represent the dark energy equation of state (EOS) parameter for a closed universe, while the dashed lines correspond to the same parameter for an open universe. For $\Delta$ values $0$ and $0.1$ the graph always lies in the quintessence region although for $\Delta$ equals $0.2$ and $0.3$ the corresponding graphs shows a transition to the phantom regime from the quintessence regeim.
\begin{table}[h!]
    \centering
    \begin{tabular}{|c||c|c|}
    \hline
         \textbf{Epochs} & \textbf{Closed universe} & \textbf{Open universe}\\
    \hline
         Dark energy-Dark matter &  0.347&0.3772\\
         Dark energy-Radiation & 22.428&21.388\\
         Dark matter-Radiation & 3100.028&2900.583\\
    \hline
    \end{tabular}
    \caption{Redshift values at different epochs for $\mathcal{Q}=-\Gamma H \rho_r$}
    \label{tab:rad density}
\end{table}
\subsection{Case when the interaction is proportional to sum of dark energy, dark matter and radiation density}
\begin{figure}[h!]
    \centering
    \includegraphics[width=0.7\linewidth]{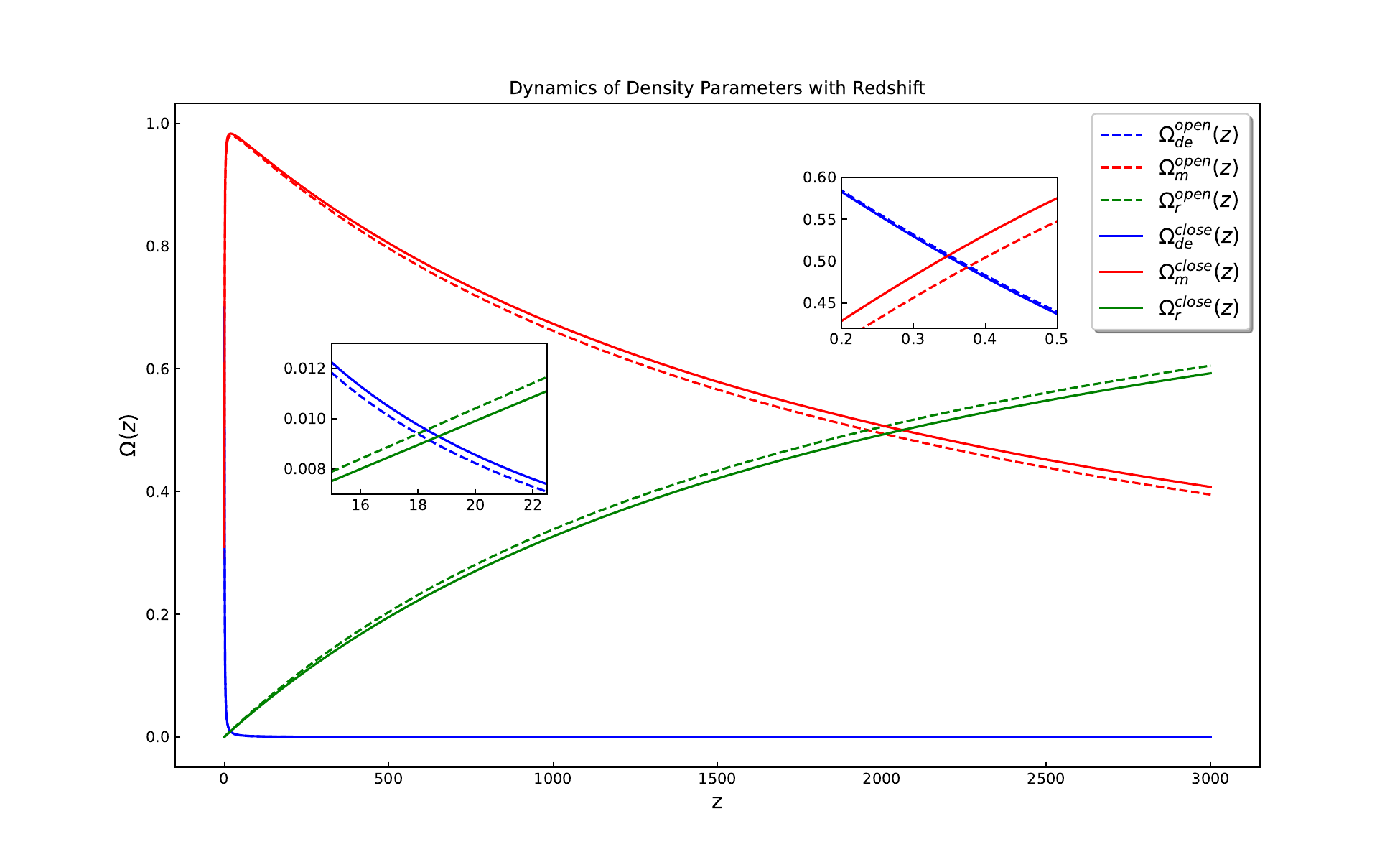}
    \caption{Plot for energy density parameters vs redshift graphs when the phenomenological interaction term in the model $\mathcal{Q}$ is proportional to sum of all the energy densities, that is $\rho_{de}+\rho_{m}+\rho_{r}$. The solid lines denotes evolution of energy density parameters for closed universe and the dashed one are for open universe. Two inset plots are also provided to show different epoch points. Interaction prop to all}
    \label{for all}
\end{figure}

\begin{figure}[h!]
    \centering
    \includegraphics[width=0.7\linewidth]{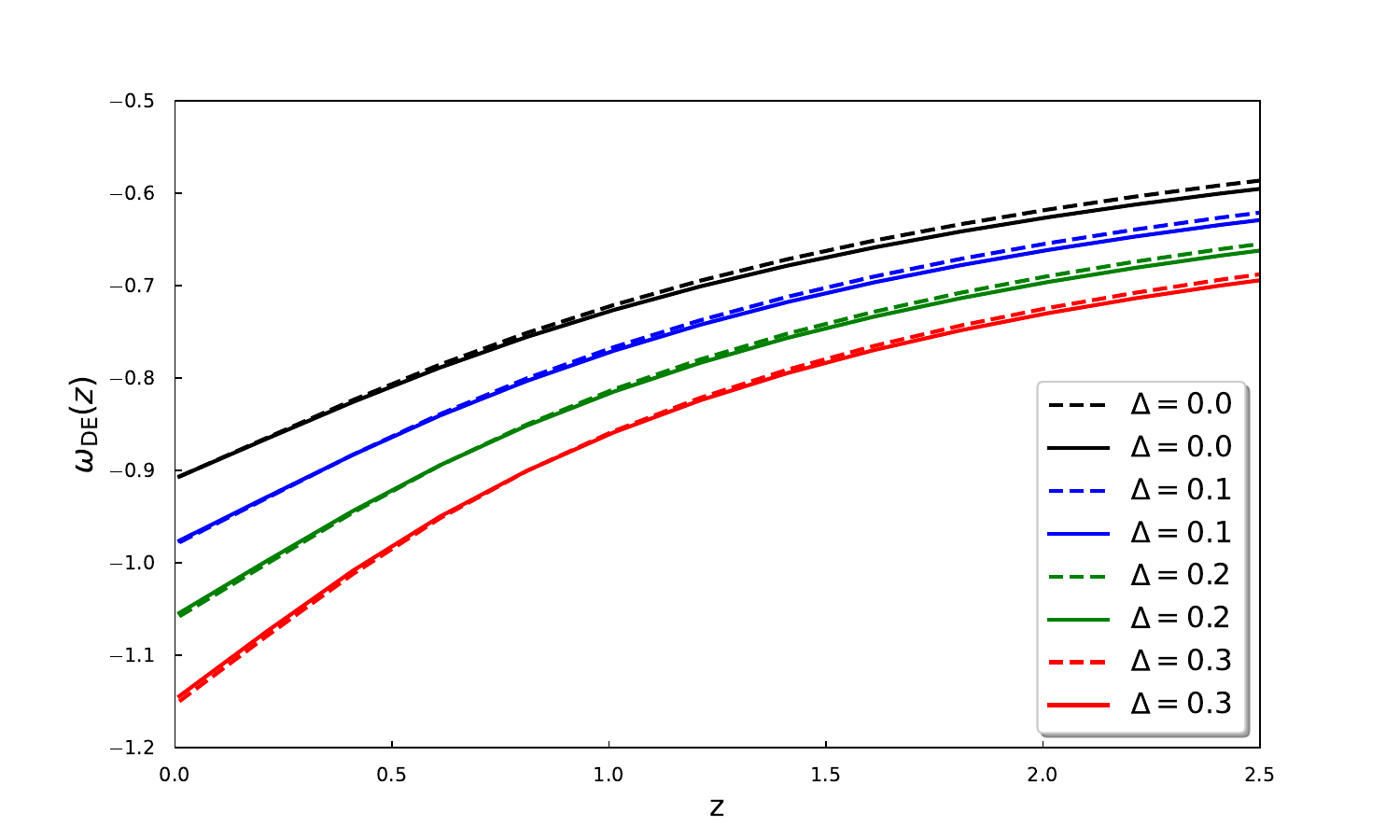}
    \caption{Plot of the dark energy equation of state with red shift for the phenomenological interaction term $\mathcal{Q}=-\Gamma H (\rho_{de}+\rho_{m}+\rho_{r})$ for both open and closed universe shown with respect to redshift $z$. Solid lines correspond to the closed universe, and the dashed lines denotes equation of state parameter for the open universe. The blue, green, red and purple curves are for barrow exponent $\Delta= 0, 0.1, 0.2 \textit{ and } 0.3$ respectively. While choosing the current values of density parameters we have taken $\ode=0.75$, $\om=0.26$, $\ol=10^{-4}$, $\Omega_{k,0}=-0.0101$ for closed universe and for open universe we choose $\ode=0.73$, $\om=0.26$, $\ol=10^{-4}$,$\Omega_{k,0}=0.0099$. $\Gamma=10^{-4}$ for both the cases. Behaviour of dark energy equation of state with redshift for the choice $\mathcal{Q}=\Gamma H \paren{\rho_{de}+\rho_{m}+\rho_{r}}$}
    \label{fig:eos_all}
\end{figure}

In this case, when the phenomenological interaction term depends on the sum of all the energy densities corresponding to dark energy, dark matter and radiation, we have graphically shown the evolution of all the energy density parameters with respect to redshift in Fig.\eqref{for all}. To show this evolution, we have used the set of coupled differential equations in subsections (3.4 and 4.4), with the same initial conditions as mentioned in the previous subsections of this section. In Fig. \eqref{for all}, the blue, red, and green curves illustrate the evolution of the energy density parameters: dark energy ($\ode$), dark matter ($\om$), and radiation ($\ol$), respectively. The solid curves represent a closed universe, while the dashed curves depict an open universe scenario. In this case, it is evident from Fig.\eqref{for all} that the dark energy and dark matter equality epoch took place at $z\sim 0.3469 $, the dark energy-radiation equality epoch happened at $z\sim 18.708$, and the dark matter-radiation equality epoch happened somewhere close to $z\sim 2061.454$, for a closed universe. For an open universe, these epochs took place at redshift values $z\sim 0.37712$, $z\sim 17.987$ and $z\sim 1957.2877$ respectively. For a comprehensive overview, Table \eqref{tab:all density} additionally presents the redshift values at different epoch points.\\
\noindent Similar to the previous cases, Fig.\eqref{fig:eos_all}, shows the variation of the EOS parameter with respect to redshift for $\Delta$ values equals $0, 0.1, 0.2, 0.3$. The curves in black, blue, green and red shows the evolution of $\omega_{de}$ with respect to redshift $z$ for $\Delta$ values equals $0, 0.1, 0.2$ and $0.3$ respectively. For each color solid lines depicts the dark energy EOS parameter for a closed universe and dashed lines depicts the dark energy EOS parameter for an open universe. Here also the EOS shows a transition from quintessence region to phantom region for $\Delta=0.2, 0.3$, for both open and closed universes. Although for $\Delta=0,0.1$, the EOS always lies in the quintessence region.
\begin{table}[h!]
    \centering
    \begin{tabular}{|c||c|c|}
    \hline
         \textbf{Epochs} & \textbf{Closed universe} & \textbf{Open universe}\\
    \hline
         Dark energy-Dark matter &  0.3469&0.37712\\
         Dark energy-Radiation & 18.708&17.987\\
         Dark matter-Radiation & 2061.454&1957.2877\\
    \hline
    \end{tabular}
    \caption{Redshift values at different epochs for $\mathcal{Q}=-\Gamma H (\rho_{de}+\rho_{m}+\rho_{r})$}
    \label{tab:all density}
\end{table}
\section{Observational Constraints}\label{Observational Constraints}
In this section, we will try to constraint various observational parameters of our interacting Barrow holographic dark energy model using various observational data sets like Cosmic chronometer (CC) \& Baryon Acoustic Oscillator (BAO) and Pantheon+ \& SHOes data sets. We start this section with a quick recap of the CC, BAO and Panteheon+ data measurements. 
\subsection{Cosmic chronometer data}
Cosmic Chronometers are some cosmological objects (for example, some special kind of galaxies ) whose dynamics are known. Observing those at different redshifts and associated changes in their evolutionary state gives the value of the Hubble parameter at that particular redshift. 
In cosmic chromatometry, the Hubble parameter H(z) at
different redshifts is usually determined through two approaches: (i) by extracting H (z) from the line of sight of the BAO data [71, 72] and (ii) by estimating H (z) by the method of
differential age (DA) of galaxies [73–76], which relies on the relation \cite{Simon:2004tf,Ratsimbazafy:2017vga}
\begin{equation}
    H(z)=-\frac{1}{1+z}\frac{dz}{dt}~.
\end{equation}
where dz/dt is approximated by determining the small time interval $\Delta t$  between two evolving galaxies that are $\Delta z $ redshift distant,  corresponding to a given z. This technique finds $H_0$, which is independent of early universe physics. The predicted values of $H_0$ more align with the recent CMB and BAO data than the SnIa data. The value of $H_0$ cannot be found only by cosmic chronometers observation because of the presence of background degeneracy between $H_0$ and $\Omega_{m,0}$. To resolve this one have to combine different observations.
[75]. The function $\chi^2$ is given by
\begin{equation}
    \chi_{CC}^2=\sum_{i=1}^{31}\frac{\Big(H_{th}(z_i)-H^{(CC)}_{obs}(z_i)\Big)^2}{\sigma^2_{H_{CC}}(z_i)}~.
\end{equation}
\begin{table}[h!]
\centering
\begin{tabular}{|c|c|c|c||c|c|c|c|}
\hline
\multicolumn{8}{|c|}{\textbf{CC}} \\
\hline
$z$ & $H(z)$ & $\sigma_H$ & Refs & $z$ & $H(z)$ & $\sigma_H$ & Refs \\
\hline
0.070 & 69 & 19.6 & \cite{ref81}& 0.4783 & 80.9 & 9 & \cite{ref82} \\
0.090 & 69 & 12 & \cite{ref78}& 0.480 & 97 & 62 & \cite{ref83} \\
0.120 & 68.6 & 26.2 & \cite{ref81}& 0.593 & 104 & 13 & \cite{ref84} \\
0.1791 & 83 & 8 & \cite{ref78}& 0.6797 & 92 & 8 & \cite{ref84} \\
0.1791 & 75 & 4 & \cite{ref84} & 0.7812 & 105 & 12 & \cite{ref84} \\
0.1993 & 75 & 5 & \cite{ref84} & 0.8754 & 125 & 17 & \cite{ref84} \\
0.200 & 72.9 & 29.6 & \cite{ref81}& 0.880 & 90 & 40 & \cite{ref83} \\
0.270 & 77 & 14 & \cite{ref78}& 0.900 & 117 & 23 & \cite{ref78}\\
0.280 & 88.8 & 36.6 & \cite{ref81}& 1.037 & 154 & 20 & \cite{ref84} \\
0.3519 & 83 & 14 & \cite{ref84} & 1.300 & 168 & 17 & \cite{ref78}\\
0.3802 & 83 & 13.5 & \cite{ref82}& 1.363 & 160 & 33.6 & \cite{ref85}\\
0.4004 & 77 & 10.2 & \cite{ref82}& 1.430 & 177 & 18 & \cite{ref78}\\
0.4247 & 87.1 & 11.2 & \cite{ref82}& 1.530 & 140 & 14 & \cite{ref84}\\
0.4497 & 92.8 & 12.9 & \cite{ref84}& 1.750 & 202 & 40 & \cite{ref84}\\
0.470 & 89 & 34 & \cite{ref79}& 1.965 & 186.5 & 50.4 & \cite{ref85} \\
\hline
\end{tabular}
\caption{Cosmic chronometer (CC) data for $H(z)$ measurements and their uncertainties.}
\end{table}
We have taken the CC data for the red shift range $0.070\leq z \leq 1.965$, consisting with $31$ data points.
\subsection{Baryon acoustic oscillator data}
In the very early time in the history of the universe up to the time of recombination, it was a hot plasma sea, where photons and baryons were coupled it oscillating between photon and baryon as a spherical sound wave, and the driving force was the pressure of the photon. But as the universe expanded, the density of the radiation falls faster than the baryons, and the coupling broke and that was the first time when hydrogen was produced as the first atom\cite{Peebles:1968ja}. The decoupled photons then travel freely. The baryonic spherical sound wave shell became frozen to stamp its signature in the angular power spectrum\cite{2dFGRS:2005yhx,SDSS:2005xqv} of the Cosmic Microwave Background Radiation spectra and on Large scale structure at the level of the sound horizon, that is, the distance travelled by the sound wave at recombination time. Analyzing the peaks of the angular power spectrum, one can use it as a standard distance. The relation of the angular scale($\theta_s$) with the comoving  angular diameter distance($d_A$) is given by following\cite{Peebles:1980yev}
\begin{equation}
    \theta_s= \frac{r_s}{d_A}~.
\end{equation}
Again $d_A$ can be related to physical distance $D_A$ as 
\begin{equation}
    d_A=(1+z)D_A
\end{equation}
Thus knowing the $\theta_s$ of the peaks of the angular power spectrum can leads to determining the Hubble constant using corelation function of the power spectrum. 

\begin{table}[h!]
\centering
\begin{tabular}{|c|c|c|c||c|c|c|c|}
\hline
\multicolumn{8}{|c|}{\textbf{BAO}} \\
\hline
$z$ & $H(z)$ & $\sigma_H$ & Refs & $z$ & $H(z)$ & $\sigma_H$ & Refs \\
\hline
0.24 & 79.69 & 2.99 & \cite{Gaztanaga:2008xz}& 0.57 & 96.8  & 3.4  & \cite{BOSS:2013rlg}\\
0.30 & 81.7  & 6.22 & \cite{Oka:2013cba}& 0.59 & 98.48 & 3.18 & \cite{Oka:2013cba}\\
0.31 & 78.18 & 4.74 & \cite{BOSS:2016zkm}& 0.60 & 87.9  & 6.1  & \cite{Blake:2012pj}\\
0.34 & 83.8  & 3.66 & \cite{Gaztanaga:2008xz}& 0.61 & 97.3  & 2.1  & \cite{BOSS:2016wmc}\\
0.35 & 82.7  & 9.1  & \cite{Chuang:2012qt}& 0.64 & 98.82 & 2.98 & \cite{Oka:2013cba}\\
0.36 & 79.94 & 3.38 & \cite{Oka:2013cba}& 0.73 & 97.3  & 7.0  & \cite{Blake:2012pj}\\
0.38 & 81.5  & 1.9  & \cite{BOSS:2016wmc}& 2.30 & 224   & 8.6  & \cite{BOSS:2012gof}\\
0.40 & 82.04 & 2.03 & \cite{Oka:2013cba}& 2.33 & 224   & 8.0  & \cite{BOSS:2017fdr}\\
0.43 & 86.45 & 3.97 & \cite{Gaztanaga:2008xz}& 2.34 & 222   & 8.5  & \cite{BOSS:2014hwf}\\
0.44 & 82.6  & 7.8  & \cite{Blake:2012pj}& 2.36 & 226   & 9.3  & \cite{BOSS:2013igd}\\
0.44 & 84.81 & 1.83 & \cite{Oka:2013cba}&      &       &      & \\
0.48 & 87.79 & 2.03 & \cite{Oka:2013cba}&      &       &      & \\
0.51 & 90.4  & 1.9  & \cite{BOSS:2016wmc}&      &       &      & \\
0.52 & 94.35 & 2.64 & \cite{Oka:2013cba}&      &       &      & \\
0.56 & 93.34 & 2.3  & \cite{Oka:2013cba}&      &       &      & \\
0.57 & 87.6  & 7.8  & \cite{BOSS:2013mwe}&      &       &      & \\
\hline
\end{tabular}
\caption{Baryon Acoustic Oscillator (BAO) data for $H(z)$ measurements and their uncertainties.}
\end{table}
We have taken the $BAO$ data for the red shift range $0.24\leq z\leq 2.36$, consisting of $26$ data points which are given in the table(2).
Similarly the $\chi^2$ function for the BAO dataset is given by
\begin{equation}
    \chi_{BAO}^2=\sum_{i=1}^{26}\frac{\Big(H_{th}(z_i)-H^{(BAO)}_{obs}(z_i)\Big)^2}{\sigma^2_{H_{BAO}}(z_i)}~.
\end{equation}
For the joint set of $CC+BAO$ data, the $\chi^2$ function is simply given by
\begin{equation}
    \chi^2=\chi^{2}_{CC}+\chi^{2}_{BAO}
\end{equation}
It should be mentioned that, as there is no overlap between the CC and BAO data sets, the cross-correlation is zero between them.
Now in order to get the correlation between several model parameters and to obtain best fit points in both the open and closed universe, we have taken help of standard Bayesian statistics and have run Markov Chain Monte Carlo (MCMC) using a dedicated package in python named "Getdist" \footnote{\href{Getdist}{https://getdist.readthedocs.io/en/latest/}}. Before running the MCMC, we have chosen different initial guesses, and also the range corresponds to different variables. 
\begin{figure}[htbp]
    \begin{subfigure}[b]{0.5\textwidth}
        \includegraphics[width=\textwidth]{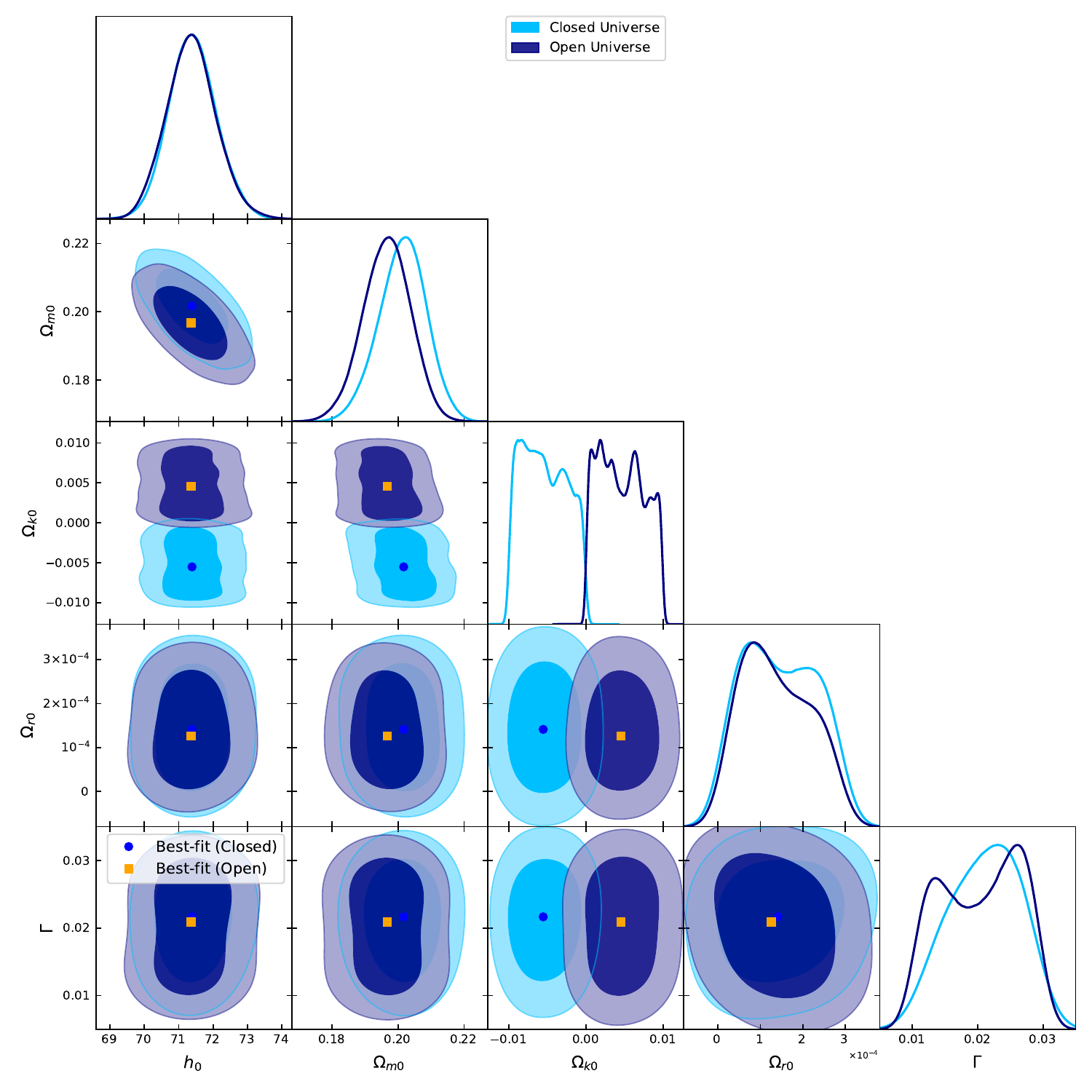}
        \caption{$\mathcal{Q}\propto\rho_{de}$}
        \label{fig:sub1}
    \end{subfigure}
    \hspace{-0.06\textwidth} 
    \begin{subfigure}[b]{0.5\textwidth}
        \includegraphics[width=\textwidth]{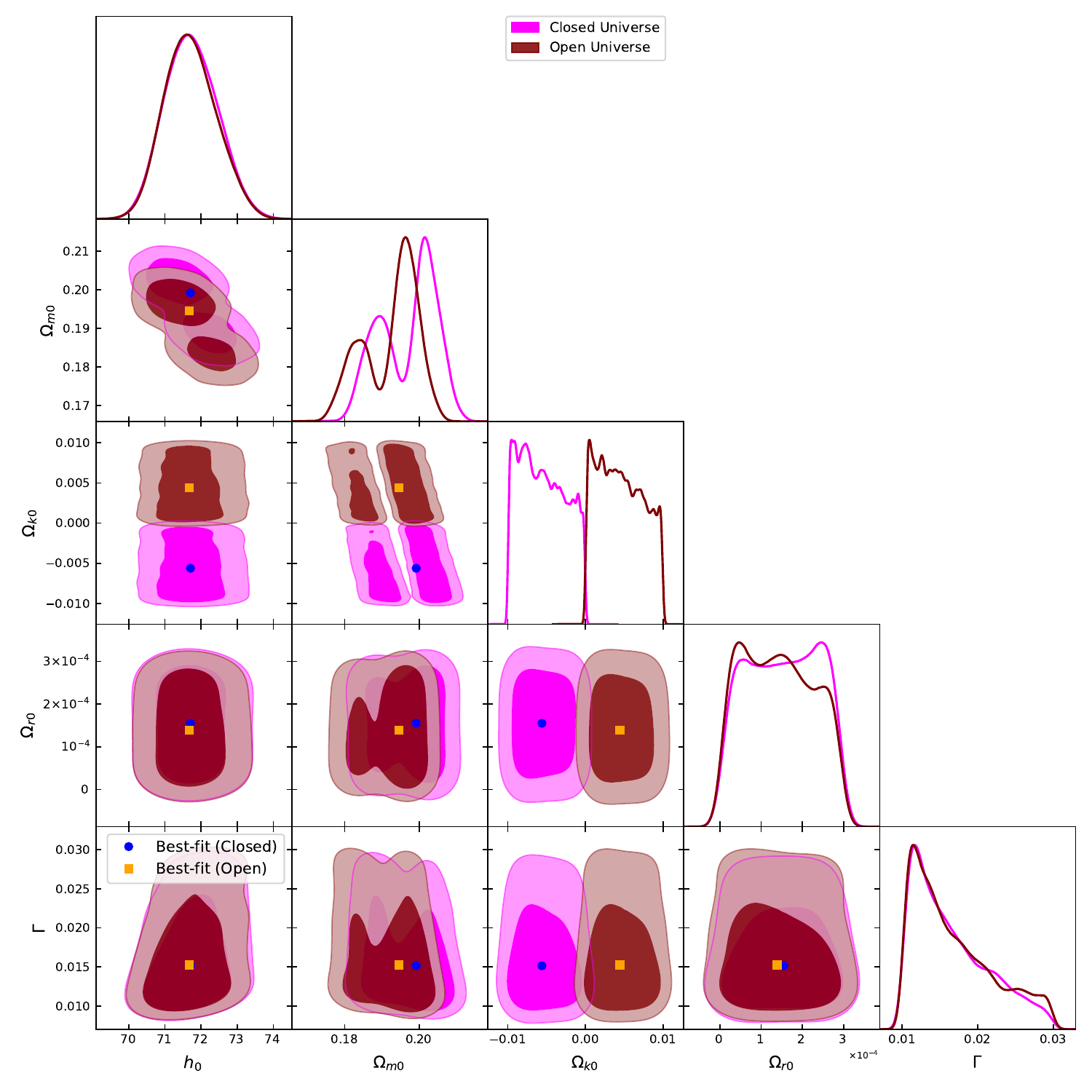}
        \caption{$\mathcal{Q}\propto\rho_{m}$}
        \label{fig:su2}
    \end{subfigure}
     \begin{subfigure}[b]{0.5\textwidth}
        \includegraphics[width=\textwidth]{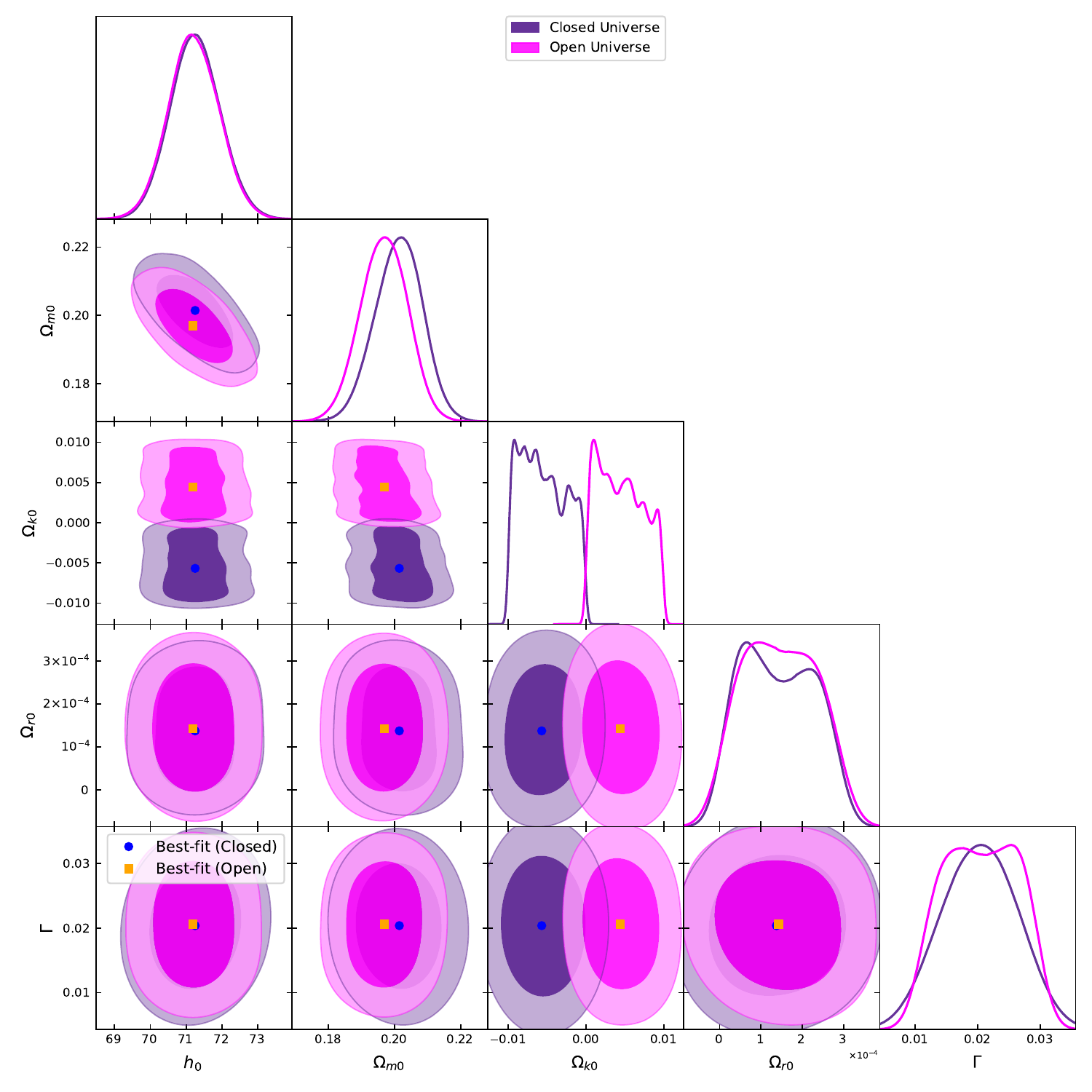}
        \caption{$\mathcal{Q}\propto\rho_{r}$}
        \label{fig:su4}
    \end{subfigure}
    \hspace{-0.01\textwidth}
     \begin{subfigure}[b]{0.5\textwidth}
        \includegraphics[width=\textwidth]{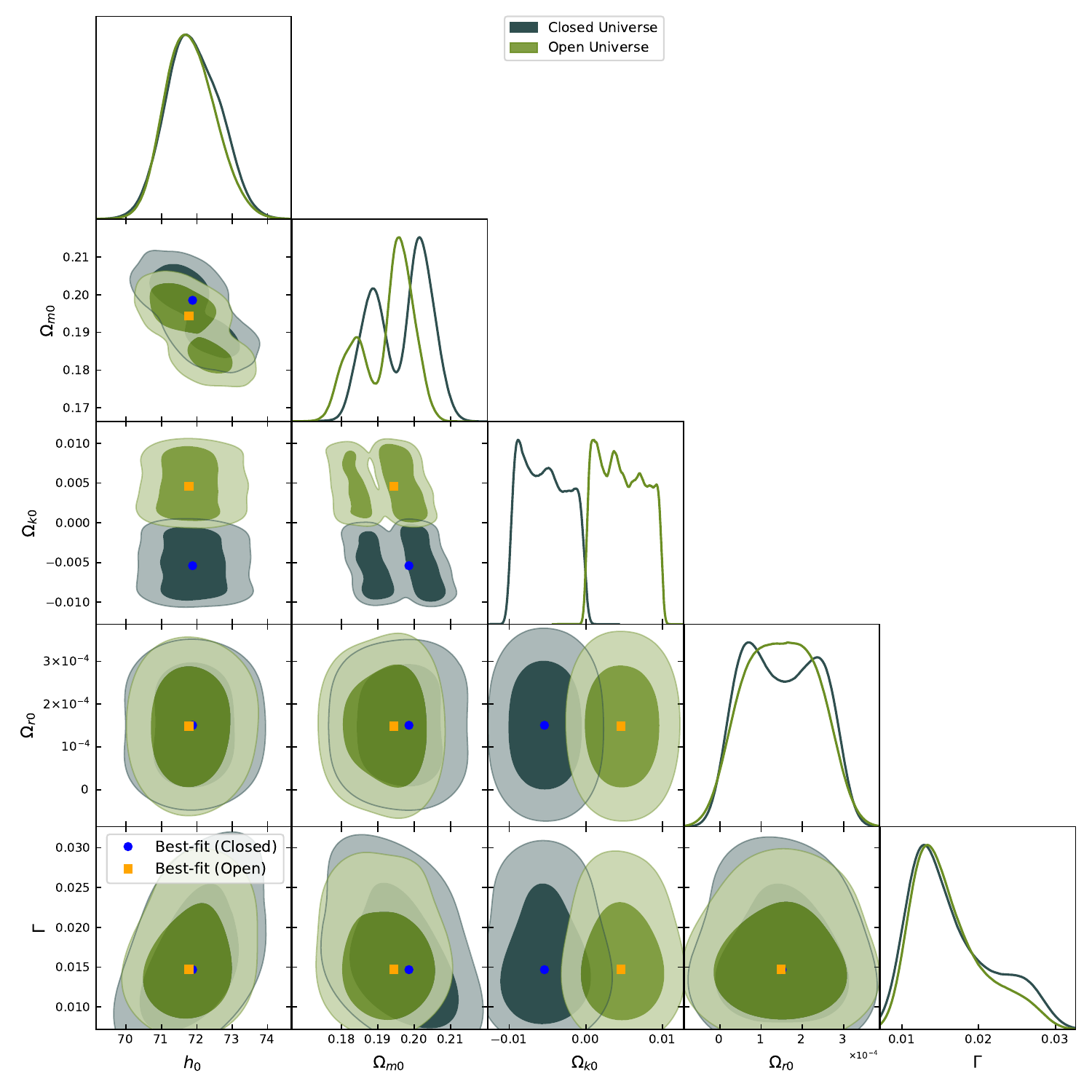}
        \caption{$\mathcal{Q}\propto\paren{\rho_{de}+\rho_{m}+\rho_{r}}$}
    \end{subfigure}
    \caption{Contour plots for all four different kinds of phenomenological interaction terms for CC \& BAO data set. Panel (a), (b), (c) and (d) shows the corner plots when the interaction term ($\mathcal{Q}$)is proportional to $\rho_{de}$, $\rho_{m}$, $\rho_{r}$ and $\rho_{de}+\rho_{m}+\rho_{r}$ respectively. For each panel, contours representing various model parameters are displayed alongside the best-fit points corresponding to observational data, are displayed for both open and closed universe scenarios.}
    \label{fig:approx fit CC combined}
\end{figure}


\subsection{Pantheon+ data}
Supernovae Type Ia which is a member of standard candles and also the most luminous among all supernovae, is a great tool to measure the luminosity distances. Supernovae are relatively easy to detect due to their extremely powerful luminosity. The peak of the luminosity very much depends on the light curve and the red shift of the galaxy to which it belongs. 6 difference surveys build up the Pantheon dataset which consists of data related to 1048 different redshifts. However the latest dataset Pantheon+ includes 1701 light curve with 1624 data points by 18 separately done surveys in the redshift range $0.01<z<2.3$. The main observable is the distance modulus $\mu^{obs}(z)$. 
The $\chi^2$ function for the Pantheon data set is given by
\begin{equation}
    \chi^2_{Pantheon}=\sum_{i,j=1}^{1071} (\mu^{th}-\mu^{obs})_{i}(C_{Pantheon})_{ij}^{-1}(\mu^{th}-\mu^{obs})_{j}
\end{equation}
Here the theoretical distance modulus is given by
\begin{equation}
    \mu_{th}(z)=5log_{10}\frac{d_{L}(z)}{1Mpc}+25
\end{equation}
where $d_{L}(z)$ represents the luminosity distance which can be obtained by integrating the expression 
\begin{equation}
    d_{L}(z,\theta_s)=(1+z)\int_{0}^{z}\frac{dz^{\prime}}{H(z^{\prime},\theta_s)}\footnote{The speed of light c should be multiplied in the right-hand side of the equation while doing numerical analysis, however, as throughout the article we have worked taking $c=1$.}
\end{equation}
where $H(z)$ is the expression of the Hubble parameter for the cosmological model and $\theta_s$ is the parameter space of the cosmological model.
\subsection*{Phenomenological fitting}
\begin{figure}[htbp]
    \begin{subfigure}[b]{0.5\textwidth}
        \includegraphics[width=\textwidth]{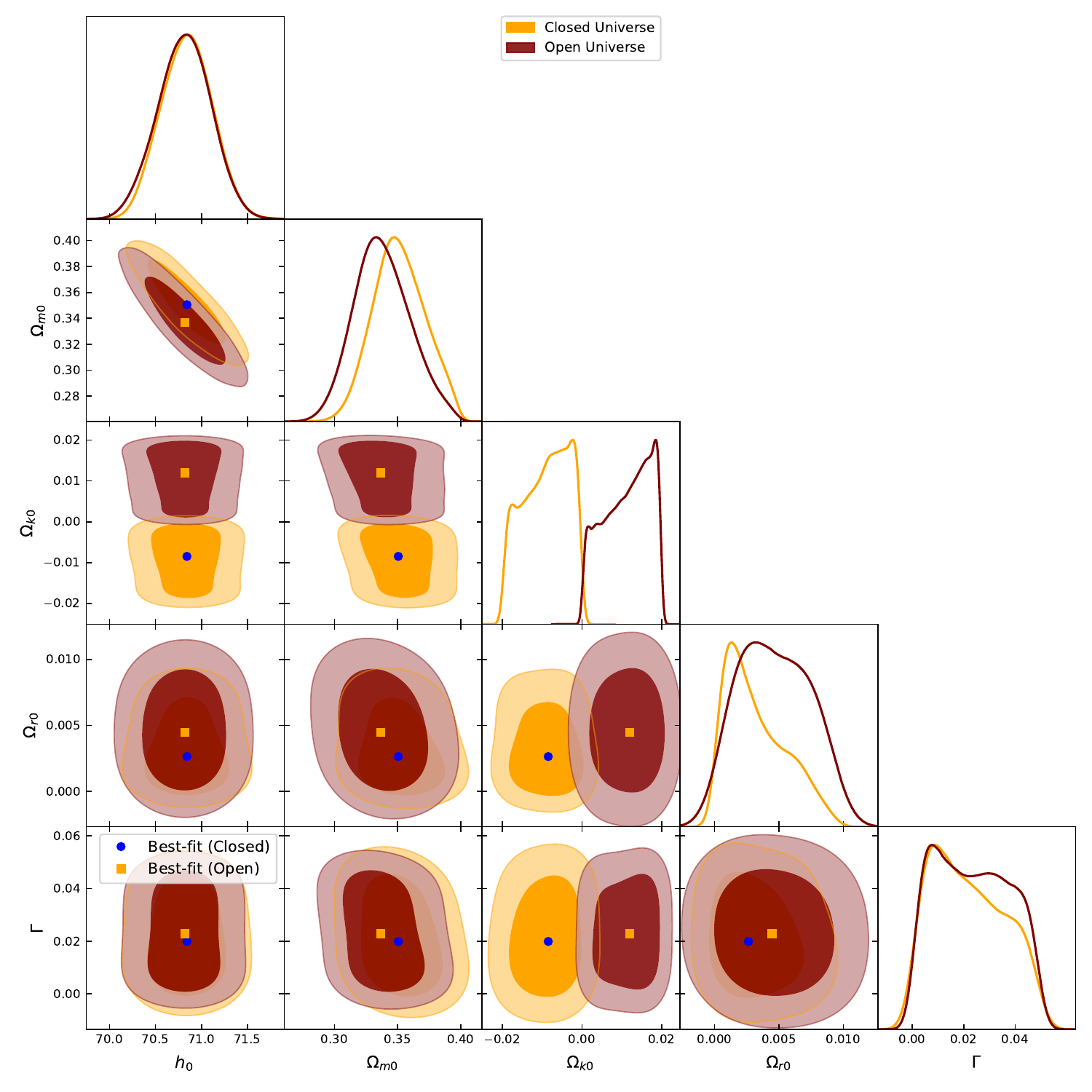}
        \caption{$\mathcal{Q}\propto\rho_{de}$}
    \end{subfigure}
    \hspace{-0.06\textwidth} 
    \begin{subfigure}[b]{0.5\textwidth}
        \includegraphics[width=\textwidth]{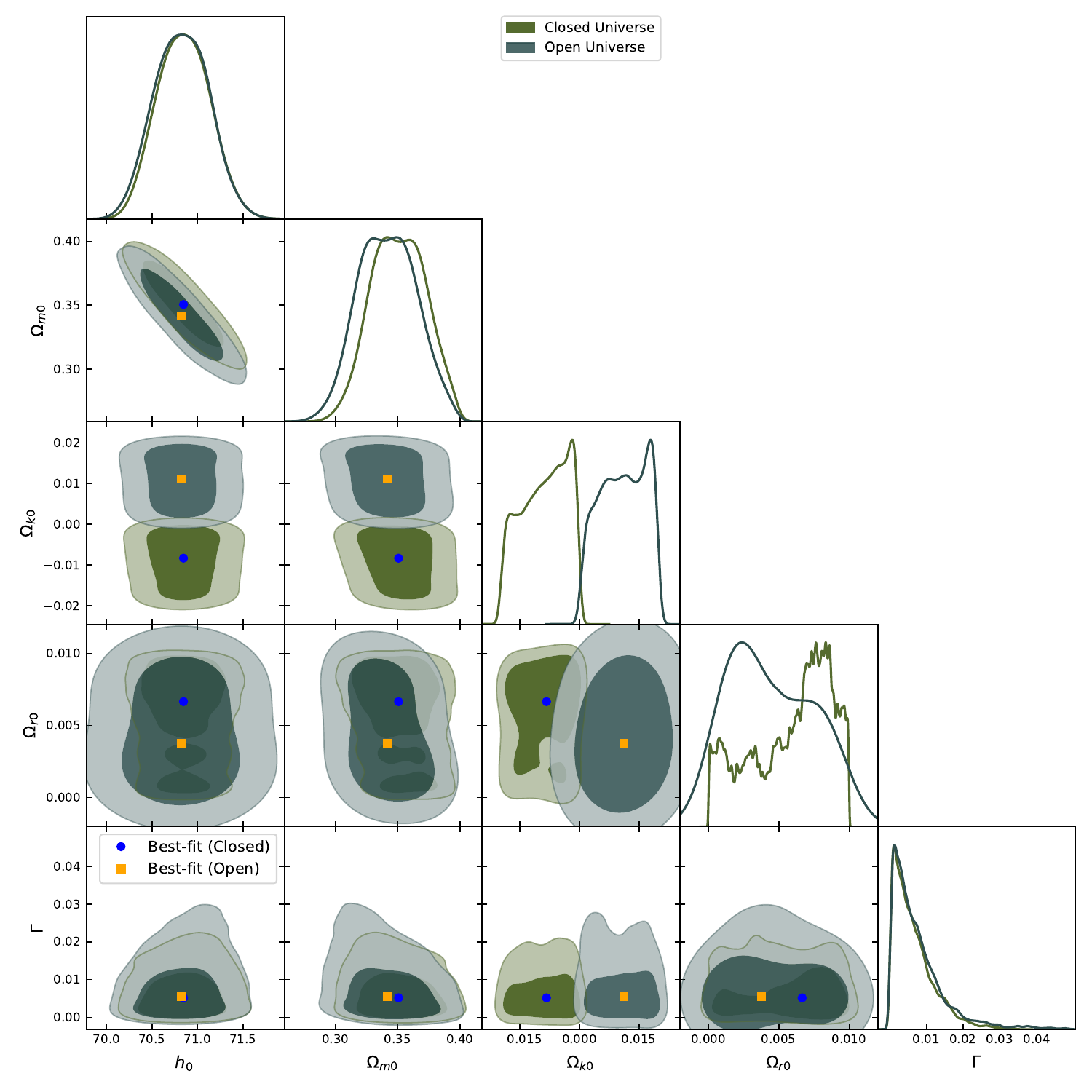}
        \caption{$\mathcal{Q}\propto\rho_{m}$}
    \end{subfigure}
     \begin{subfigure}[b]{0.5\textwidth}
        \includegraphics[width=\textwidth]{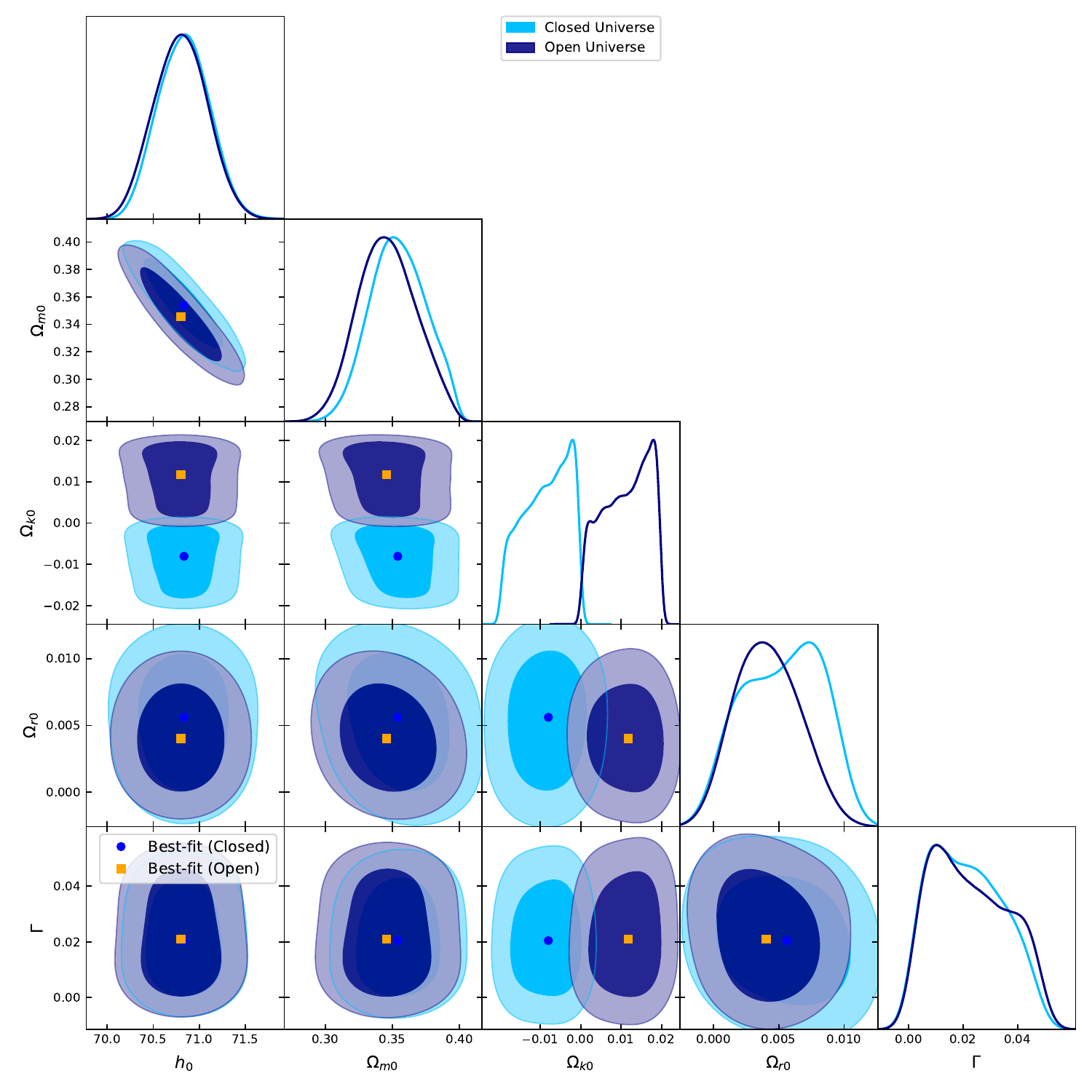}
        \caption{$\mathcal{Q}\propto\rho_{r}$}
    \end{subfigure}
    \hspace{-0.01\textwidth}
     \begin{subfigure}[b]{0.5\textwidth}
        \includegraphics[width=\textwidth]{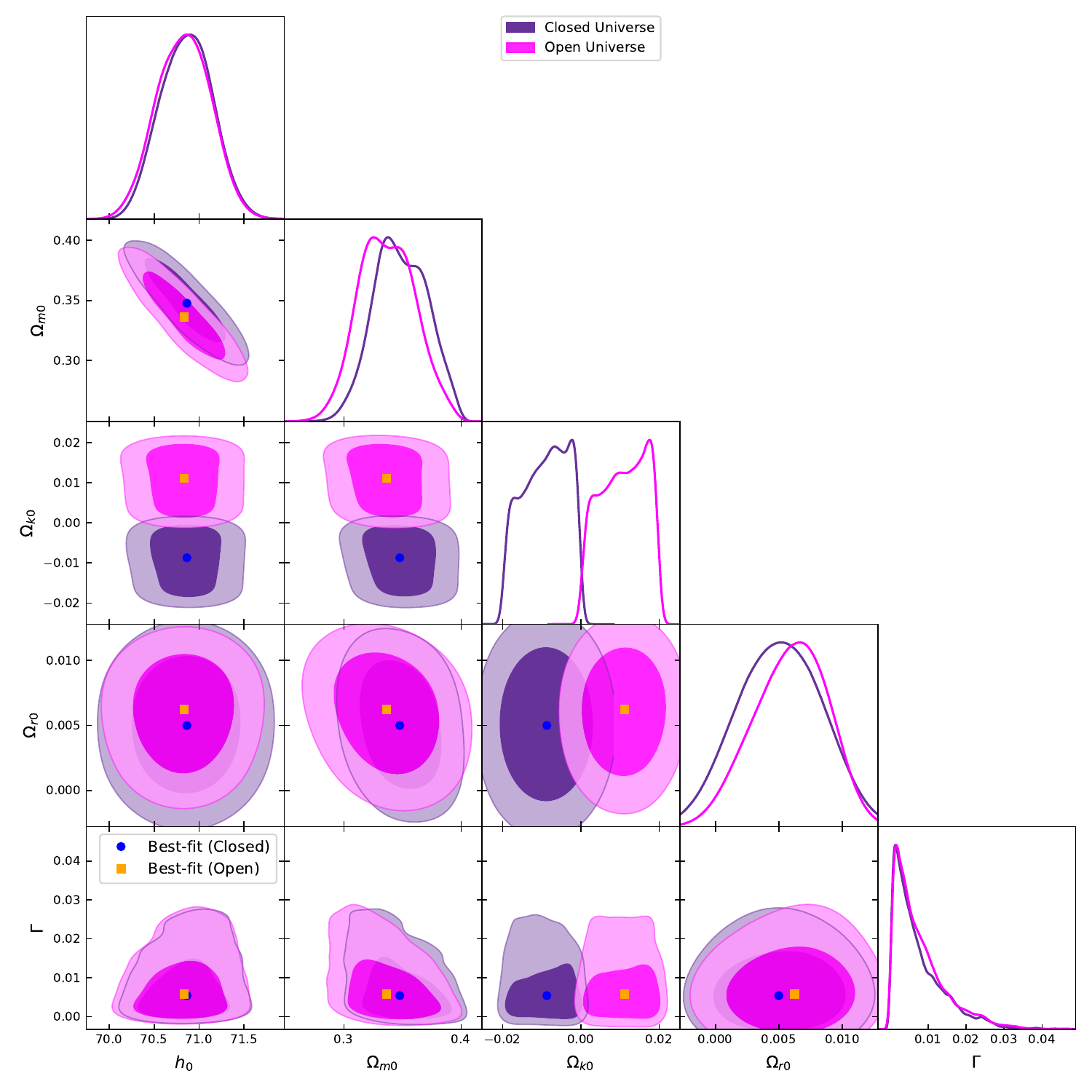}
        \caption{$\mathcal{Q}\propto\paren{\rho_{de}+\rho_{m}+\rho_{r}}$}
    \end{subfigure}
    \caption{Contour plots for all four different kinds of phenomenological interaction terms for Pantheon+ data set. Panel (a), (b), (c) and (d) shows the corner plots when the interaction term ($\mathcal{Q}$)is proportional to $\rho_{de}$, $\rho_{m}$, $\rho_{r}$ and $\rho_{de}+\rho_{m}+\rho_{r}$ respectively. For each panel, contours representing various model parameters are displayed alongside the best-fit points corresponding to observational data, are displayed for both open and closed universe scenarios.}
    \label{fig:approx fit Pantheon combined}
\end{figure}

Before constraining our actual model, we will try to constrain a phenomenological model where matter and radiation evolve independently, but the dynamics of dark energy depend on matter and radiation. In this model, we will consider the expression of the Hubble parameter such that matter, radiation, dark energy all evolve with the redshift parameter similar to the $\Lambda$CDM although the dark energy equation of state parameter is not $-1$ but it's value depends upon various energy density parameters. To be clear the expression of $\omega_{de}$ has been taken from sections (3) and (4). One should remember that, in this model we are not solving the coupled differential equations corresponding to $\ode$, $\om$ and $\ol$, we are just taking a modified dark energy EOS parameter provided every density parameter in the model evolve as the $\Lambda$CDM model. Therefore, one can say this phenomenological model is nothing but a small correction over the $\Lambda$CDM model for interacting Barrow holographic dark energy model. Here we will do all of our analysis with a Hubble parameter, which is given by the following expression 
\begin{equation}\label{approx hubble}
    H(z)=H_0\sqrt{{\om}_{,0}\paren{1+z}^3+{\ol}_{,0}\paren{1+z}^4+{\ok}_{,0}\paren{1+z}^2+{\ode}_{,0}(1+z)^{3\paren{1+\omega_{de}}}}~,
\end{equation}
where $\omega_{de}$ is the equation of state for the dynamical dark energy for this model. $H_{0}$, ${\ode}_{,0}$, ${\om}_{,0}$, ${\ol}_{,0}$ and ${\ok}_{,0}$ are present values corresponding the Hubble parameter, dark energy density parameter, dark matter density parameter, radiation density parameter, curvature density parameter of the universe respectively.\\
We will now proceed to constrain various parameters in this approximate model using CC \& BAO and Pantheon+ \& SHOes data sets. In order to constrain these model parameters, we need to perform $\chi^2$ minimization technique. We will use the expression of the Hubble parameter in eq.\eqref{approx hubble} to calculate the $\chi^2$ function for CC \& BAO and Pantheon+ \& SHOes data sets for both open and closed universe scenarios. We have calculated $\chi^2$ for all four different interaction models. Then we have used the Montecarlo-Hestings algorithm to minimize the $\chi^2$ function for all different interaction models to obtain the best fit values of different model parameters to match with observational data sets. Before doing the MCMC analysis, we have to set the prior values for different model parameters. For MCMC analysis, we have chosen the flat priors on different cosmological parameters of our model, as follows: $H_0 \in [50,80]$, ${\om}_{,0}\in [0.1,0.3]$, ${\ol}_{,0}\in [0,0.0003]$, $|{\ok}_{,0}|\in [0,0.01]$ and $\Gamma\in [0.01,0.03]$. It should be mentioned that for the approximate model while doing the MCMC analysis, we have set $\Delta=0.1$ and $\alpha=1$.\\
In Fig.\eqref{fig:approx fit CC combined}, we have shown the $1\sigma$ and $2\sigma$ likelihood contours (along with best fit points) for this approximate model corresponding to CC \& BAO dataset. For all four different phenomenological interaction terms, each panel in Fig.\eqref{fig:approx fit CC combined} shows likelihood contours for both closed and open universe scenarios. In Table \eqref{tab:pheno_interaction_params_CC_BAO} the best fit values of different cosmological model parameters along with their error values are provided.\\
Similarly Fig.\eqref{fig:approx fit Pantheon combined}, the $1\sigma$ and $2\sigma$ likelihood contours (along with best fit points) are plotted for this approximate model corresponding to the Pantheon+ \& SH0ES data set. Each panel of Fig.\eqref{fig:approx fit Pantheon combined} contains likelihood contours corresponding to both open and closed universe scenarios. The best fit parameters for all four different interaction models for both oepn and closed universe are provided in Table \eqref{tab:pheno_interaction_params_Pantheon_Shoes}.\\
For both data sets across all four interaction models, the optimal fit values of the Hubble parameter exceed those obtained in the $\Lambda$CDM model\footnote{For the $\Lambda$CDM model the value of the Hubble parameter is approximately $68$ km/s/Mpc.}. Based on the obtained results, it can be concluded that this approximate model has the potential to offer a viable solution to the Hubble tension problem. Also the results from Table \eqref{tab:pheno_interaction_params_CC_BAO} and Table \eqref{tab:pheno_interaction_params_Pantheon_Shoes} shows non zero values corresponding to the parameters ${\ok}_{,0}$ and $\Gamma$ for both open and closed universes in all four different interaction models. Non-zero values of ${\ok}_{,0}$ indicates presence of curvature in our universe also having a non-zero $\Gamma$ value shows interaction between dark energy, dark matter and radiation in this approximate model.

\begin{table}[h!]
    \centering
\renewcommand{\arraystretch}{1.5}
\begin{tabular}{|c||c|c||c|c||c|c||c|c|}
\hline
\textbf{Params} 
& \multicolumn{2}{c||}{$\mathcal{Q} = -\Gamma H \rho_{de}$} 
& \multicolumn{2}{c||}{$\mathcal{Q} = -\Gamma H \rho_{m}$} 
& \multicolumn{2}{c||}{$\mathcal{Q} = -\Gamma H \rho_{r}$} 
& \multicolumn{2}{c|}{$\mathcal{Q} = -\Gamma H(\rho_{de} + \rho_m + \rho_r)$} \\
\hline
         & $k$ & Value & $k$ & Value & $k$ & Value & $k$ & Value \\
\hline
$H_0$    & 1   & $71.694^{+0.801}_{-0.702}$ 
         & 1   & $71.706^{+0.780}_{-0.721}$ 
         & 1   & $71.250^{+0.709}_{-0.669}$ 
         & 1   & $71.880^{0.799}_{-0.706}$ \\
\cline{2-9}
         & -1  & $71.358^{+0.709}_{-0.705}$ 
         & -1  & $71.676^{+0.774}_{-0.691}$ 
         & -1  & $71.193^{+0.731}_{-0.668}$ 
         & -1  & $71.776^{0.765}_{-0.662}$ \\
\hline
${\om}_{,0}$    & 1   & $0.198^{+0.0057}_{-0.0113}$ 
         & 1   & $0.199^{+0.0053}_{-0.0115}$ 
         & 1   & $0.201^{+0.0067}_{-0.0074}$ 
         & 1   & $0.198^{+0.0057}_{-0.0112}$ \\
\cline{2-9}
         & -1  & $0.196^{+0.0071}_{-0.0073}$
         & -1  & $0.194^{+0.0047}_{-0.0114}$

         & -1  & $0.196^{+0.0069}_{-0.0072}$
 
         & -1  & $0.194^{+0.0049}_{-0.0109}$ \\
\hline
${\ol}_{,0}$    & 1   & $0.00014^{+0.00011}_{-0.00009}$ 
         & 1   & $0.00016^{+0.00010}_{-0.00010}$ 
         & 1   & $0.00014^{+0.00011}_{-0.00010}$ 
         & 1   & $0.00015^{+0.00011}_{-0.00010}$ \\
\cline{2-9}
         & -1  & $0.00013^{+0.00011}_{-0.00008}$ 
         & -1  & $0.00014^{+0.00011}_{-0.00010}$ 
         & -1  & $0.00014^{+0.00011}_{-0.00010}$ 
         & -1  & $0.00015^{+0.00010}_{-0.00010}$ \\
\hline
${\ok}_{,0}$    & 1   & $-0.0056^{+0.0038}_{-0.0029}$ 
         & 1   & $-0.0056^{+0.0036}_{-0.0030}$ 
         & 1   &  $-0.0057^{+0.0037}_{-0.0030}$ 
         & 1   & $-0.0054^{+0.0035}_{-0.0032}$ \\
\cline{2-9}
         & -1  & $0.0045^{+0.0035}_{-0.0031}$ 
         & -1  & $0.0044^{+0.0036}_{-0.0031}$ 
         & -1  & $0.0044^{+0.0035}_{-0.0032}$ 
         & -1  & $0.0046^{+0.0037}_{-0.0032}$ \\
\hline
$\Gamma$    & 1   & $0.0157^{+0.0076}_{-0.0042}$ 
         & 1   & $0.0252^{+0.0505}_{-0.0215}$ 
         & 1   & $0.0203^{+0.0061}_{-0.0064}$ 
         & 1   & $0.0146^{+0.0083}_{-0.0035}$ \\
\cline{2-9}
         & -1  & $0.0208^{+0.0067}_{-0.0082}$ 
         & -1  & $0.0152^{+0.0074}_{-0.0038}$ 
         & -1  & $0.0205^{+0.0070}_{-0.0070}$ 
         & -1  & $0.0147^{+0.0066}_{-0.0033}$ \\
\hline
\end{tabular}
     \caption{Best-fit cosmological parameters for various interaction terms and spatial curvature values for CC+BAO data. }
    \label{tab:pheno_interaction_params_CC_BAO}
\end{table}
\begin{table}[h!]
    \centering
\renewcommand{\arraystretch}{1.5}
\begin{tabular}{|c||c|c||c|c||c|c||c|c|}
\hline
\textbf{Params} 
& \multicolumn{2}{c||}{$\mathcal{Q} = -\Gamma H \rho_{de}$} 
& \multicolumn{2}{c||}{$\mathcal{Q} = -\Gamma H \rho_{m}$} 
& \multicolumn{2}{c||}{$\mathcal{Q} = -\Gamma H \rho_{r}$} 
& \multicolumn{2}{c|}{$\mathcal{Q} = -\Gamma H(\rho_{de} + \rho_m + \rho_r)$} \\
\hline
         & $k$ & Value & $k$ & Value & $k$ & Value & $k$ & Value \\
\hline
$H_0$    & 1   & $70.841^{+0.275}_{-0.283}$ 
         & 1   & $70.844^{+0.294}_{-0.293}$ 
         & 1   & $70.834^{+0.280}_{-0.286}$ 
         & 1   & $70.864^{0.288}_{-0.305}$ \\
\cline{2-9}
         & -1  & $70.818^{+0.282}_{-0.297}$ 
         & -1  & $70.823^{+0.304}_{-0.307}$ 
         & -1  & $70.800^{+0.288}_{-0.296}$ 
         & -1  & $70.833^{+0.301}_{-0.311}$ \\
\hline
${\om}_{,0}$    & 1   & $0.350^{+0.022}_{-0.019}$ 
         & 1   & $0.350^{+0.022}_{-0.022}$ 
         & 1   & $0.353^{+0.023}_{-0.020}$ 
         & 1   & $0.347^{+0.024}_{-0.021}$ \\
\cline{2-9}
         & -1  & $0.336^{+0.024}_{-0.020}$
         & -1  & $0.341^{+0.024}_{-0.023}$

         & -1  & $0.345^{+0.023}_{-0.021}$
 
         & -1  & $0.336^{+0.025}_{-0.023}$ \\
\hline
${\ol}_{,0}$    & 1   & $0.0026^{+0.0033}_{-0.0019}$ 
         & 1   & $0.0066^{+0.0020}_{-0.0043}$ 
         & 1   & $0.0056^{+0.0031}_{-0.0041}$ 
         & 1   & $0.0049^{+0.0035}_{-0.0034}$ \\
\cline{2-9}
         & -1  & $0.0044^{+0.0034}_{-0.0030}$ 
         & -1  & $0.0037^{+0.0045}_{-0.0026}$ 
         & -1  & $0.0040^{+0.0027}_{-0.0025}$ 
         & -1  & $0.0062^{+0.0027}_{-0.0035}$ \\
\hline
${\ok}_{,0}$    & 1   & $-0.0085^{+0.0058}_{-0.0073}$ 
         & 1   & $-0.0083^{+0.0059}_{-0.0074}$ 
         & 1   & $-0.0080^{+0.0057}_{-0.0071}$ 
         & 1   & $-0.0087^{+0.0060}_{-0.0071}$ \\
\cline{2-9}
         & -1  & $0.0119^{+0.0057}_{-0.0076}$ 
         & -1  & $0.0111^{+0.0064}_{-0.0068}$ 
         & -1  & $0.0117^{+0.0058}_{-0.0074}$ 
         & -1  & $0.0111^{+0.0063}_{-0.0071}$ \\
\hline
$\Gamma$    & 1   & $0.0199^{+0.0190}_{-0.0145}$ 
         & 1   & $0.0051^{+0.0065}_{-0.0037}$ 
         & 1   & $0.0204^{+0.0163}_{-0.0142}$ 
         & 1   & $0.0053^{+0.0086}_{-0.0039}$ \\
\cline{2-9}
         & -1  & $0.0228^{+0.0175}_{-0.0162}$ 
         & -1  & $0.0055^{+0.0074}_{-0.0040}$ 
         & -1  & $0.0209^{+0.0190}_{-0.0144}$ 
         & -1  & $0.0058^{+0.0083}_{-0.0042}$ \\
\hline
\end{tabular}
     \caption{Best-fit cosmological parameters for various interaction terms and spatial curvature values for Pantheon+ \& SH0ES data.}
    \label{tab:pheno_interaction_params_Pantheon_Shoes}
\end{table}
\subsection*{Fitting the exact model}
\begin{figure}[htbp]
    \centering

    \begin{subfigure}[b]{0.48\textwidth}
        \includegraphics[width=\textwidth]{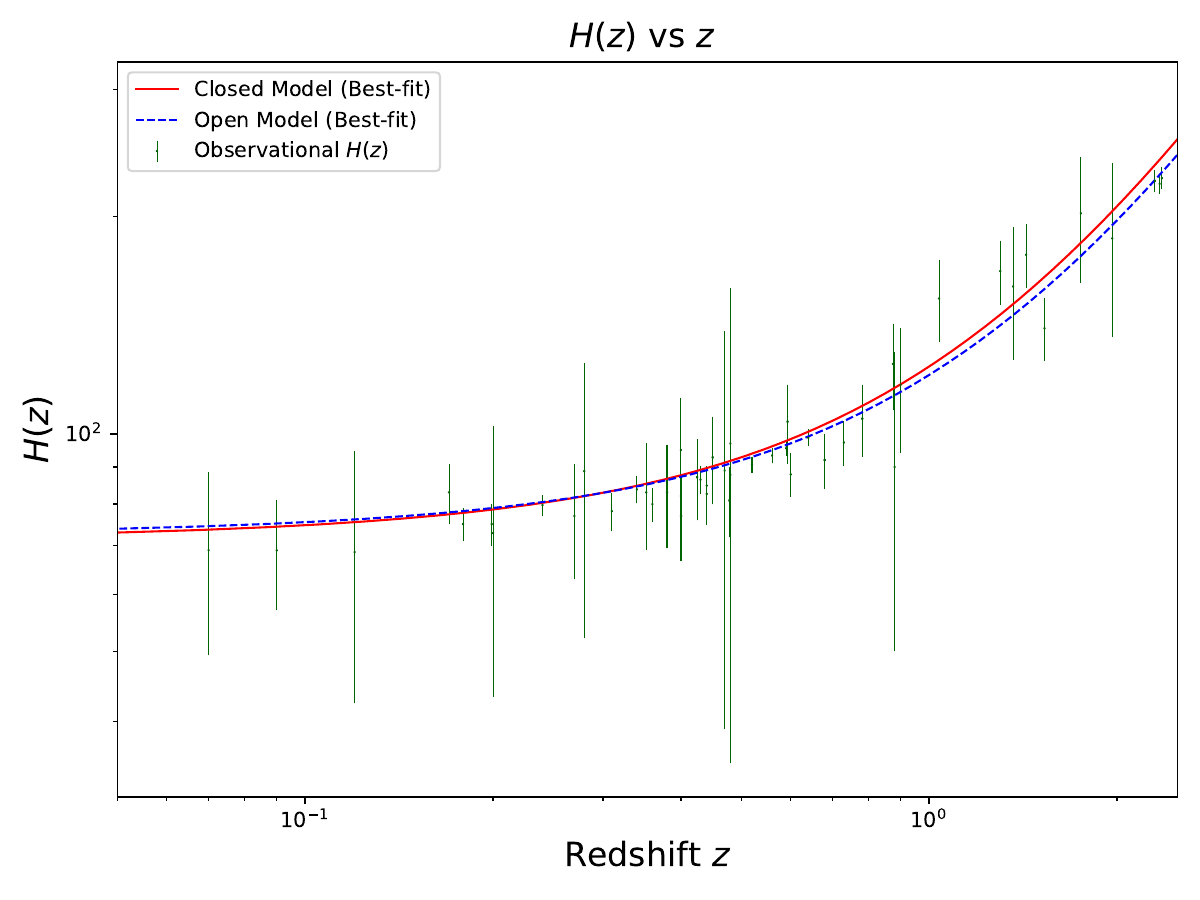}
        \caption{$\mathcal{Q}\propto\rho_{de}$}
        \label{fig:CC de}
    \end{subfigure}
    \hfill
    \begin{subfigure}[b]{0.48\textwidth}
        \includegraphics[width=\textwidth]{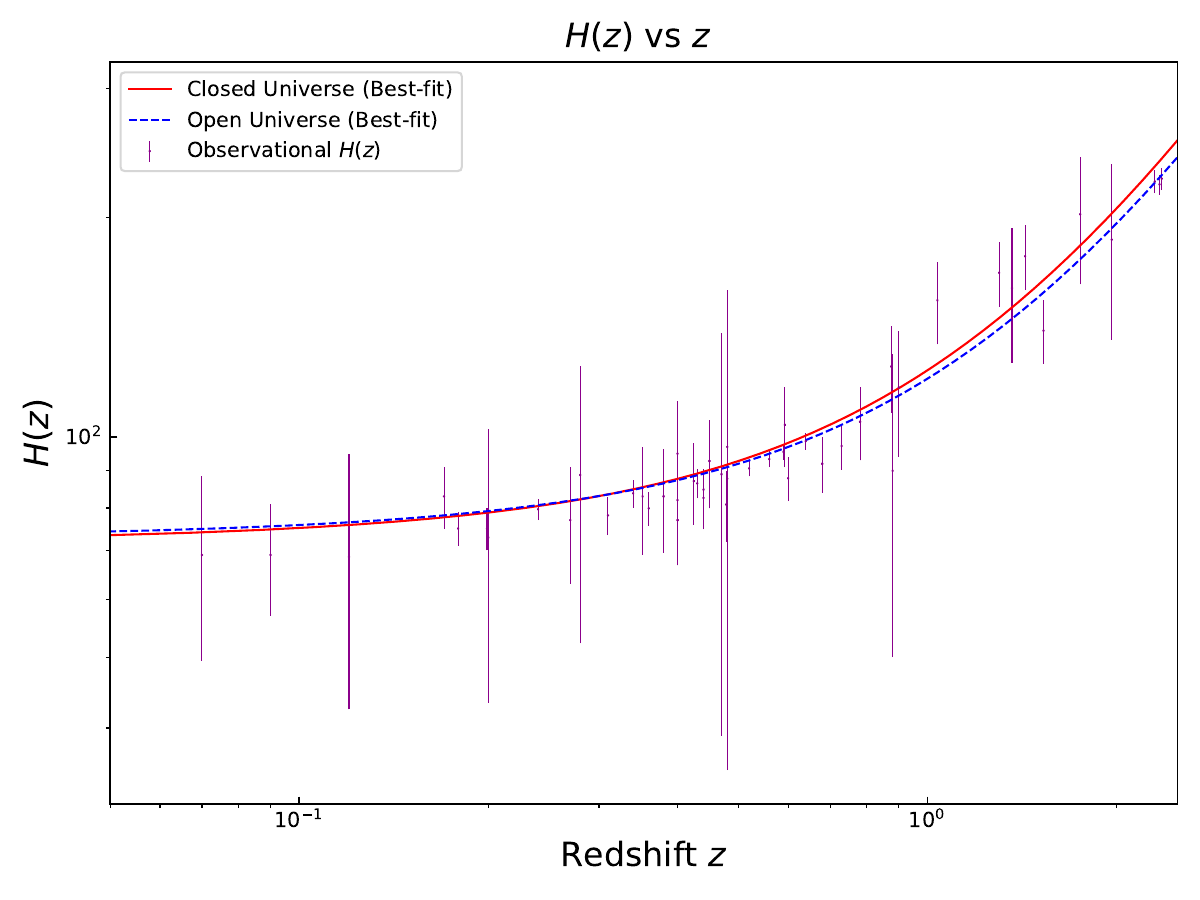}
        \caption{$\mathcal{Q}\propto\rho_{m}$}
        \label{fig:CC m}
    \end{subfigure}

    \vskip\baselineskip 

    \begin{subfigure}[b]{0.48\textwidth}
        \includegraphics[width=\textwidth]{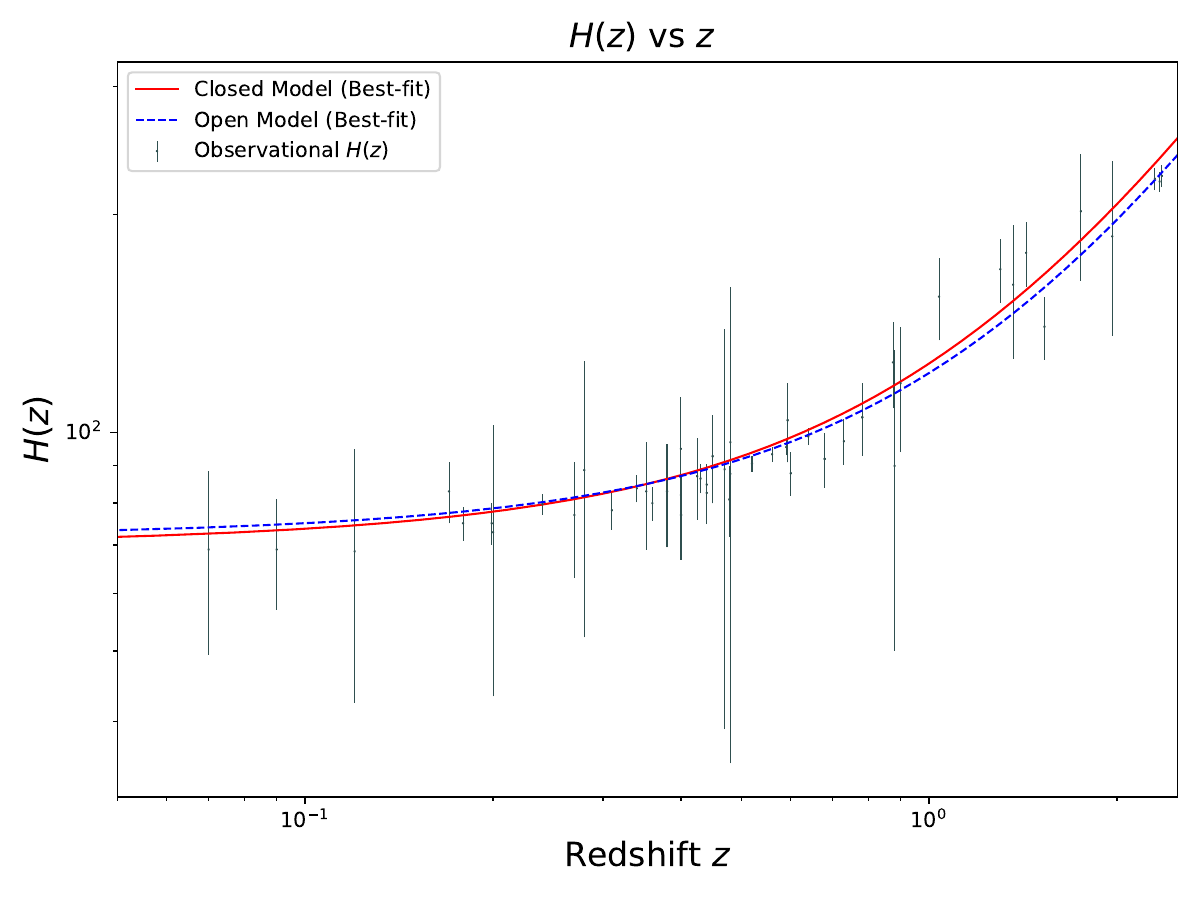}
        \caption{$\mathcal{Q}\propto\rho_{r}$}
        \label{fig:CC r}
    \end{subfigure}
    \hfill
    \begin{subfigure}[b]{0.48\textwidth}
        \includegraphics[width=\textwidth]{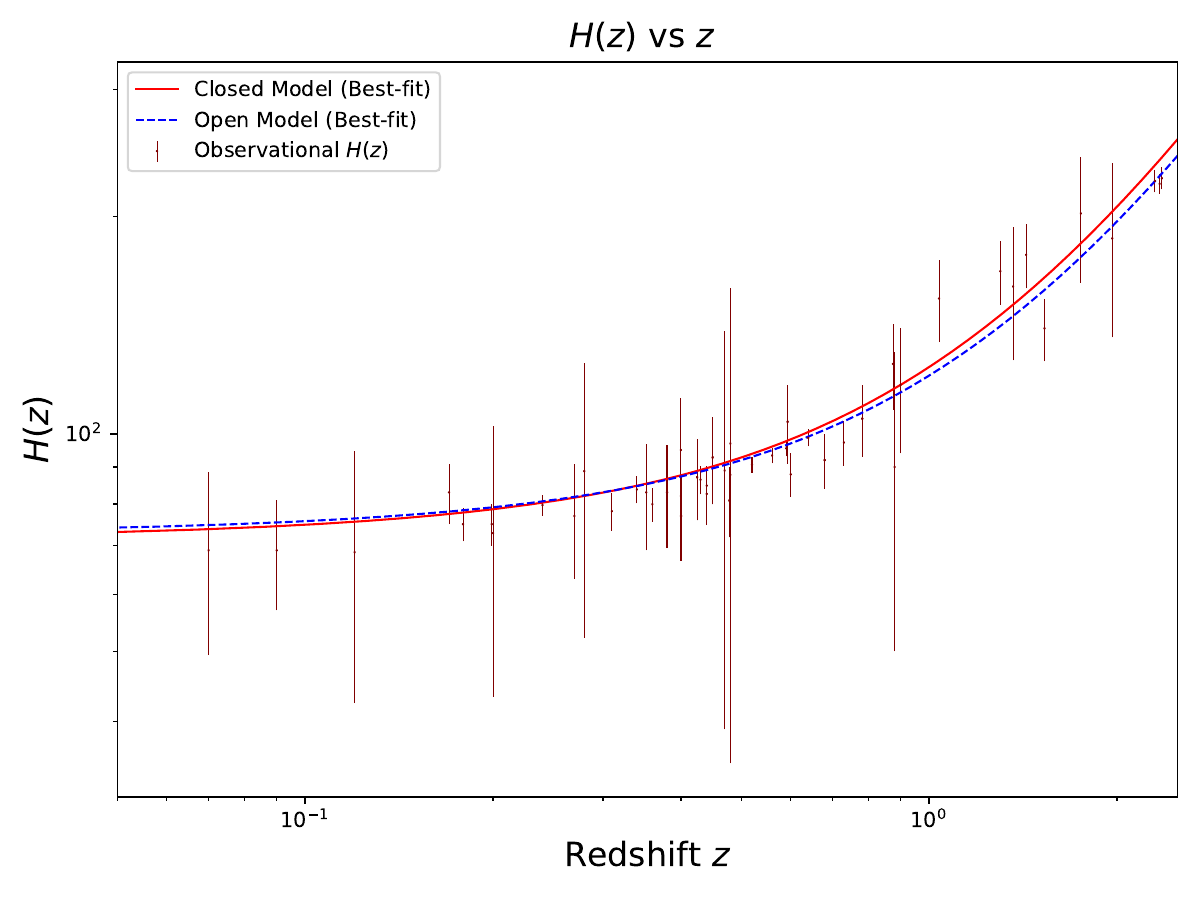}
        \caption{$\mathcal{Q}\propto\left(\rho_{de}+\rho_{m}+\rho_{r}\right)$}
        \label{fig:CC all}
    \end{subfigure}

    \caption{The Hubble parameter versus redshift graphs for all four distinct interaction models are presented. In panels (a), (b), (c), and (d), the Hubble parameter \(H(z)\) is plotted against redshift \(z\) for models where the interaction term \(\mathcal{Q}\) is proportional to \(\rho_{de}\), \(\rho_{m}\), \(\rho_{r}\), and the combined density \(\rho_{de} + \rho_{m} + \rho_{r}\), respectively, using CC \& BAO data. In each panel, solid and dashed lines represent the best-fit results from our model for closed and open universe configurations, respectively. Additionally, we have also plotted the observational data points with their associated error bars for CC \& BAO data set.}
    \label{fig:combined H vs z}
\end{figure}

\begin{figure}[h!]
    \begin{subfigure}[b]{0.5\textwidth}
        \includegraphics[width=\textwidth]{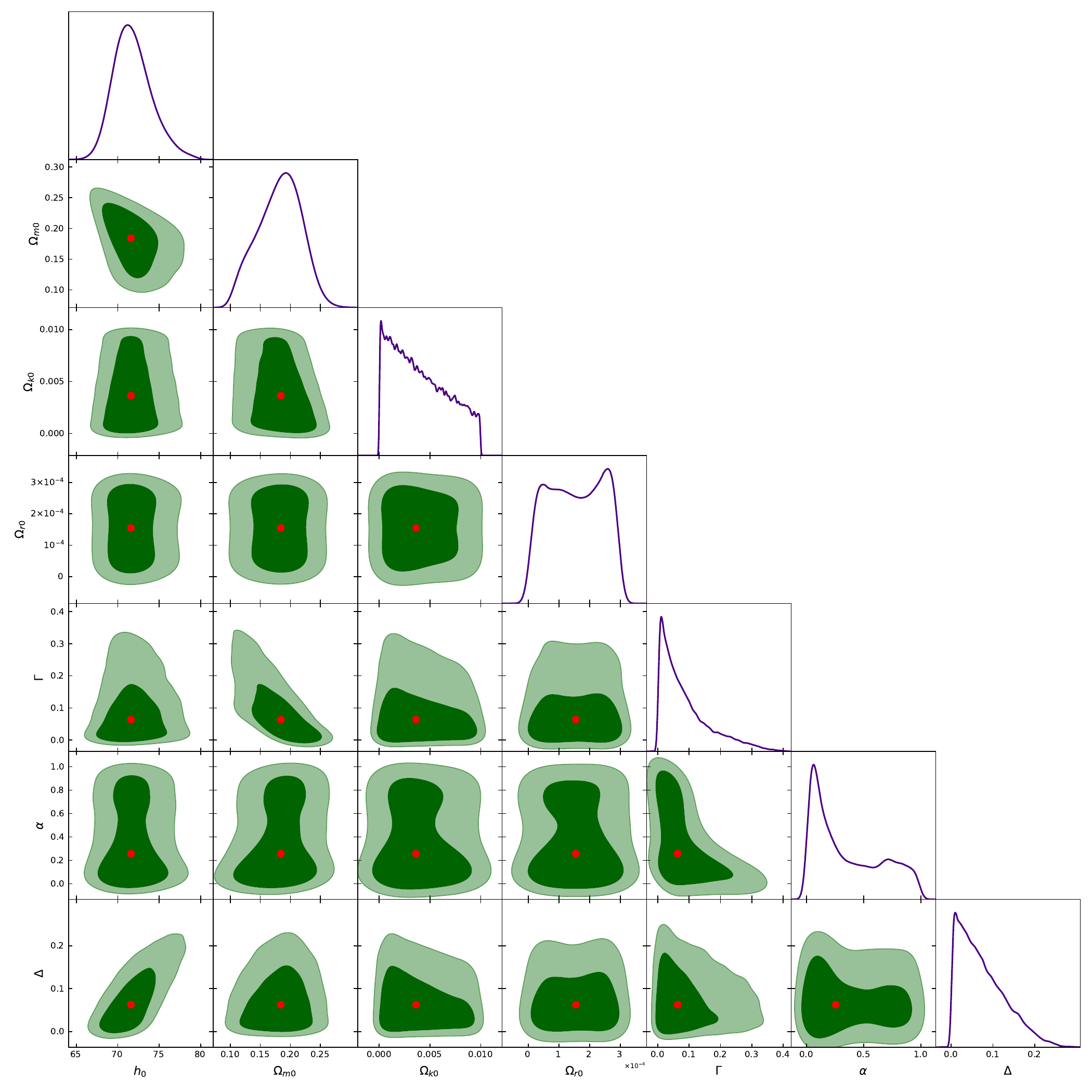}
        \caption{$\mathcal{Q}\propto\rho_{de}$}
    \end{subfigure}
    \hspace{-0.06\textwidth} 
    \begin{subfigure}[b]{0.5\textwidth}
        \includegraphics[width=\textwidth]{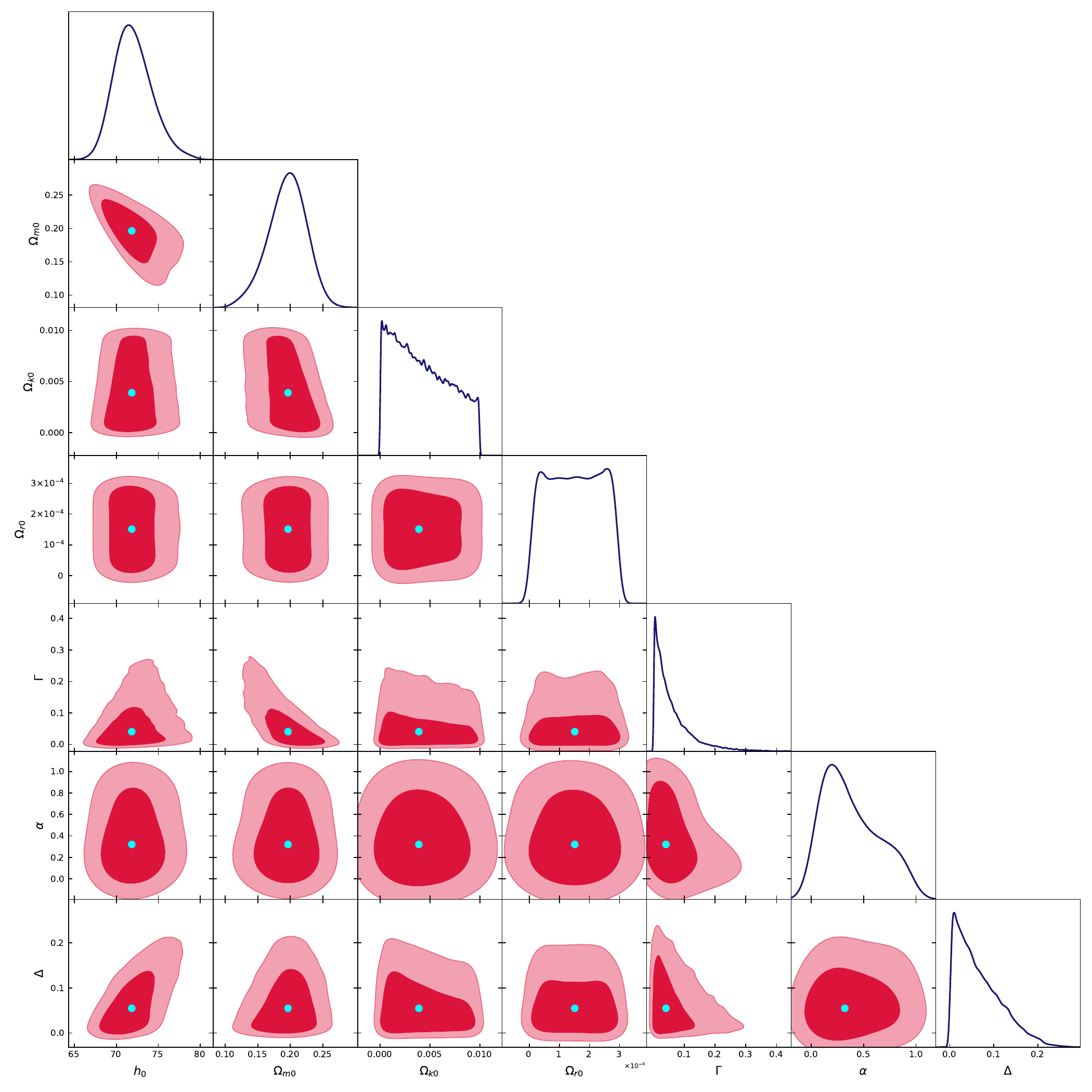}
        \caption{$\mathcal{Q}\propto\rho_{m}$}
    \end{subfigure}
     \begin{subfigure}[b]{0.5\textwidth}
        \includegraphics[width=\textwidth]{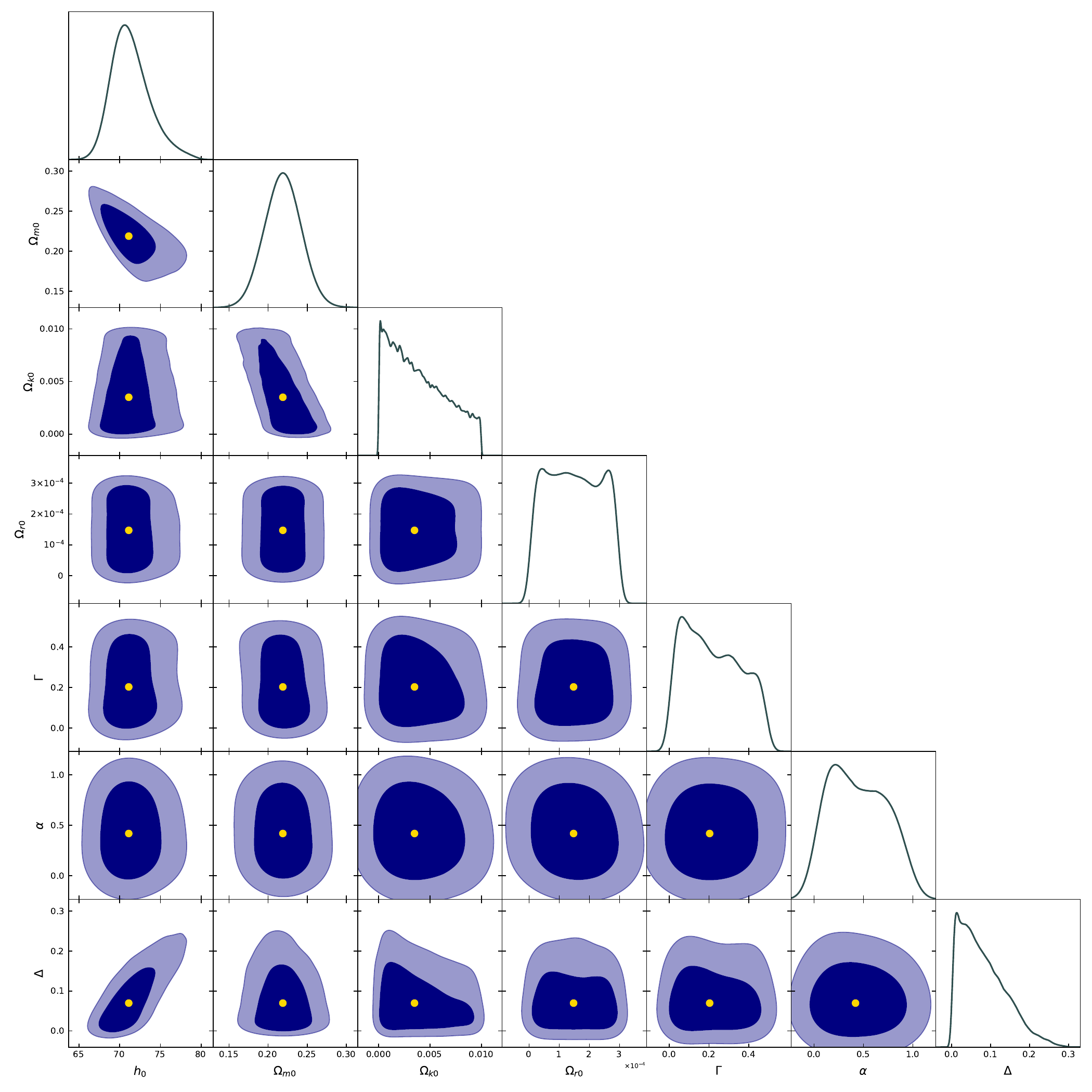}
        \caption{$\mathcal{Q}\propto\rho_{r}$}
    \end{subfigure}
    \hspace{-0.01\textwidth}
     \begin{subfigure}[b]{0.5\textwidth}
        \includegraphics[width=\textwidth]{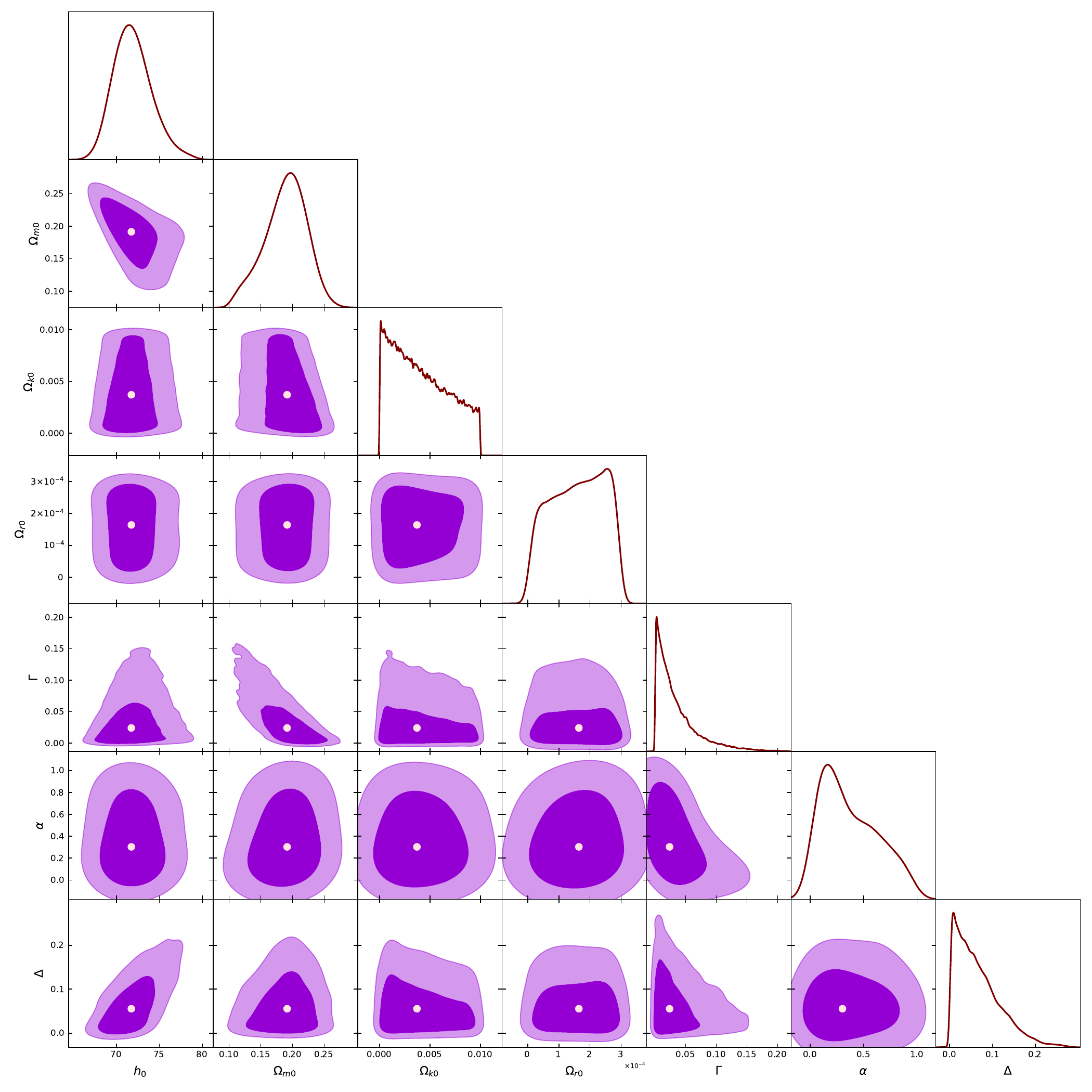}
        \caption{$\mathcal{Q}\propto\paren{\rho_{de}+\rho_{m}+\rho_{r}}$}
    \end{subfigure}
    \caption{Contour plots for all four different kind of phenomenological interaction terms for CC+BAO data set for open universe. Panel (a), (b), (c) and (d) shows the corner plots when the interaction term ($\mathcal{Q}$)is proportional to $\rho_{de}$, $\rho_{m}$, $\rho_{r}$ and $\rho_{de}+\rho_{m}+\rho_{r}$ respectively. For each panel, contours representing various model parameters are displayed alongside the best-fit points corresponding to observational data, are displayed for both open and closed universe scenarios.}
    \label{fig:exact open CC+BAO combined}
\end{figure}

\begin{figure}[htbp]
    \begin{subfigure}[b]{0.5\textwidth}
        \includegraphics[width=\textwidth]{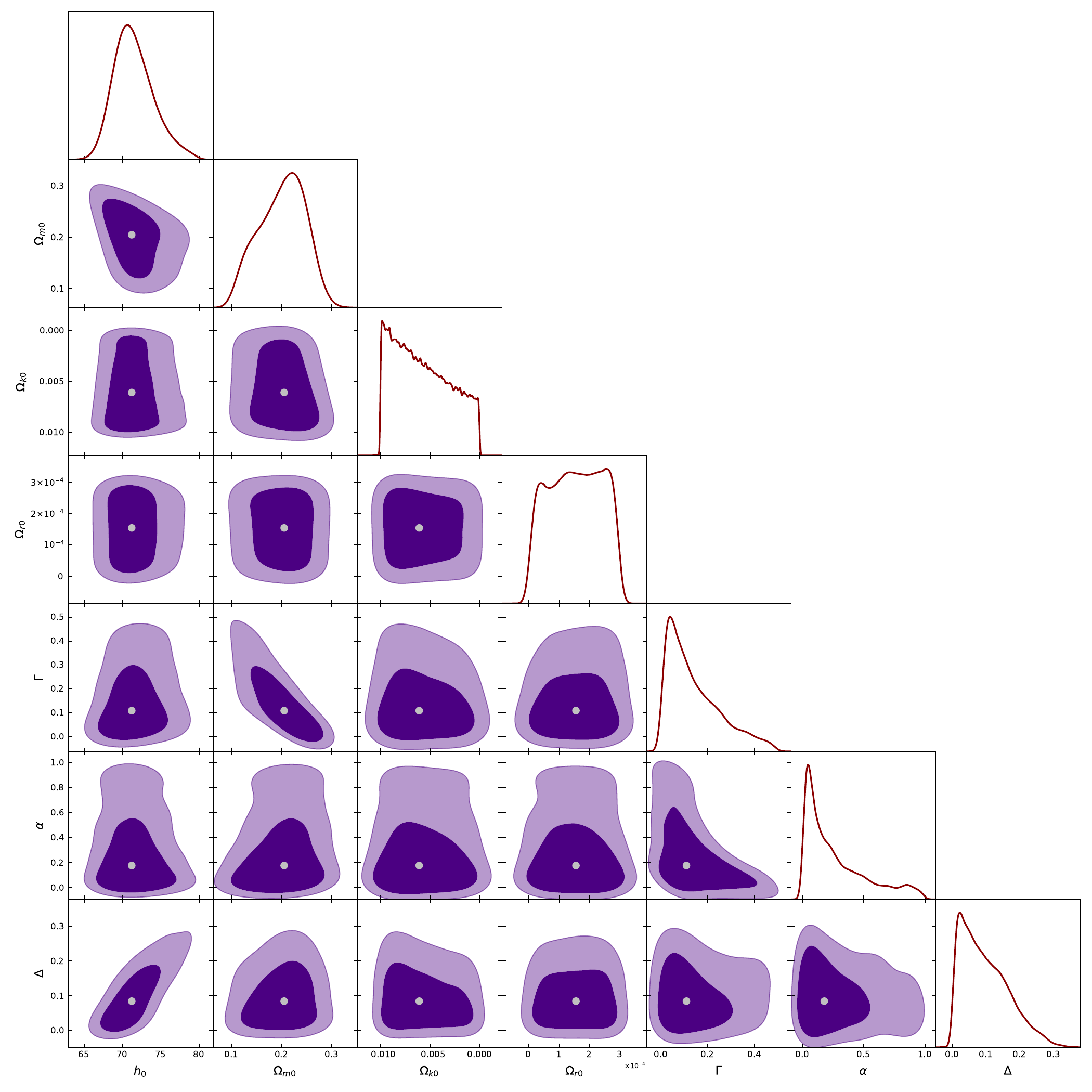}
        \caption{$\mathcal{Q}\propto\rho_{de}$}
    \end{subfigure}
    \hspace{-0.06\textwidth} 
    \begin{subfigure}[b]{0.5\textwidth}
        \includegraphics[width=\textwidth]{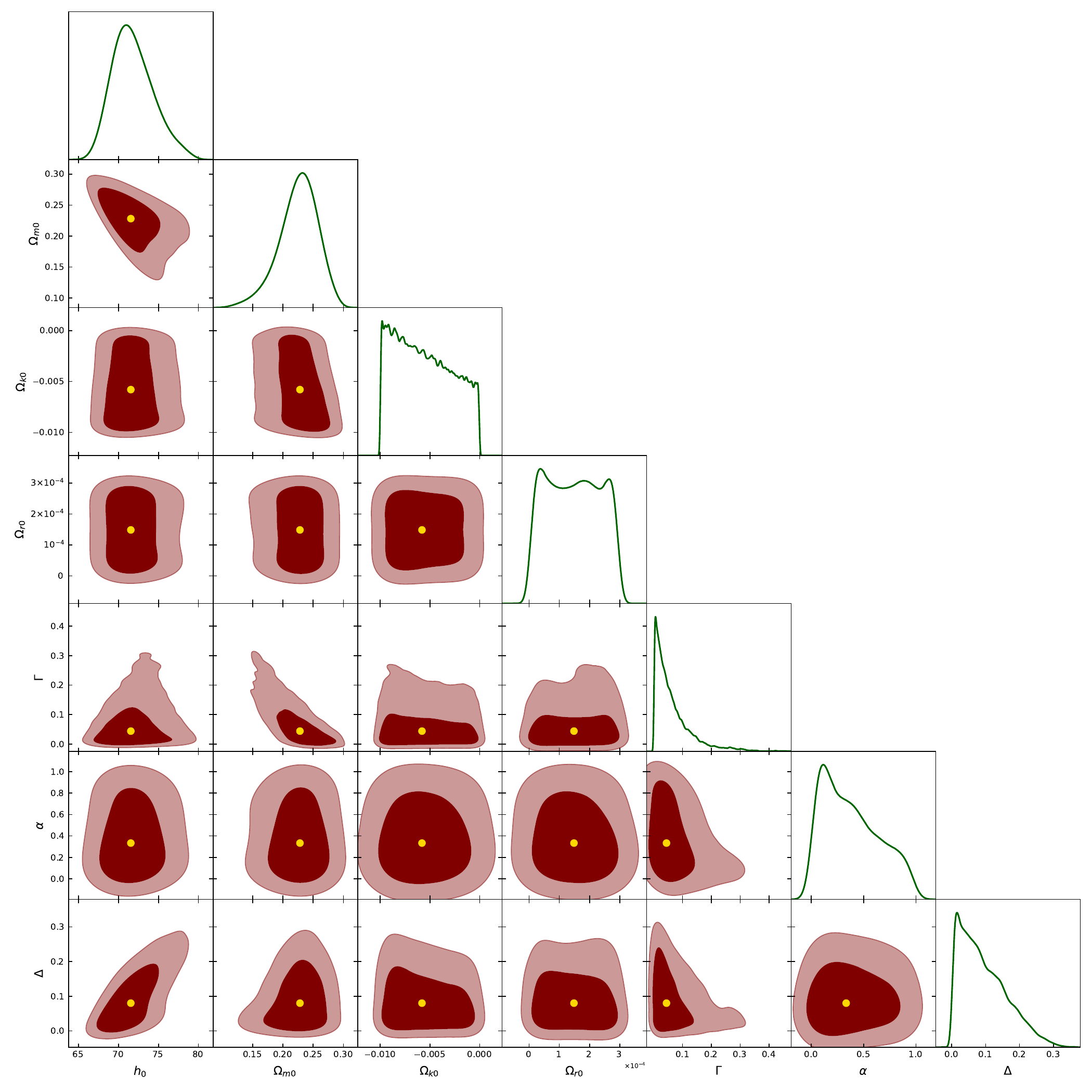}
        \caption{$\mathcal{Q}\propto\rho_{m}$}
    \end{subfigure}
     \begin{subfigure}[b]{0.5\textwidth}
        \includegraphics[width=\textwidth]{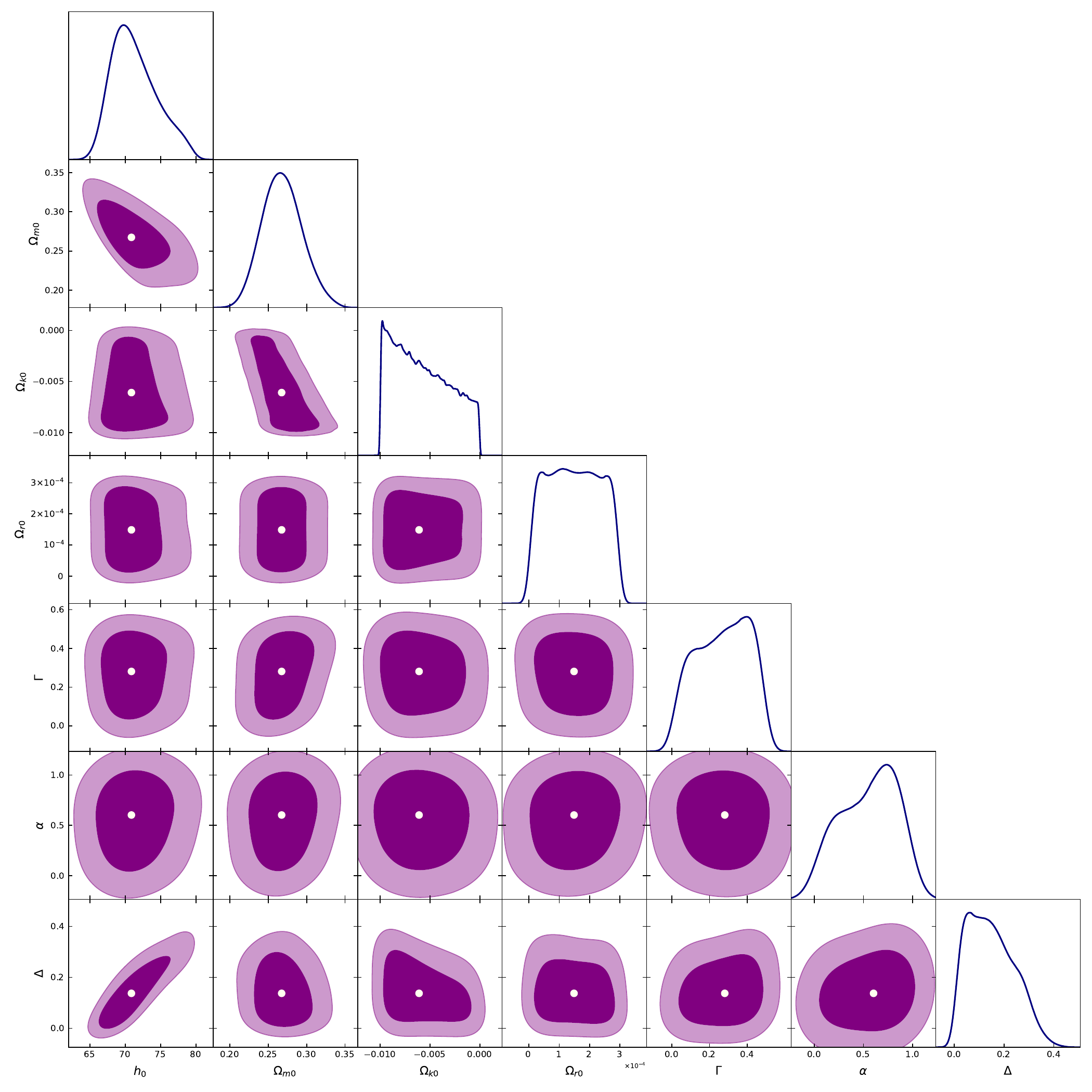}
        \caption{$\mathcal{Q}\propto\rho_{r}$}
    \end{subfigure}
    \hspace{-0.01\textwidth}
     \begin{subfigure}[b]{0.5\textwidth}
        \includegraphics[width=\textwidth]{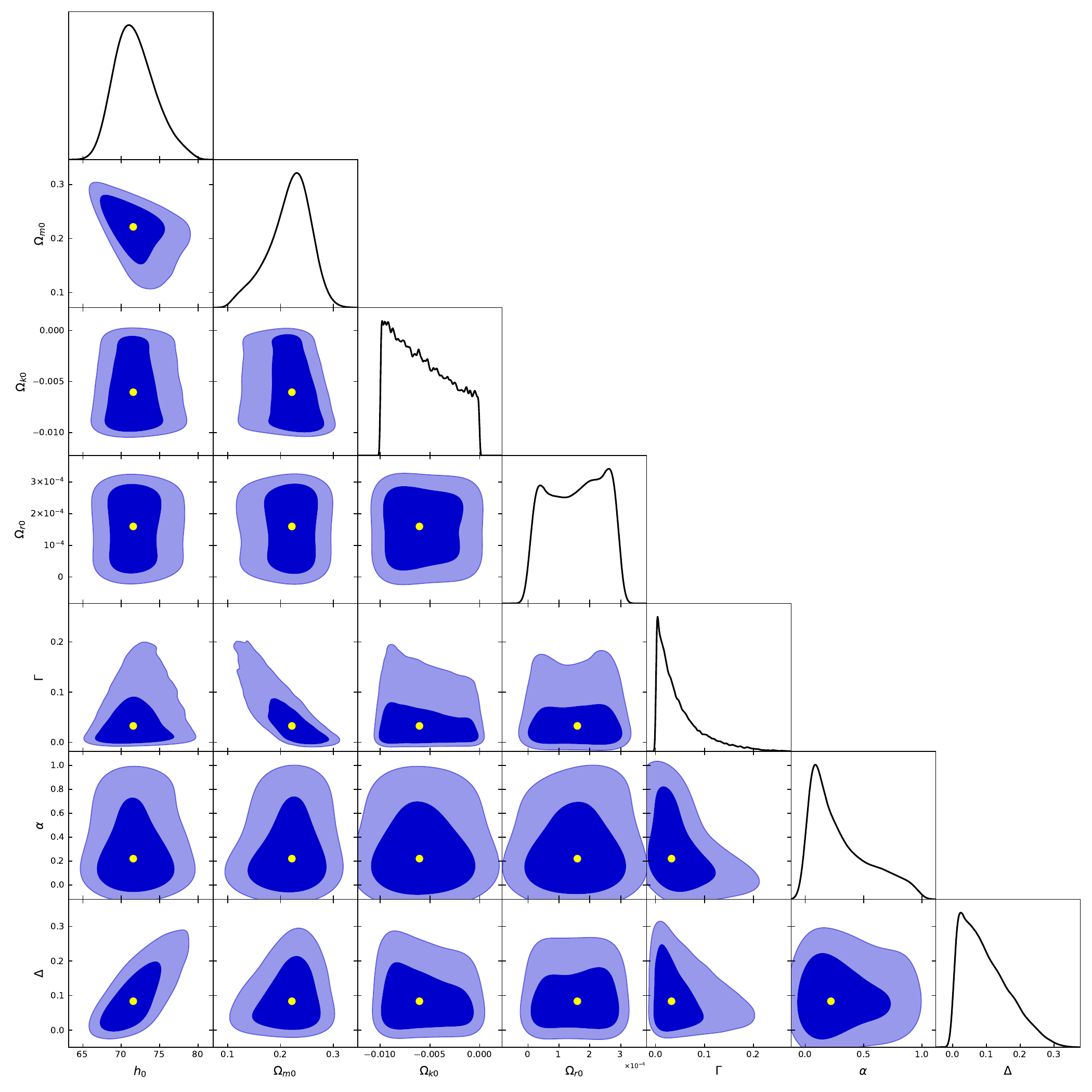}
        \caption{$\mathcal{Q}\propto\paren{\rho_{de}+\rho_{m}+\rho_{r}}$}
    \end{subfigure}
    \caption{Contour plots for all four different kind of phenomenological interaction terms for CC+BAO data set for closed universe. Panel (a), (b), (c) and (d) shows the corner plots when the interaction term ($\mathcal{Q}$)is proportional to $\rho_{de}$, $\rho_{m}$, $\rho_{r}$ and $\rho_{de}+\rho_{m}+\rho_{r}$ respectively. For each panel, contours representing various model parameters are displayed alongside the best-fit points corresponding to observational data, are displayed for both open and closed universe scenarios.}
    \label{fig:exact closed CC+BAO combined}
\end{figure}

Now we will move forward to obtain the best fit parameters for our model by solving the coupled differential equations corresponding to $\ode$, $\om$ and $\ol$ for all four different kinds of interaction parameters and for both open and closed universes. The solution of these equations is used to obtain the value of $\chi^2$ function for both CC+BAO and Pantheon+Shoes data sets. We have used the following expression of the Hubble parameter to calculate the $\chi^2$ function for each data sets
\begin{equation}
    H(z)=H_0\sqrt{{\om}+{\ol}+{\ok}_{,0}\paren{1+z}^2+{\ode}}~,
\end{equation}
where $\om,\ol$ and $\ode$ come from the numerical solution of the coupled differential equations for a particular interaction model. $H_0$, ${\ok}_{,0}$ are present time values of the Hubble parameter and curvature density parameter.\\
Then this $\chi^2$ function is minimized using the Montecarlo-Hestings  algorithm to obtain the best fit values of our model parameters. We would like to mention that for the approximate model, we have not treated $\Delta$ and $\alpha$ to be free parameters and their values were fixed. Although while studying the exact model, we will treat them as free parameters. Before doing the $\chi^2$ minimization for both CC \& BAO and Pantheon+ \& SH0ES data sets, we have to set prior values for our model parameters ($H_0$, ${\om}_{,0}$, ${\ol}_{,0}$, ${\ok}_{,0}$, $\Gamma$, $\Delta$ and $\alpha$). Here also for doing the MCMC analysis, we have chosen the flat priors on different cosmological parameters of our model, as follows: $H_0 \in [50,80]$, ${\om}_{,0}\in [0.1,0.35]$, ${\ol}_{,0}\in [0,0.0003]$, $|{\ok}_{,0}|\in [0,0.01]$, $\Gamma\in [0.0001,0.5]$, $\Delta\in [0,1]$ and $\alpha\in [0,1]$.\\
In Fig.\eqref{fig:combined H vs z}, we present the best-fit graphs illustrating the variation of our model's Hubble parameter, $H(z)$, with respect to redshift $z$, based on the CC and BAO datasets. Fig.(s)(\eqref{fig:CC de},\eqref{fig:CC m},\\ \eqref{fig:CC r},\eqref{fig:CC all}) corresponds to the phenomenological interaction terms $\mathcal{Q}=-\Gamma H \rho_{de}$, $\mathcal{Q}=-\Gamma H \rho_{de}$, $\mathcal{Q}=-\Gamma H \rho_{m}$, $\mathcal{Q}=-\Gamma H \rho_{r}$ and $\mathcal{Q}=-\Gamma H (\rho_{de}+\rho_{m}+\rho_{r})$ respectively. For each panel in Fig.\eqref{fig:combined H vs z}, the solid lines are for closed universe scenario and the dashed lined are for open universe scenario. In each panel, to validate our fitted curve, we have additionally displayed the observational data points with their corresponding error bars for the CC \& BAO datasets. Fig.(s)(\eqref{fig:exact closed CC+BAO combined},\eqref{fig:exact open CC+BAO combined}), shows the $1\sigma$ and $2\sigma$ likelihood contours for different model parameters along with their best fit points for closed and open universe respectively. 
In Table \eqref{tab:original_interaction_params_CC_BAO}, we have provided the best fit values of our model parameters which are obtained from CC \& BAO data set. One can see that the constrained values of Hubble parameter for all four different interaction models have a higher value compared to the $\lambda$CDM model. Thus this model could lead to a potential solution for the Hubble tension problem. 

\begin{figure}[htbp]
    \begin{subfigure}[b]{0.55\textwidth}
        \includegraphics[width=\textwidth]{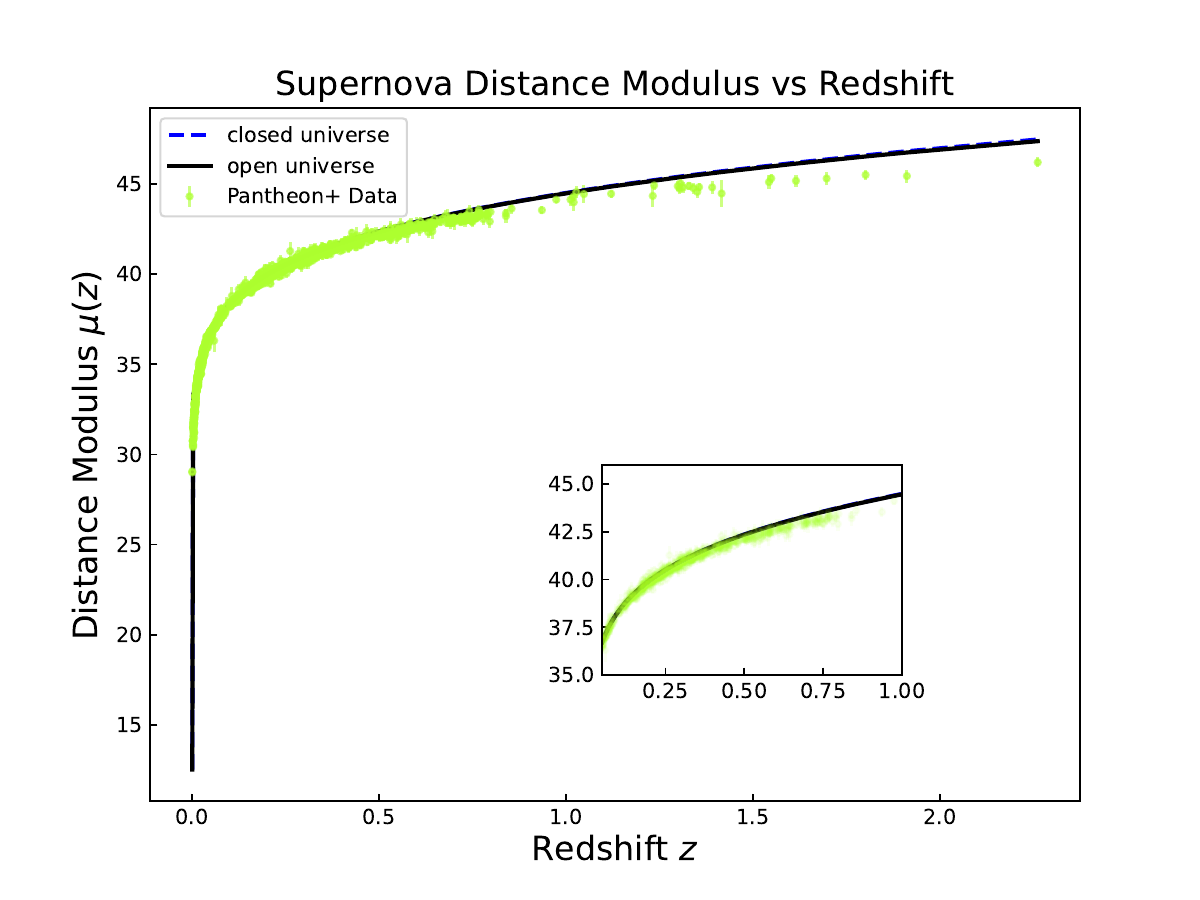}
        \caption{$\mathcal{Q}\propto\rho_{de}$}
        \label{fig:panth de}
    \end{subfigure}
    \hspace{-0.06\textwidth} 
    \begin{subfigure}[b]{0.55\textwidth}
        \includegraphics[width=\textwidth]{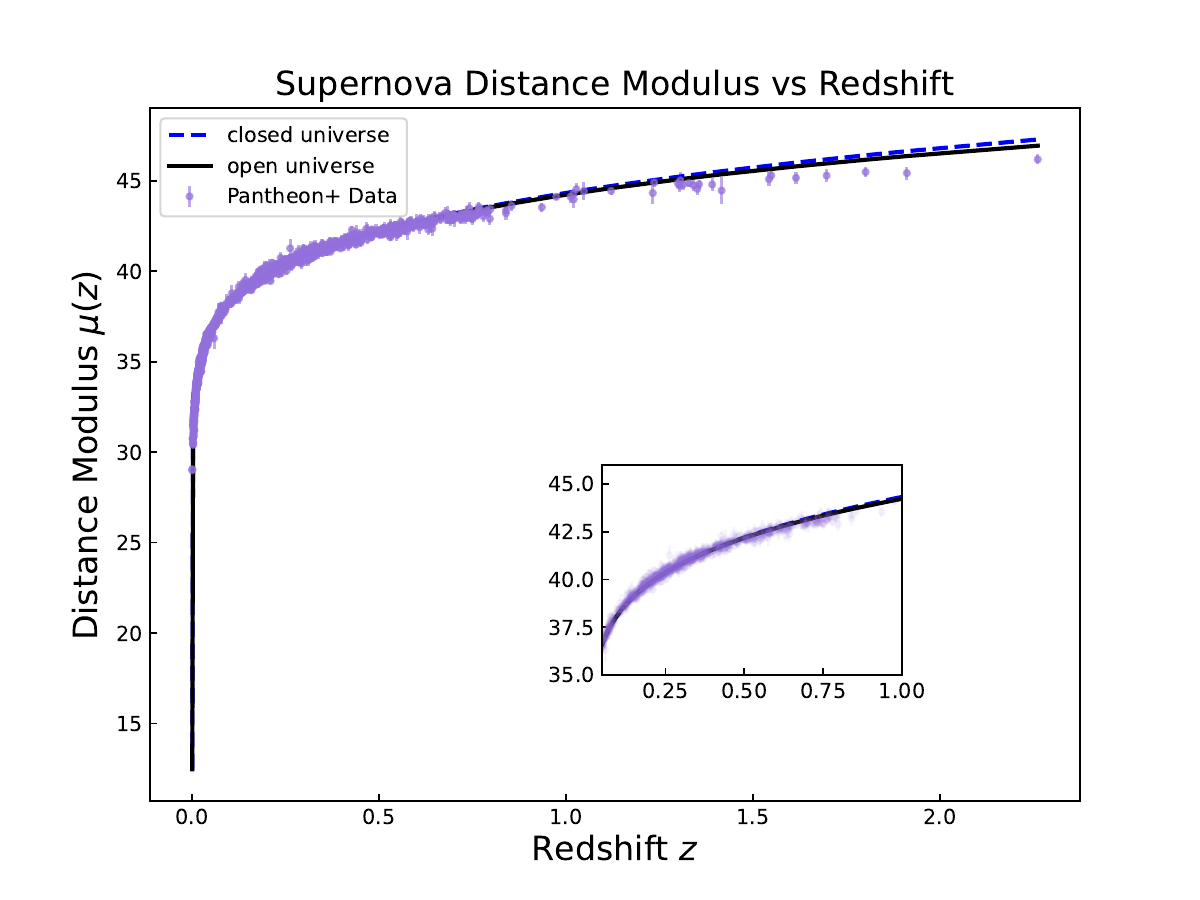}
        \caption{$\mathcal{Q}\propto\rho_{m}$}
        \label{fig:panth m}
    \end{subfigure}
     \begin{subfigure}[b]{0.55\textwidth}
        \includegraphics[width=\textwidth]{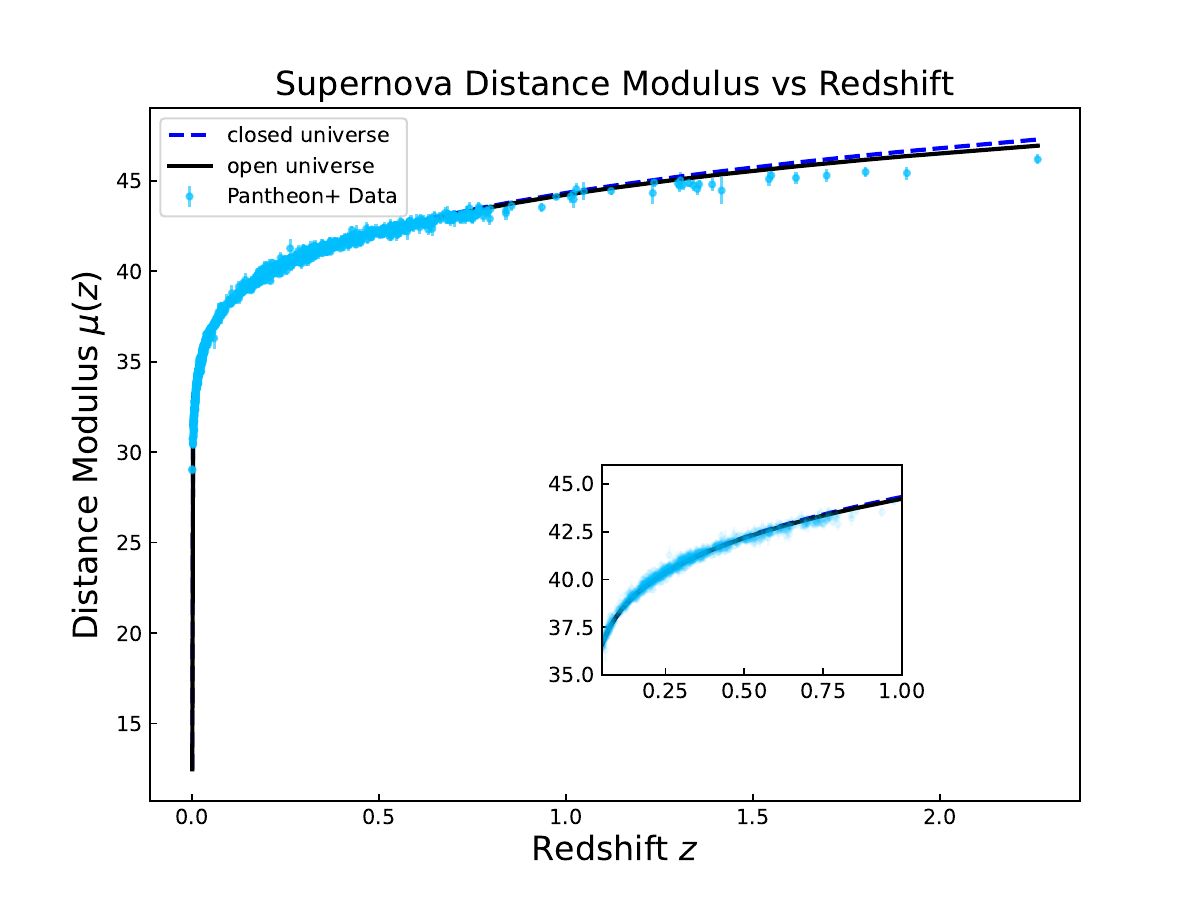}
        \caption{$\mathcal{Q}\propto\rho_{r}$}
        \label{fig:panth r}
    \end{subfigure}
    \hspace{-0.06\textwidth}
     \begin{subfigure}[b]{0.55\textwidth}
        \includegraphics[width=\textwidth]{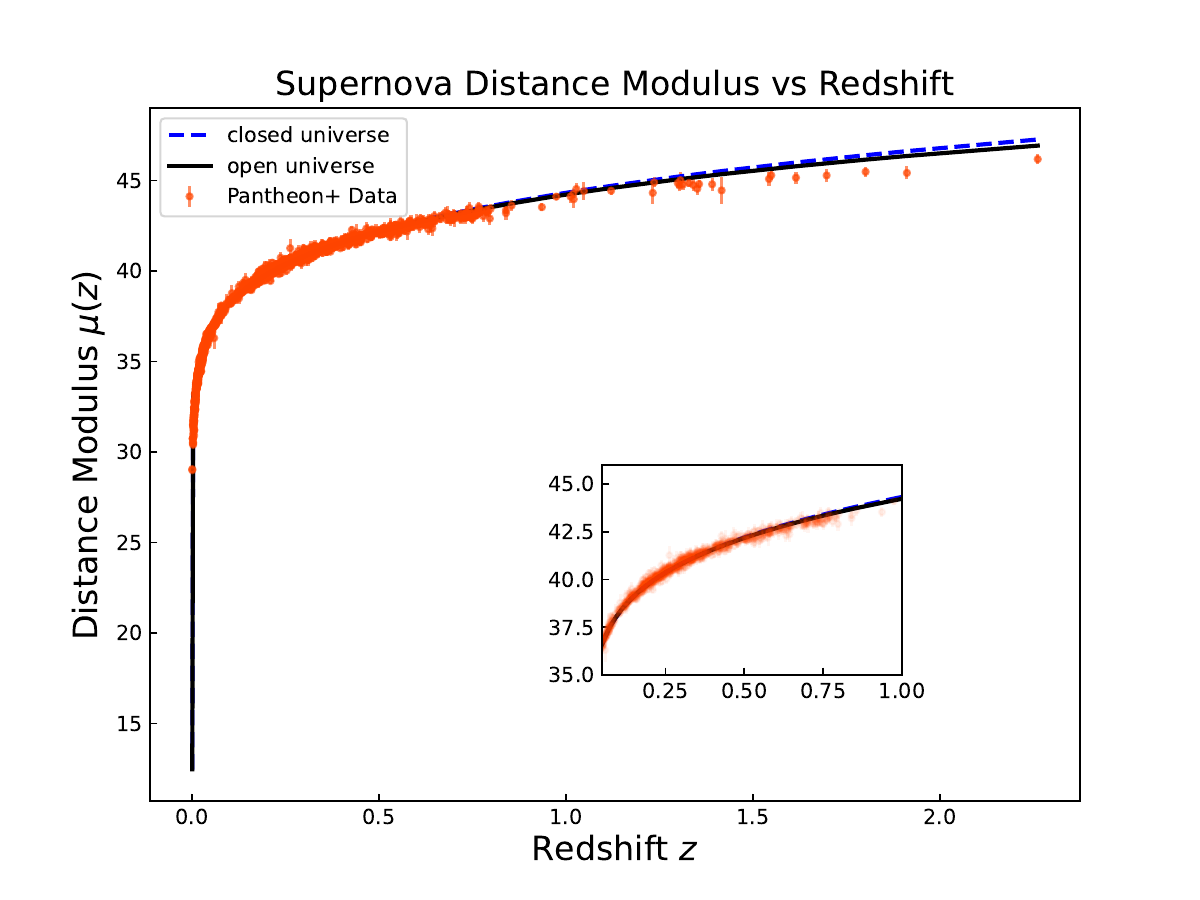}
        \caption{$\mathcal{Q}\propto\paren{\rho_{de}+\rho_{m}+\rho_{r}}$}
        \label{fig:panth all}
    \end{subfigure}
    \caption{The distance modules ($\mu(z)$) versus redshift ($z$) graphs for all four distinct interaction models are plotted. In panels (a), (b), (c), and (d), the Hubble parameter $\mu(z)$ is plotted against redshift \(z\) for models where the interaction term \(\mathcal{Q}\) is proportional to \(\rho_{de}\), \(\rho_{m}\), \(\rho_{r}\), and the combined density \(\rho_{de} + \rho_{m} + \rho_{r}\), respectively, using the Pantheon+ \& SH0ES data. In each panel, solid and dashed lines represent the best-fit results from our model for closed and open universe configurations, respectively. Additionally, we have also plotted the observational data points with their associated error bars for Pantheon+ \& SH0ES data set.}
    \label{fig:combined mu vs z}
\end{figure}

\begin{figure}[htbp]
    \begin{subfigure}[b]{0.5\textwidth}
        \includegraphics[width=\textwidth]{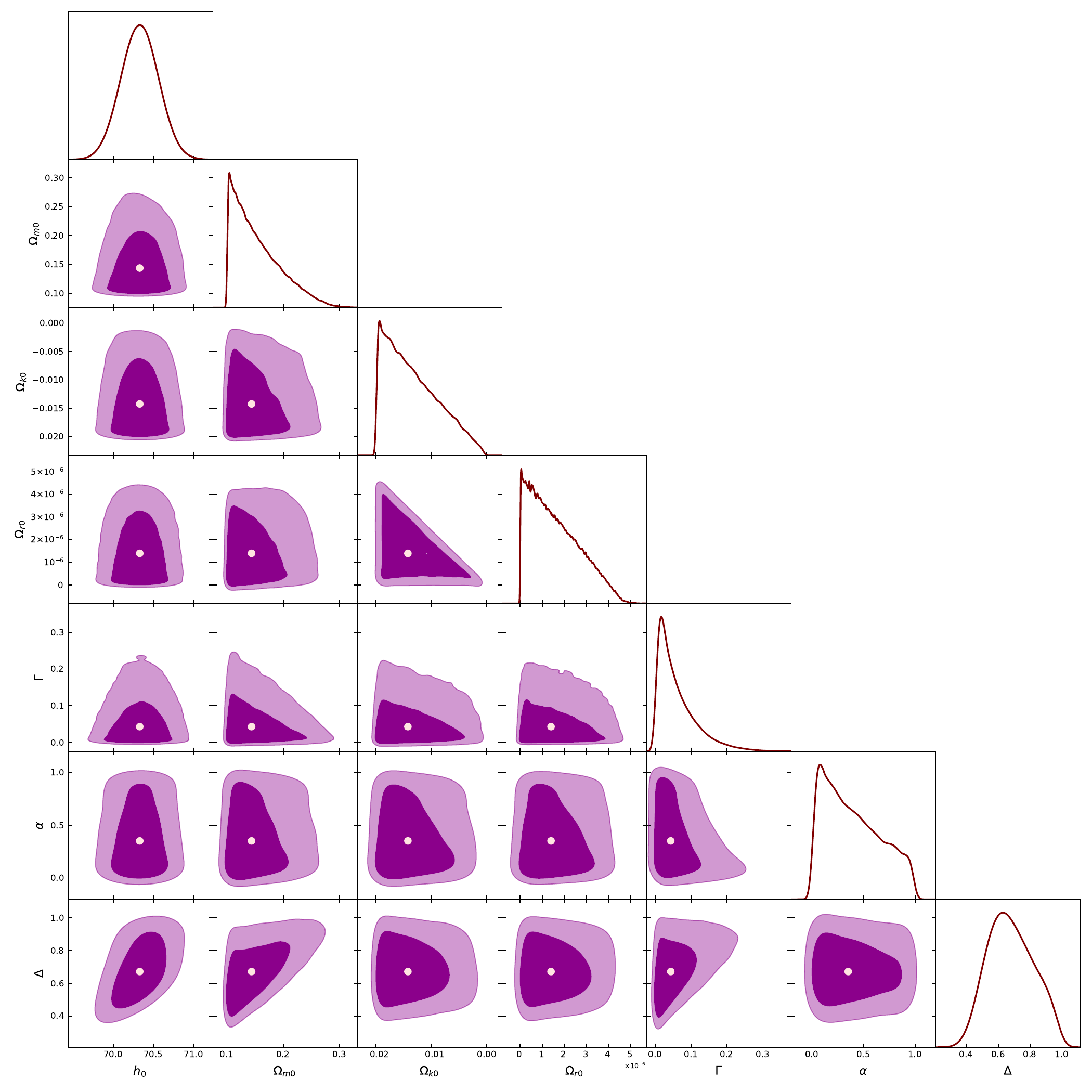}
        \caption{$\mathcal{Q}\propto\rho_{de}$}
    \end{subfigure}
    \hspace{-0.06\textwidth} 
    \begin{subfigure}[b]{0.5\textwidth}
        \includegraphics[width=\textwidth]{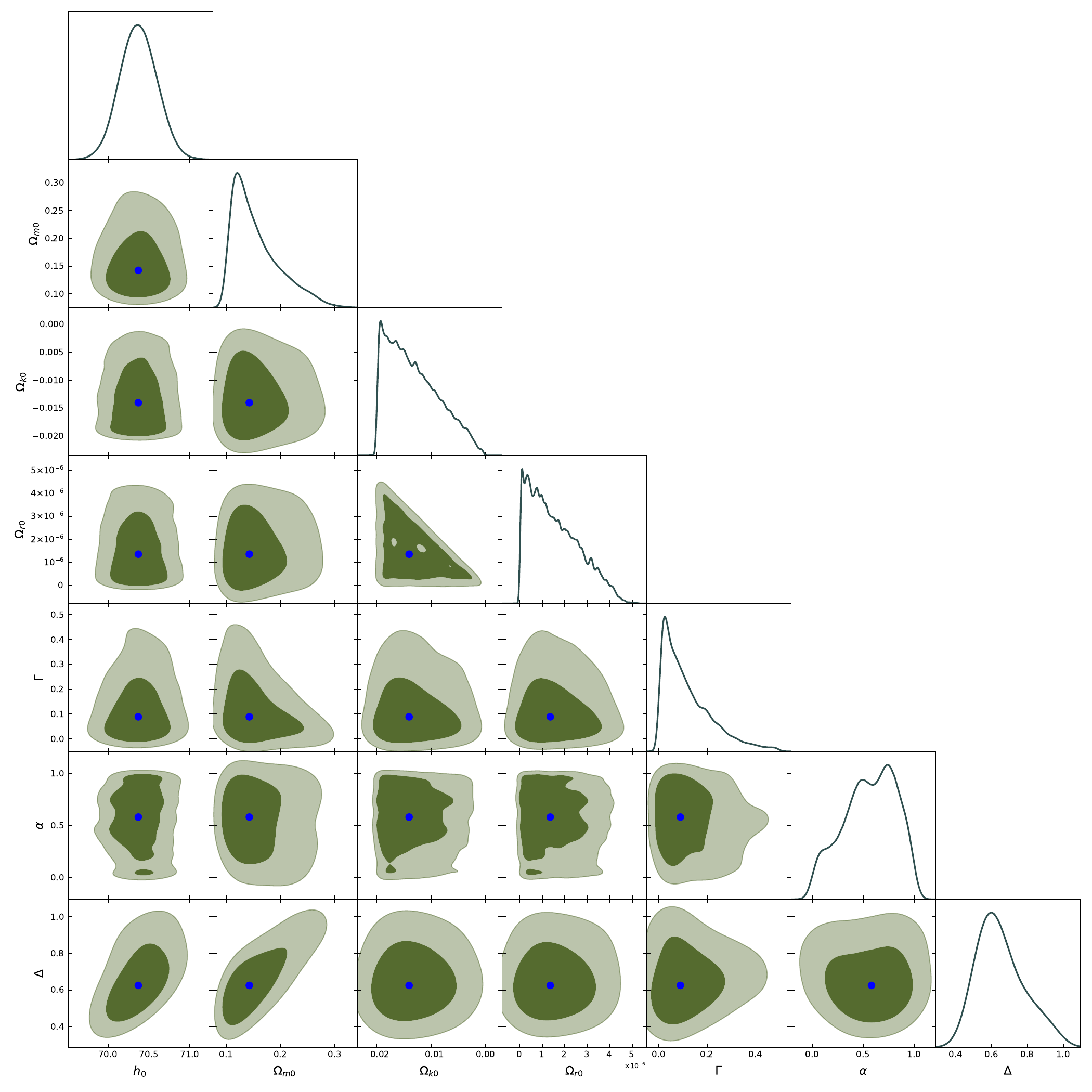}
        \caption{$\mathcal{Q}\propto\rho_{m}$}
    \end{subfigure}
     \begin{subfigure}[b]{0.5\textwidth}
        \includegraphics[width=\textwidth]{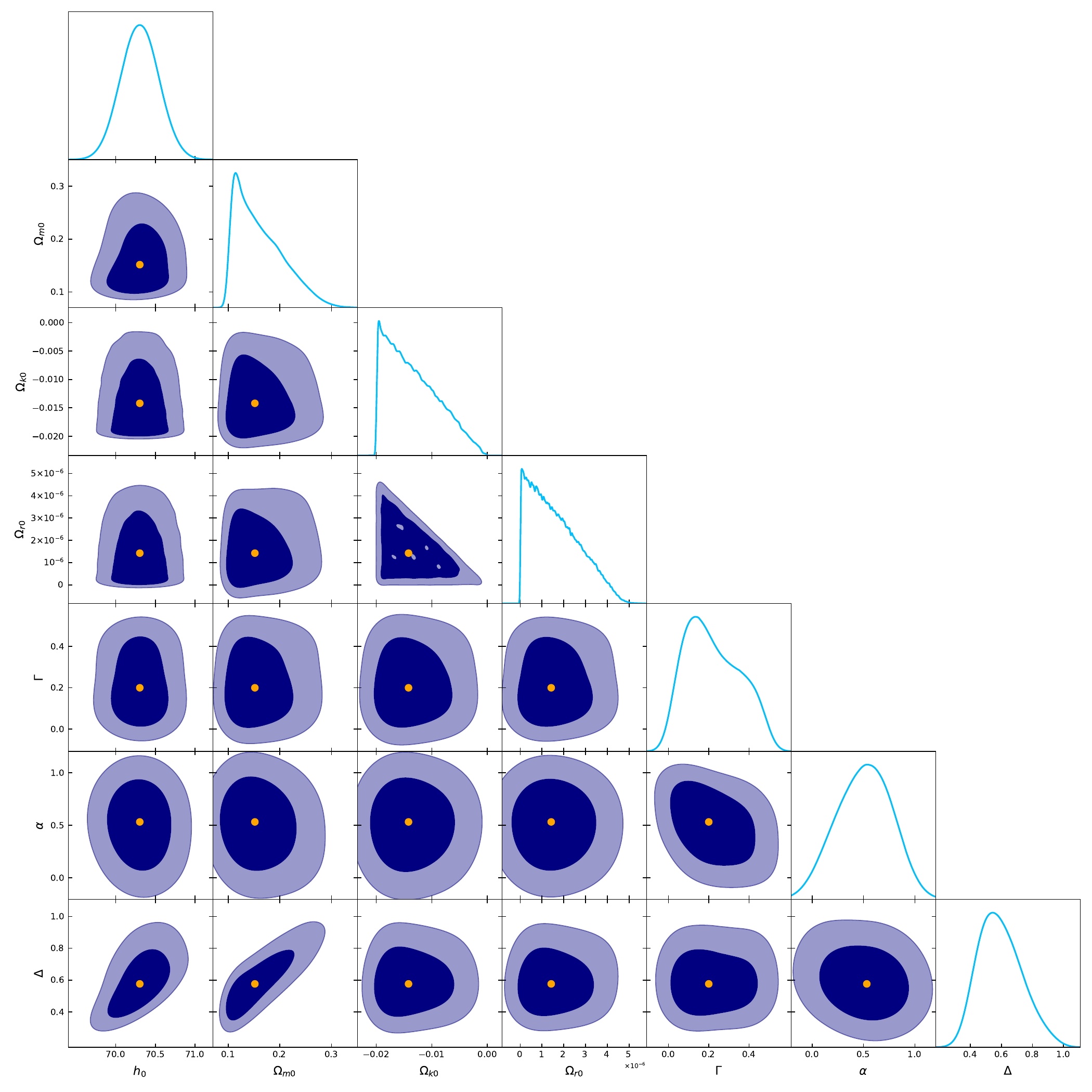}
        \caption{$\mathcal{Q}\propto\rho_{r}$}
    \end{subfigure}
    \hspace{-0.01\textwidth}
     \begin{subfigure}[b]{0.5\textwidth}
        \includegraphics[width=\textwidth]{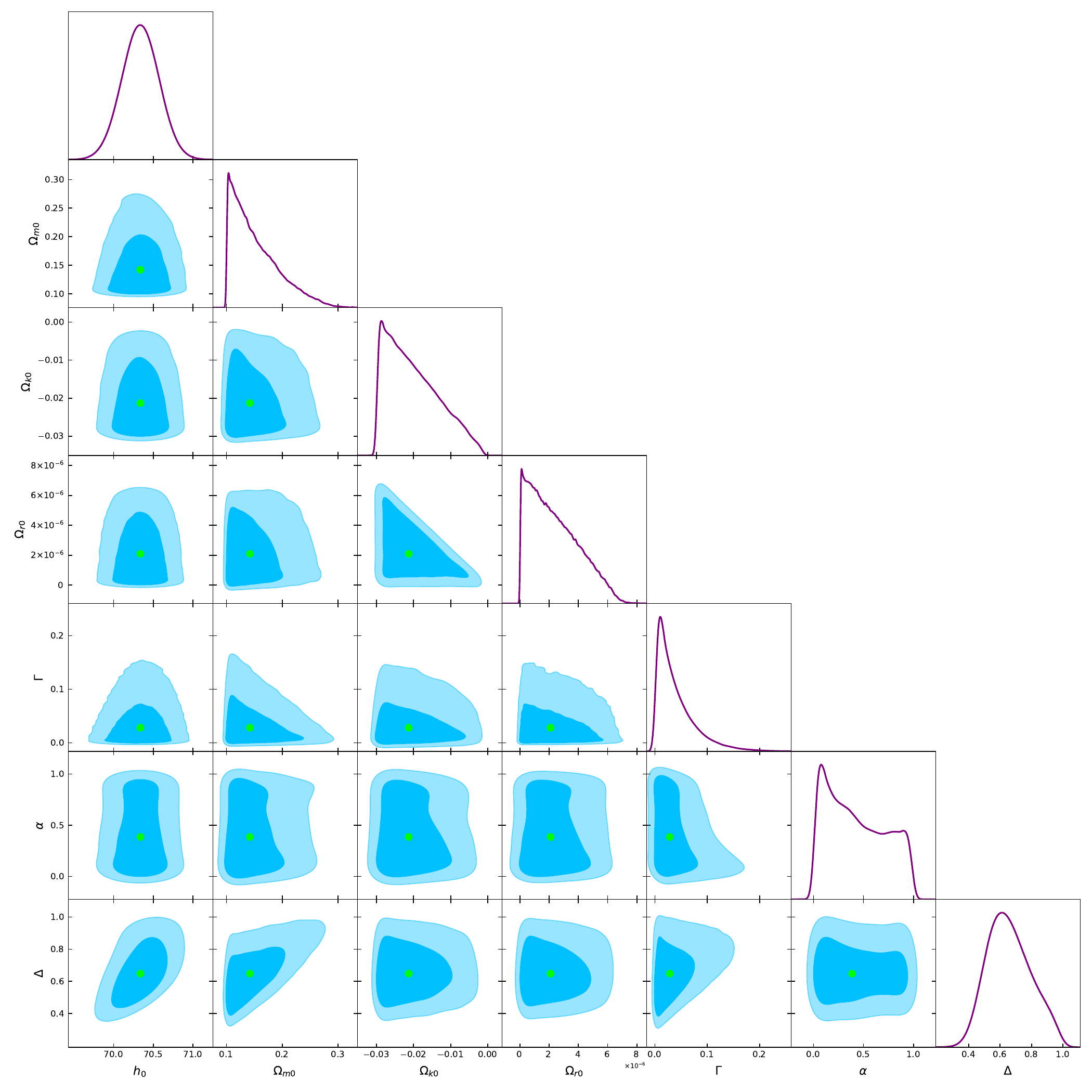}
        \caption{$\mathcal{Q}\propto\paren{\rho_{de}+\rho_{m}+\rho_{r}}$}
    \end{subfigure}
    \caption{Contour plots for all four different kind of phenomenological interaction terms for Pantheon+ data set for closed universe. Panel (a), (b), (c) and (d) shows the corner plots when the interaction term ($\mathcal{Q}$)is proportional to $\rho_{de}$, $\rho_{m}$, $\rho_{r}$ and $\rho_{de}+\rho_{m}+\rho_{r}$ respectively. For each panel, contours representing various model parameters are displayed alongside the best-fit points corresponding to observational data, are displayed for both open and closed universe scenarios.}
    \label{fig:exact closed Pantheon combined}
\end{figure}

\begin{figure}[htbp]
    \begin{subfigure}[b]{0.5\textwidth}
        \includegraphics[width=\textwidth]{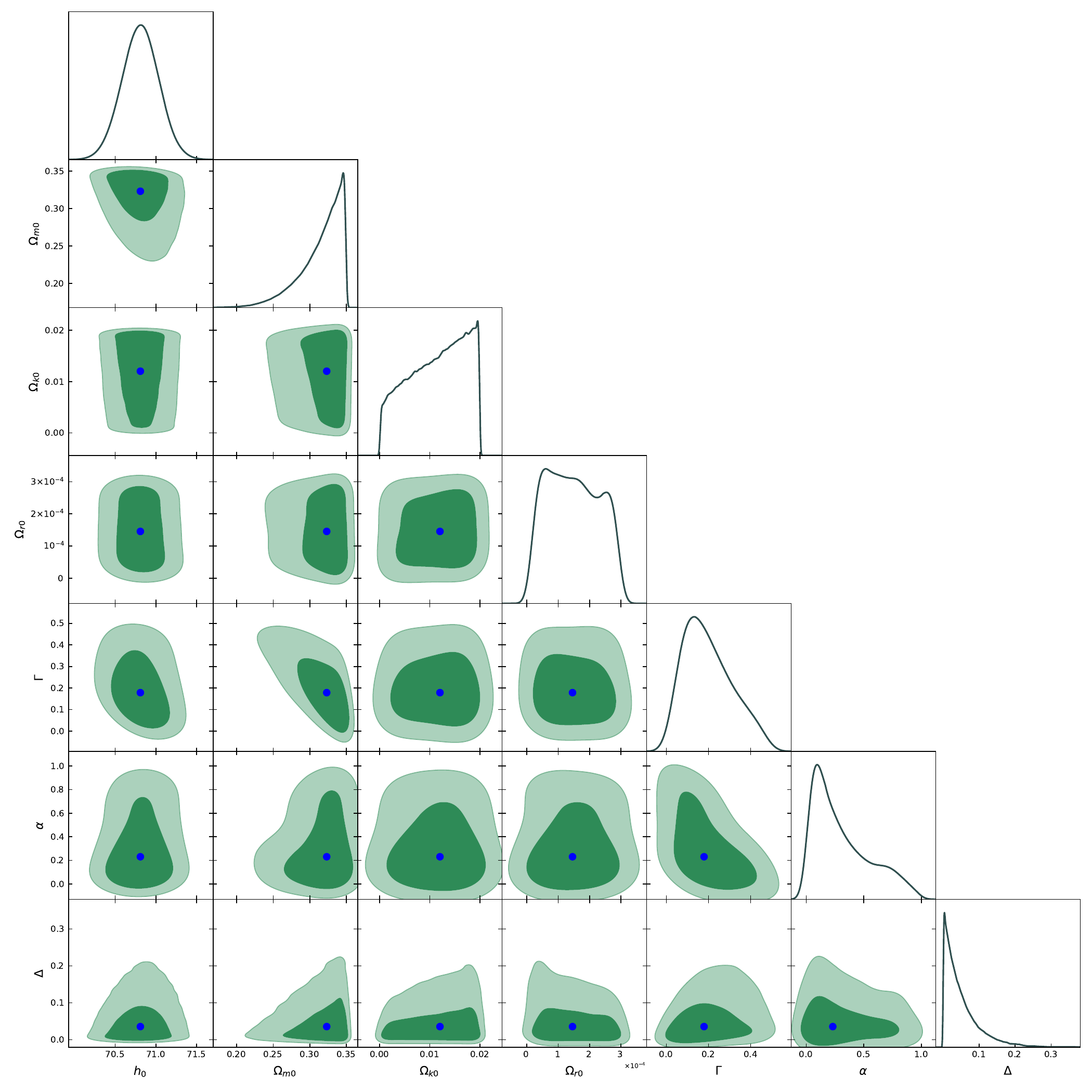}
        \caption{$\mathcal{Q}\propto\rho_{de}$}
    \end{subfigure}
    \hspace{-0.06\textwidth} 
    \begin{subfigure}[b]{0.5\textwidth}
        \includegraphics[width=\textwidth]{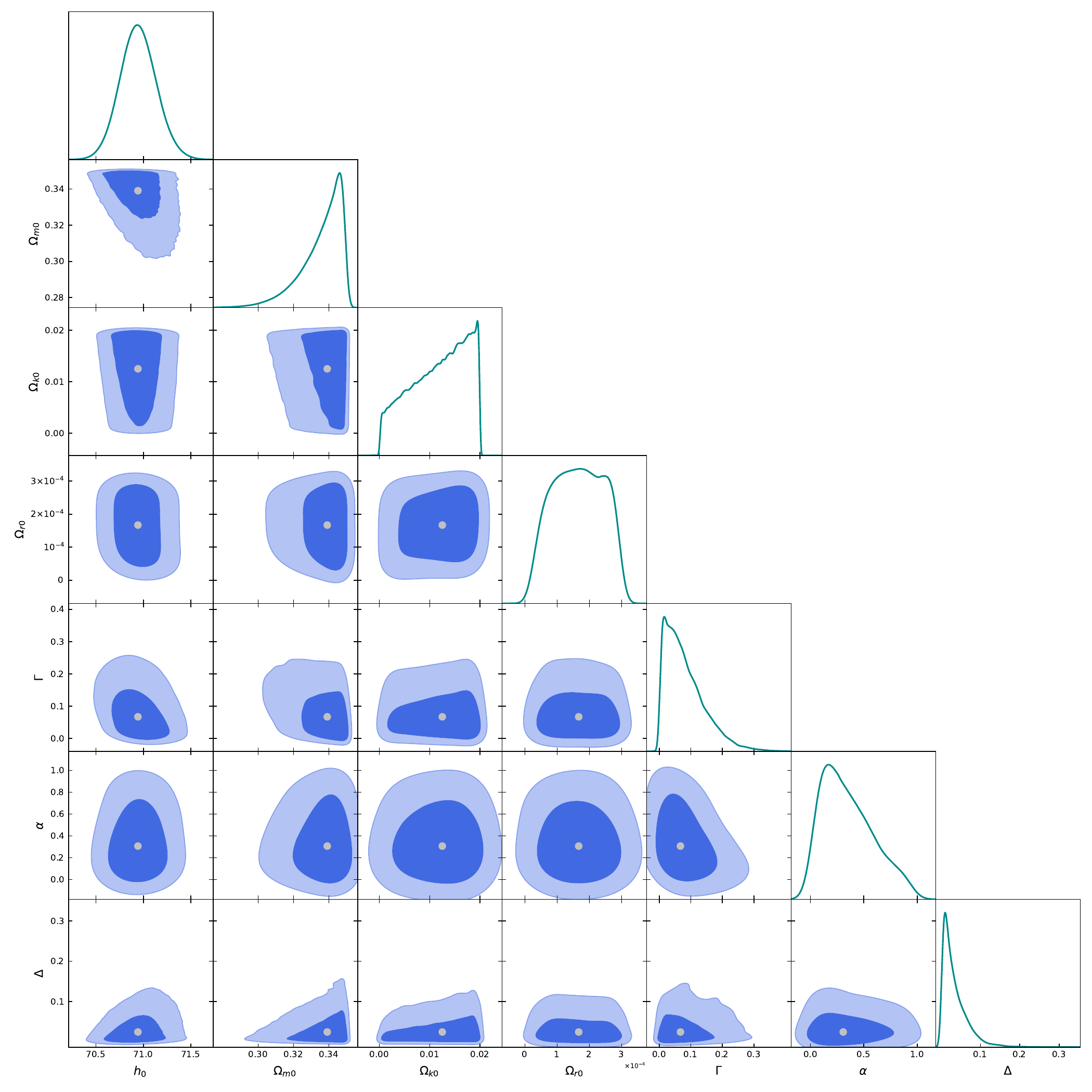}
        \caption{$\mathcal{Q}\propto\rho_{m}$}
    \end{subfigure}
     \begin{subfigure}[b]{0.5\textwidth}
        \includegraphics[width=\textwidth]{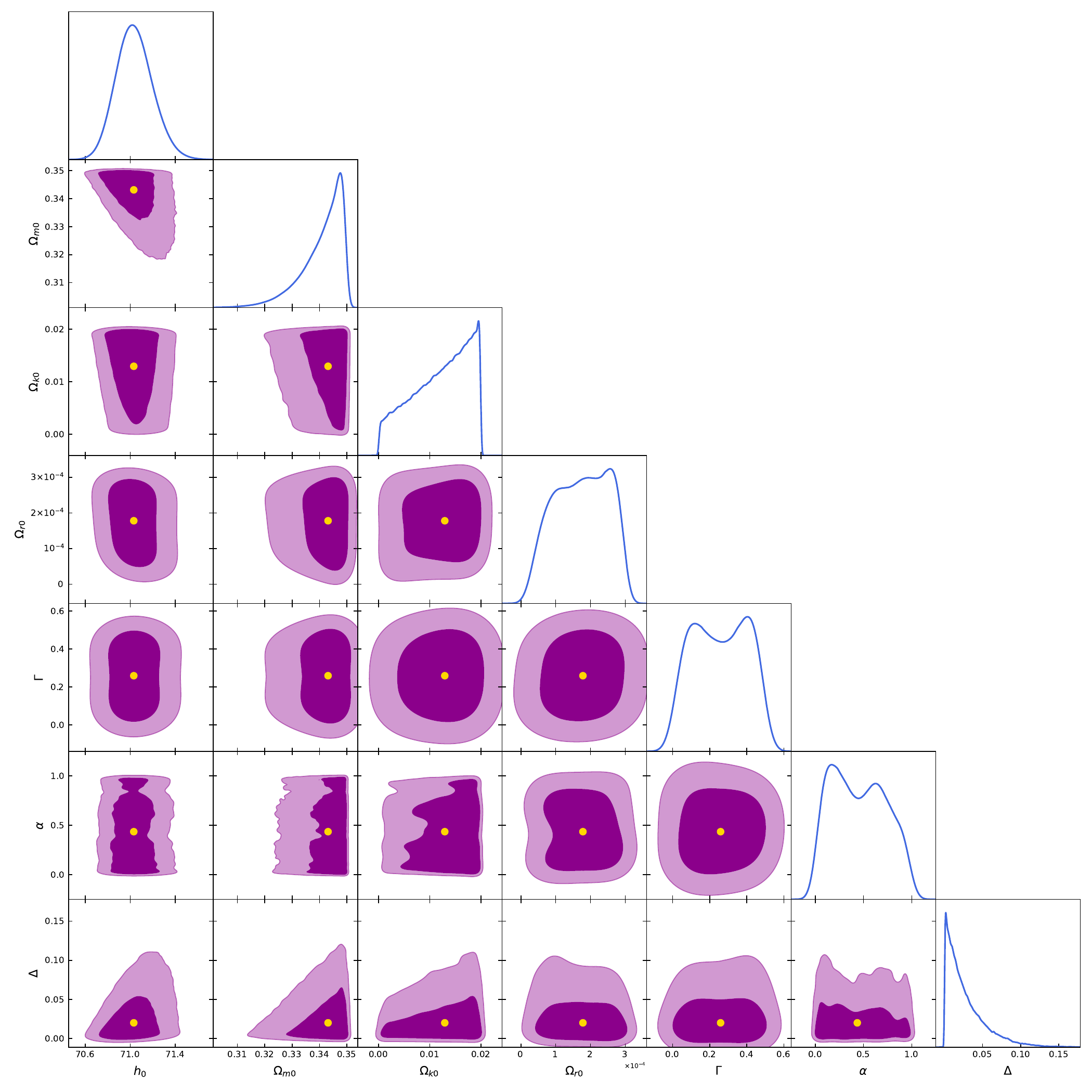}
        \caption{$\mathcal{Q}\propto\rho_{r}$}
    \end{subfigure}
    \hspace{-0.01\textwidth}
     \begin{subfigure}[b]{0.5\textwidth}
        \includegraphics[width=\textwidth]{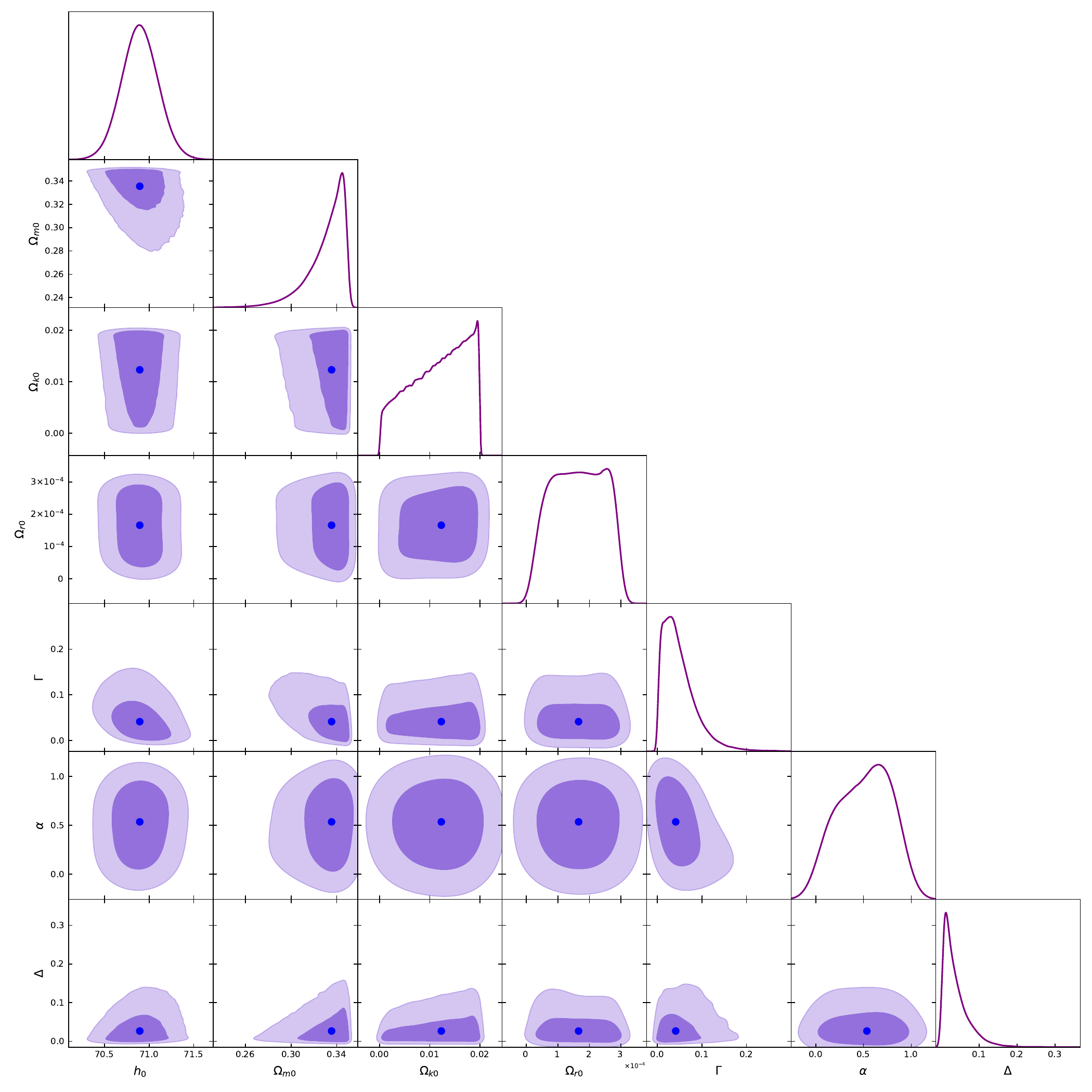}
        \caption{$\mathcal{Q}\propto\paren{\rho_{de}+\rho_{m}+\rho_{r}}$}
    \end{subfigure}
    \caption{Contour plots for all four different kind of phenomenological interaction terms for Pantheon+ data set for open universe. Panel (a), (b), (c) and (d) shows the corner plots when the interaction term ($\mathcal{Q}$)is proportional to $\rho_{de}$, $\rho_{m}$, $\rho_{r}$ and $\rho_{de}+\rho_{m}+\rho_{r}$ respectively. For each panel, contours representing various model parameters are displayed alongside the best-fit points corresponding to observational data, are displayed for both open and closed universe scenarios.}
    \label{fig:exact open Pantheon combined}
\end{figure}

\noindent Fig.\eqref{fig:combined mu vs z} presents a graphical depiction of the best-fit theoretical distance modulus, $\mu(z)$, derived from our model as a function of redshift $z$, using the Pantheon+ and SH0ES datasets. Fig.(s)(\eqref{fig:panth de}, \eqref{fig:panth m}, \eqref{fig:panth r} and \eqref{fig:panth all}) shows the best-fit graphs for phenomenological interaction terms $\mathcal{Q}=-\Gamma H \rho_{de}$, $\mathcal{Q}=-\Gamma H \rho_{de}$, $\mathcal{Q}=-\Gamma H \rho_{m}$, $\mathcal{Q}=-\Gamma H \rho_{r}$ and $\mathcal{Q}=-\Gamma H (\rho_{de}+\rho_{m}+\rho_{r})$ respectively. For each panel in Fig.\eqref{fig:combined mu vs z}, the solid curves denotes closed universe and the dashed curve refers to an open universe. We have also provided inset plots in each panel to visualize the fitting in the low redshift regime. In each panel, we have also shown the observational data points along with their respective error bars for the Panntheon+ \& SH0ES datasets to validate our fitted curve. In Fig.(s)(\eqref{fig:panth de}, \eqref{fig:panth m}, \eqref{fig:panth r} and \eqref{fig:panth all}) the observational data points along with the corresponding error bars are shown in green, purple, skyblue and orange respectively. Fig.(s)(\eqref{fig:exact closed Pantheon combined},\eqref{fig:exact open Pantheon combined}) for Pantheon+ \& SH0ES data display the \(1\sigma\) and \(2\sigma\) likelihood contours for various model parameters, along with their corresponding best-fit points, for closed and open universe scenarios, respectively.

\noindent Table \eqref{tab:original_interaction_params_Panth_SH0ES}, contains the best-fit values of our model parameters which are obtained using Pantheon+ \& SH0ES data set. We have provided the best-fit model parameters for all four different interaction models and for both open and closed universe scenarios. From the constrained values of the Hubble parameter in Table \eqref{tab:original_interaction_params_Panth_SH0ES}, one can clearly see that the Hubble parameter has larger values compared to the $\Lambda$CDM model, therefore indicating towards a possible resolution to the Hubble tension problem. \\
From Table \eqref{tab:original_interaction_params_CC_BAO} and Table \eqref{tab:original_interaction_params_Panth_SH0ES}, we can see ${\ok}_{,0}$, $\Gamma$ and $\alpha$ have nonzero values corresponding to all four different interaction models. The nonzero value of ${\ok}_{,0}$ indicates the nonzero spatial curvature of our universe, nonzero $\Gamma$ values indicate the presence of interaction between dark energy, dark matter and radiation and nonzero $\alpha$ values suggests continuous exchange of energy between dark energy, dark matter and radiation.

\begin{table}[h!]
\centering
\renewcommand{\arraystretch}{1.5}
\setlength{\tabcolsep}{8pt}
\begin{tabular}{|c|c||c||c||c||c|}
\hline
\textbf{Params} & \textbf{$k$} 
& $\mathcal{Q} = -\Gamma H \rho_{de}$ 
& $\mathcal{Q} = -\Gamma H \rho_{m}$ 
& $\mathcal{Q} = -\Gamma H \rho_{r}$ 
& $\mathcal{Q} = -\Gamma H(\rho_{de} + \rho_m + \rho_r)$ \\
\hline

\multirow{2}{*}{$H_0$} 
& 1   & $71.1919^{+2.9028}_{-2.2521}$ & $71.5439^{+2.9244}_{-2.3076}$ & $70.8523^{+3.9370}_{-2.7438}$ & $71.5506^{+2.9822}_{-2.4010}$ \\
\cline{2-6}
& -1  & $71.5997^{+2.5065}_{-2.0482}$ & $71.8395^{+2.4396}_{-2.0661}$ & $71.1084^{+2.6118}_{-2.0095}$ & $71.7289^{+2.3738}_{-2.0651}$ \\
\hline

\multirow{2}{*}{$\Omega_{m,0}$} 
& 1   & $0.2050^{+0.0431}_{-0.0579}$ & $0.2281^{+0.0290}_{-0.0340}$ & $0.2674^{+0.0271}_{-0.0271}$ & $0.2215^{+0.0333}_{-0.0462}$ \\
\cline{2-6}
& -1  & $0.1842^{+0.0340}_{-0.0422}$ & $0.1964^{+0.0277}_{-0.0309}$ & $0.2189^{+0.0232}_{-0.0234}$ & $0.1915^{+0.0300}_{-0.0367}$ \\
\hline

\multirow{2}{*}{$\Omega_{r,0}$} 
& 1   & \smaller$1.55^{+0.1}_{-0.1} \times 10^{-4}
$ & \smaller$1.48^{+1.04}_{-1.04} \times 10^{-4}$
 & \smaller $1.49^{+1.00}_{-0.99} \times 10^{-4}
$
 & \smaller$1.60^{+0.98}_{-1.10} \times 10^{-4}
$ \\
\cline{2-6}
& -1  & \smaller $1.54^{+1.03}_{-1.05} \times 10^{-4}
 $ & \smaller$1.51^{+1.02}_{-1.03} \times 10^{-4}$
 & \smaller$1.47^{+1.06}_{-1.00} \times 10^{-4}$
 & \smaller $1.64^{+0.92}_{-1.05} \times 10^{-4}
$ \\
\hline

\multirow{2}{*}{$\Omega_{k,0}$} 
& 1   & \smaller$-6.07^{+3.74}_{-2.81} \times 10^{-3}
$ & \smaller$-5.80^{+3.67}_{-2.97} \times 10^{-3}
$ & \smaller$-6.09^{+3.71}_{-2.81} \times 10^{-3}$ & \smaller$-6.05^{+3.73}_{-2.81} \times 10^{-3}$ \\
\cline{2-6}
& -1  & \smaller$3.63^{+3.75}_{-2.60} \times 10^{-3}$ & \smaller$3.90^{+3.73}_{-2.77} \times 10^{-3}$ & \smaller$3.50^{+3.70}_{-2.53} \times 10^{-3}$ & \smaller$3.71^{+3.75}_{-2.66} \times 10^{-3}
$ \\
\hline

\multirow{2}{*}{$\Gamma$} 
& 1   & $0.1084^{+0.1544}_{-0.0791}$ & $0.0441^{+0.0746}_{-0.0328}$ & $0.2810^{+0.1519}_{-0.1809}$ & $0.0327^{+0.0574}_{-0.0243}$ \\
\cline{2-6}
& -1  & $0.0635^{+0.1045}_{-0.0476}$ & $0.0404^{+0.0719}_{-0.0303}$ & $0.2029^{+0.1895}_{-0.1467}$ & $0.0239^{+0.0407}_{-0.0178}
$ \\
\hline

\multirow{2}{*}{$\Delta$} 
& 1   & $0.0842^{+0.0864}_{-0.0599}
$ & $0.0801^{+0.0881}_{-0.0575}
$ & $0.1372^{+0.1145}_{-0.0918}
$ & $0.0834^{+0.0895}_{-0.0580}
$ \\
\cline{2-6}
& -1  & $0.0625^{+0.0708}_{-0.0446}
$ & $0.0548^{+0.0674}_{-0.0395}
$ & $0.0697^{+0.0742}_{-0.0490}
$ & $0.0554^{+0.0639}_{-0.0401}
$ \\
\hline

\multirow{2}{*}{$\alpha$} 
& 1   & $0.1768^{+0.3600}_{-0.1394}
$ & $0.3338^{+0.3693}_{-0.2546}
$ & $0.6027^{+0.2809}_{-0.4233}
$ & $0.2196^{+0.3790}_{-0.1698}
$ \\
\cline{2-6}
& -1  & $0.2567^{+0.4870}_{-0.2034}
$ & $0.3212^{+0.3873}_{-0.2318}
$ & $0.4189^{+0.3682}_{-0.3103}
$ & $0.3026^{+0.3790}_{-0.2251}
$ \\
\hline

\end{tabular}
\caption{Best-fit values for different interaction models and curvatures ($k=1$ and $k=-1$ shown separately) for exact model corresponding to CC+BAO data set.}
\label{tab:original_interaction_params_CC_BAO}
\end{table}
\begin{table}[h!]
\centering
\renewcommand{\arraystretch}{1.5}
\setlength{\tabcolsep}{8pt}
\begin{tabular}{|c|c||c||c||c||c|}
\hline
\textbf{Params} & \textbf{$k$} 
& $\mathcal{Q} = -\Gamma H \rho_{de}$ 
& $\mathcal{Q} = -\Gamma H \rho_{m}$ 
& $\mathcal{Q} = -\Gamma H \rho_{r}$ 
& $\mathcal{Q} = -\Gamma H(\rho_{de} + \rho_m + \rho_r)$ \\
\hline

\multirow{2}{*}{$H_0$} 
& 1   & $70.3295^{+0.2322}_{-0.2328}$ & $70.3690^{+0.2309}_{-0.2246}$ & $70.3038^{+0.2385}_{-0.2372}$ & $70.3371^{+0.2320}_{-0.2324}$ \\
\cline{2-6}
& -1  & $70.8092^{+0.2240}_{-0.2261}$ & $70.9429^{+0.1926}_{-0.1855}$ & $71.0311^{+0.1680}_{-0.1537}$ & $70.8947^{+0.2040}_{-0.2012}$ \\
\hline

\multirow{2}{*}{$\Omega_{m,0}$} 
& 1   & $0.1436^{+0.0553}_{-0.0319}$ & $0.1423^{+0.0586}_{-0.0302}$ & $0.1516^{+0.0598}_{-0.0381}$ & $0.1421^{+0.0545}_{-0.0307}$ \\
\cline{2-6}
& -1  & $0.3232^{+0.0195}_{-0.0362}$ & $0.3391^{+0.0079}_{-0.0145}$ & $0.3431^{+0.0050}_{-0.0093}$ & $0.3355^{+0.0105}_{-0.0200}$ \\
\hline

\multirow{2}{*}{$\Omega_{r,0}$} 
& 1   & ${10^{-6}}^{~+10^{-6}}_{~-10^{-6}}$ & \smaller$1.35^{+1.43}_{-0.97} \times 10^{-6}$
 & \smaller $1.42^{+1.45}_{-1.02} \times 10^{-6}$
 & \smaller$2.09^{+2.15}_{-1.49} \times 10^{-6}$ \\
\cline{2-6}
& -1  & \smaller $1.45^{+1.03}_{-0.92} \times 10^{-4}$ & \smaller$1.67^{+0.90}_{-0.89} \times 10^{-4}$
 & \smaller$1.78^{+0.85}_{-0.92} \times 10^{-4}$
 & \smaller $1.66^{+9.25}_{-9.27} \times 10^{-4}$ \\
\hline

\multirow{2}{*}{$\Omega_{k,0}$} 
& 1   & $-0.0142^{+0.0062}_{-0.0041}$ & $-0.0140^{+0.0061}_{-0.0042}$ & $-0.0142^{+0.0060}_{-0.0041}$ & $-0.0213^{+0.0091}_{-0.0062}$ \\
\cline{2-6}
& -1  & $0.0120^{+0.0056}_{-0.0072}$ & $0.0125^{+0.0053}_{-0.0072}$ & $0.0129^{+0051}_{-0073}$ & $0.0123^{+0.0055}_{-0.0072}$ \\
\hline

\multirow{2}{*}{$\Gamma$} 
& 1   & $0.0431^{+0.0675}_{-0.0319}$ & $0.0889^{+0.1306}_{-0.0674}$ & $0.1997^{+0.1813}_{-0.1280}$ & $0.028^{+0.0444}_{-0.0209}$ \\
\cline{2-6}
& -1  & $0.1786^{+0.1488}_{-0.1036}$ & $0.0672^{+0.0761}_{-0.0470}$ & $0.2590^{+0.1771}_{-0.1783}$ & $0.0412^{+0.0423}_{-0.0266}$ \\
\hline

\multirow{2}{*}{$\Delta$} 
& 1   & $0.6714^{+0.1717}_{-0.1403}$ & $0.6247^{+0.1775}_{-0.1099}$ & $0.5761^{+0.1593}_{-0.1289}$ & $0.6481^{+0.1646}_{-0.1301}$ \\
\cline{2-6}
& -1  & $0.0358^{+0.0582}_{-0.0267}$ & $0.0241^{+0.0382}_{-0.0179}$ & $0.0120^{+0.0318}_{-0.0148}$ & $0.0267^{+0.0411}_{-0.0198}$ \\
\hline

\multirow{2}{*}{$\alpha$} 
& 1   & $0.3511^{+0.3742}_{-0.2540}$ & $0.5791^{+0.2566}_{-0.3092}$ & $0.5320^{+0.2500}_{-0.3282}$ & $0.3858^{+0.4071}_{-0.2850}$ \\
\cline{2-6}
& -1  & $0.2310^{+0.3713}_{-0.1739}$ & $0.3069^{+0.3045}_{-0.2202}$ & $0.4354^{+0.3383}_{-0.3132}$ & $0.5354^{+0.2753}_{-0.3674}$ \\
\hline

\end{tabular}
\caption{Best-fit values for different interaction models and curvatures ($k=1$ and $k=-1$ shown separately) for exact model corresponding to Pantheon+ \& SH0ES data set.}
\label{tab:original_interaction_params_Panth_SH0ES}
\end{table}
\section{Conclusion}\label{Conclusion}
Now we will summarize our findings. In this paper, we have studied the interacting Barrow holographic dark energy model in the non-flat universe (both for open and closed universes) in the presence of both dark matter and radiation. In many previous studies, only the interaction between dark energy and dark matter is considered in the IBHDE model. Although CMB observations confirm that at the present time still there is presence of some amount of radiation. This fact motivates us to consider the study of the interaction between the dark energy, dark matter and radiation in the IBHDE model. Due to the presence of interaction, the right-hand side of the continuity equations corresponding to dark energy, dark matter and radiation are non-zero; this non-zero term is the phenomenological interaction, and its expression is obtained from dimensional analysis and it depends on the Hubble parameter, some energy density and an interaction parameter. One can form nine distinct linear combinations of energy densities of the dark energy, dark matter and radiation density. Although in this paper we have only considered four different kind of phenomenological interaction terms where the interaction term is proportional to a dark energy, dark matter, radiation density and sum of all these three. We have derived the evolution equations for the energy density parameters associated with dark energy, dark matter, and radiation across four distinct models, considering both open and closed universe geometries.  We have then numerically solved the evolution equations and plotted the results with respect to redshift parameters. For all four interaction models and for both open universe scenarios, we have obtained different epoch point corresponding to dark energy-dark matter, dark energy-radiation and dark matter-radiation respectively. Our obtained epoch points are also consistent with the thermal history of the universe. We have also illustrated graphically how the dark energy equation of state parameter varies with different values of the Barrow exponent, considering all four interaction models and both open and closed universe scenarios. From the plots of dark energy EOS parameter, we have seen that for Barrow exponent $\Delta=0,0.1$ the EOS parameter always lies in the quintessence regime. However, increasing the Barrow exponent values ($\Delta=0.2, 0.3$) induces a transition of the dark energy equation-of-state (EOS) parameter into the phantom regime, moving away from the quintessence regime at lower redshift values. It is also shown that due to the presence of interaction between dark energy, dark matter and radiation, the equation of state parameter of dark energy is no longer equal to $-1$. We would also like to mention that, due to interaction, it is ideal not to consider the equation of state parameters of dark matter and radiation to be $0$ and $1/3$ respectively; one should consider them to be unknown and calculate them explicitly from the model. Although for simplicity, only the dark energy equation of state parameter is considered to be unknown. From the numerical solution, we have found that the dark energy equation of state parameter is less than $-1$ in the present time (for $\Delta=0.2, 0.3$); this can be justified in a way that, in the IBHDE model, there is always a continuous transformation of radiation and matter into dark energy taking place. It is also clear from the graph of $\omega_{de}$ vs the redshift that for increasing values of the Barrow exponent, $\omega_{de}$ takes a more negative value for both open and closed universes. This suggests the quantum gravitational corrections make the equation of state parameter more negative at the present time. \\
We have also performed MCMC analysis to obtain various observational constraints in our model. In the observational constraint part of this paper, we have considered both Cosmic Chronometer (CC), which consists of $31$ data points in the redshift range $0.070<z<1.965$ and Baryon Acoustic Oscillator (BAO) data, which is a collection of $26$ data points in the redshift range $0.24<z<2.36$. Also, there is no overlapping between these two data sets, indicating zero cross-correlation between them. We utilized a combined dataset consisting of $57$ data points from cosmic chronometers (CC) and baryon acoustic oscillations (BAO). We have also used the Pantheon+ data set, which contains $1701$ data points, to constrain different model parameters. 
In this paper, for simplicity before constraining the model parameters of our exact cosmological model, we have used an approximate model. In this model, we assume that the Hubble parameter's evolution follows a similar pattern to the $\Lambda$CDM model, with matter, radiation, and dark energy all evolving with the redshift parameter. However, unlike the standard model where the dark energy equation of state parameter is fixed at $-1$, here it varies depending on multiple energy density parameters. Then we have constrained cosmological model parameters for our exact model using various observational data sets. 
For both the models and for all four different interaction terms the Markov Chain Monte Carlo (MCMC) sampling was performed using the Metropolis–Hastings algorithm, and the resulting chains were analyzed and visualized using the GetDist Python package to generate the corresponding $1\sigma$ and $2\sigma$ contours are shown in the corner plots. This is done for both open and closed universe scenarios. We would like to mention that for the approximate cosmological model while running MCMC, we have taken the Barrow exponent and the $\alpha$ parameter to be fixed, although for our exact model all the model parameters are taken as free. Before doing the MCMC analysis, we have set flat priors for both the models. For the approximate and exact model, we have provided different best fit model parameters corresponding to CC \& BAO and Pantheon+ \& SH0ES data, that appear in different tables in this paper. The best fit values of the Hubble parameter for all four different interaction models (for both open and closed universe scenario) has larger values compared to the $\Lambda$CDM model. Therefore, our model can be possible candidate to resolve the Hubble tension problem. From best fit values of different model parameters, we can see ${\ok}_{,0}$, $\Gamma$ and $\alpha$ have nonzero values corresponding to all four different interaction models. The nonzero value of ${\ok}_{,0}$ indicates the nonzero spatial curvature of our universe, nonzero $\Gamma$ values indicate the presence of interaction between dark energy, dark matter and radiation and nonzero $\alpha$ values suggests continuous exchange of energy between dark energy, dark matter and radiation. For the specific cosmological model under consideration, we have generated the best-fit plots of the Hubble parameter \( H(z) \) versus redshift \( z \), as well as the distance modulus \( \mu(z) \) versus \( z \), using the CC \& BAO data sets and the Pantheon+ \& SH0ES data sets, respectively. In this graphs, for higher redshift values that is in early time, a deviation between the best-fit graphs of open and closed universe have be seen. This shows a domination of spatial curvature effects on the early time of our cosmological model. As a future direction of this work, one can study inflation in the IBHDE model in the presence of radiation. Also an interacting stiff-matter component can be added in the model to study the evolution of stiff matter and find different epochs. Another possible direction could be the study of our cosmological model with interacting Tsallis model of holographic dark energy (also for other dark energy models like Renyi holographic dark energy, Kaniadakis holographic dark energy etc.) instead of Barrow holographic dark energy. Also one can study of our model in the flat universe with zero spatial curvature. The observational constraint part of this paper can also be done for CMB and DESI data as a future work.
\begin{center}\label{appendix}
    \Large
    \textbf{Appendix}
\end{center}
Here we will very briefly give expression of differential equations of other possible linear interactions containing two density terms, which have not been discussed in the above sections for both closed and open universe. For these kind of two component linear interaction terms, the evolution equations along with the dark energy EOS parameter for closed universe is given in Appendix A and for an open universe the corresponding equations are given in Appendix B. We are leaving the numerical and data analysis part of this kind, for a future work.  
\appendix
\section*{Appendix A: Other Possible Interactions (Closed Universe)}
\phantomsection
\addcontentsline{toc}{section}{Appendix A: Other Possible Interactions (Closed Universe)}
\setcounter{equation}{0}
\renewcommand{\theequation}{A.\arabic{equation}}
\subsection*{$\mathcal{Q}_m=-\Gamma H (\rho_{de}+\rho_m)$:}
When interaction is proportional to both the energy density of dark matter and dark energy then the three differential equations for density parameters reads
\begin{align}
    \frac{\ode '}{\ode(1-\ode)}&=-\frac{1}{1-\ode}\left[2\paren{\ode+\om+\ol-1}-\ol\paren{\frac{\Gamma\alpha\paren{1+r_1}}{r_2}+4}\right.\\& \nonumber\left.-\om\paren{\frac{\Gamma \paren{1+r_1}}{r_1}+3}\right]  
    +(\Delta-2)\Big(1-\sqrt{\frac{3M_{p}^2 \Omega_{de}}{C}}L^{-\frac{-\Delta}{2}}\cos{y}\Big)~.
    \end{align} 
\begin{align}
    \frac{\om'}{\om}&=-\paren{\frac{\Gamma\paren{1+r_1}}{r_1}+3}
    -\left[2\paren{\ode+\om+\ol-1}-\ol\paren{\frac{\alpha\Gamma\paren{1+r_1}}{r_2}+4}\right.\\&\nonumber \left. -\om\paren{\frac{\Gamma\paren{1+r_1}}{r_1}+3}\right]-\ode\paren{\Delta-2}\paren{1-\sqrt{\frac{3M_P^2\ode}{C}}L^{-\frac{\Delta}{2}}\cos{y}}
\end{align}
\begin{align}
    \frac{\ol'}{\ol}&=-\paren{\frac{\alpha\Gamma\paren{1+r_1}}{r_2}+4} -\left[2\paren{\ode+\om+\ol-1}-\ol\paren{\frac{\alpha\Gamma\paren{1+r_1}}{r_2}+4}\right.\\&\nonumber\left.-\om\paren{\frac{\Gamma\paren{1+r_1}}{r_1}+3}\right]-\ode\paren{\Delta-2}\paren{1-\sqrt{\frac{3M_P^2\ode}{C}}L^{-\frac{\Delta}{2}}\cos{y}}
\end{align}
The equation of state for dark energy for this case can be written as
\begin{align}
    \omega_{de}=\frac{\paren{1+\alpha}\paren{1+r_1}\Gamma}{3}-\Big(\frac{1+\Delta}{3}\Big)+\Big(\frac{\Delta+2}{3}\Big)\sqrt{\frac{3M_{p}^2 \Omega_{de}}{C}}L^{-\frac{\Delta}{2}}\cos{y}~.
\end{align}
\subsection*{$\mathcal{Q}=-\Gamma H (\rho_{de}+\rho_r)$:}
For the interaction type where it depends on the energy density of dark energy as well as radiation, the differential equations can be written as,
\begin{align}
    \frac{\ode '}{\ode(1-\ode)}&=-\frac{1}{1-\ode}\left[2\paren{\ode+\om+\ol-1}-\ol\paren{\frac{\Gamma\alpha\paren{1+r_2}}{r_2}+4}\right.\\& \nonumber\left.-\om\paren{\frac{\Gamma \paren{1+r_2}}{r_1}+3}\right]  
    +(\Delta-2)\Big(1-\sqrt{\frac{3M_{p}^2 \Omega_{de}}{C}}L^{-\frac{-\Delta}{2}}\cos{y}\Big)~.
    \end{align} 
\begin{align}
    \frac{\om'}{\om}&=-\paren{\frac{\Gamma\paren{1+r_2}}{r_1}+3}
    -\left[2\paren{\ode+\om+\ol-1}-\ol\paren{\frac{\alpha\Gamma\paren{1+r_2}}{r_2}+4}\right.\\&\nonumber \left. -\om\paren{\frac{\Gamma\paren{1+r_2}}{r_1}+3}\right]-\ode\paren{\Delta-2}\paren{1-\sqrt{\frac{3M_P^2\ode}{C}}L^{-\frac{\Delta}{2}}\cos{y}}
\end{align}
\begin{align}
    \frac{\ol'}{\ol}&=-\paren{\frac{\alpha\Gamma\paren{1+r_2}}{r_2}+4} -\left[2\paren{\ode+\om+\ol-1}-\ol\paren{\frac{\alpha\Gamma\paren{1+r_2}}{r_2}+4}\right.\\&\nonumber\left.-\om\paren{\frac{\Gamma\paren{1+r_2}}{r_1}+3}\right]-\ode\paren{\Delta-2}\paren{1-\sqrt{\frac{3M_P^2\ode}{C}}L^{-\frac{\Delta}{2}}\cos{y}}
\end{align}
and the eos for dark energy takes the form 
\begin{align}
    \omega_{de}=\frac{\paren{1+\alpha}\paren{1+r_2}\Gamma}{3}-\Big(\frac{1+\Delta}{3}\Big)+\Big(\frac{\Delta+2}{3}\Big)\sqrt{\frac{3M_{p}^2 \Omega_{de}}{C}}L^{-\frac{\Delta}{2}}\cos{y}~.
\end{align}
\subsection*{$\mathcal{Q}=-\Gamma H (\rho_{m}+\rho_r)$:}
Another possible interaction is, it depends on the the lenear combination of radiation and matter energy density. Three differential equations for this reads,
\begin{align}
    \frac{\ode '}{\ode(1-\ode)}&=-\frac{1}{1-\ode}\left[2\paren{\ode+\om+\ol-1}-\ol\paren{\frac{\Gamma\alpha\paren{r_1+r_2}}{r_2}+4}\right.\\& \nonumber\left.-\om\paren{\frac{\Gamma \paren{r_1+r_2}}{r_1}+3}\right]  
    +(\Delta-2)\Big(1-\sqrt{\frac{3M_{p}^2 \Omega_{de}}{C}}L^{-\frac{-\Delta}{2}}\cos{y}\Big)~.
    \end{align} 
\begin{align}
    \frac{\om'}{\om}&=-\paren{\frac{\Gamma\paren{r_1+r_2}}{r_1}+3}
    -\left[2\paren{\ode+\om+\ol-1}-\ol\paren{\frac{\alpha\Gamma\paren{r_1+r_2}}{r_2}+4}\right.\\&\nonumber \left. -\om\paren{\frac{\Gamma\paren{r_1+r_2}}{r_1}+3}\right]-\ode\paren{\Delta-2}\paren{1-\sqrt{\frac{3M_P^2\ode}{C}}L^{-\frac{\Delta}{2}}\cos{y}}
\end{align}
\begin{align}
    \frac{\ol'}{\ol}&=-\paren{\frac{\alpha\Gamma\paren{r_1+r_2}}{r_2}+4} -\left[2\paren{\ode+\om+\ol-1}-\ol\paren{\frac{\alpha\Gamma\paren{r_1+r_2}}{r_2}+4}\right.\\&\nonumber\left.-\om\paren{\frac{\Gamma\paren{r_1+r_2}}{r_1}+3}\right]-\ode\paren{\Delta-2}\paren{1-\sqrt{\frac{3M_P^2\ode}{C}}L^{-\frac{\Delta}{2}}\cos{y}}
\end{align}
The EOS for the dark energy in this can be written as 
\begin{align}
    \omega_{de}=\frac{\paren{1+\alpha}\paren{r_1+r_2}\Gamma}{3}-\Big(\frac{1+\Delta}{3}\Big)+\Big(\frac{\Delta+2}{3}\Big)\sqrt{\frac{3M_{p}^2 \Omega_{de}}{C}}L^{-\frac{\Delta}{2}}\cos{y}~.
\end{align}
\section*{Appendix B: Other Possible Interactions (Open Universe)}
\phantomsection
\addcontentsline{toc}{section}{Appendix B:  Other Possible Interactions (Open Universe)}
\setcounter{equation}{0}
\renewcommand{\theequation}{A.\arabic{equation}}
Similar to the closed universe for a open universe the calculations can be done for open universe which are briefly given in this appendix.
\subsection*{$\mathcal{Q}=-\Gamma H (\rho_{de}+\rho_m)$:}
When dark energy and matter density only effects the interaction the differential equation reads, 
\begin{align}
    \frac{\ode '}{\ode(1-\ode)}&=-\frac{1}{1-\ode}\left[2\paren{\ode+\om+\ol-1}-\ol\paren{\frac{\Gamma\alpha\paren{1+r_1}}{r_2}+4}\right.\\& \nonumber\left.-\om\paren{\frac{\Gamma \paren{1+r_1}}{r_1}+3}\right]  
    +(\Delta-2)\Big(1-\sqrt{\frac{3M_{p}^2 \Omega_{de}}{C}}L^{-\frac{-\Delta}{2}}\cosh{y}\Big)~.
    \end{align} 
\begin{align}
    \frac{\om'}{\om}&=-\paren{\frac{\Gamma\paren{1+r_1}}{r_1}+3}
    -\left[2\paren{\ode+\om+\ol-1}-\ol\paren{\frac{\alpha\Gamma\paren{1+r_1}}{r_2}+4}\right.\\&\nonumber \left. -\om\paren{\frac{\Gamma\paren{1+r_1}}{r_1}+3}\right]-\ode\paren{\Delta-2}\paren{1-\sqrt{\frac{3M_P^2\ode}{C}}L^{-\frac{\Delta}{2}}\cosh{y}}
\end{align}
\begin{align}
    \frac{\ol'}{\ol}&=-\paren{\frac{\alpha\Gamma\paren{1+r_1}}{r_2}+4} -\left[2\paren{\ode+\om+\ol-1}-\ol\paren{\frac{\alpha\Gamma\paren{1+r_1}}{r_2}+4}\right.\\&\nonumber\left.-\om\paren{\frac{\Gamma\paren{1+r_1}}{r_1}+3}\right]-\ode\paren{\Delta-2}\paren{1-\sqrt{\frac{3M_P^2\ode}{C}}L^{-\frac{\Delta}{2}}\cosh{y}}
\end{align}
The EOS for dark energy has the following form,
\begin{align}
    \omega_{de}=\frac{\paren{1+\alpha}\paren{1+r_1}\Gamma}{3}-\Big(\frac{1+\Delta}{3}\Big)+\Big(\frac{\Delta+2}{3}\Big)\sqrt{\frac{3M_{p}^2 \Omega_{de}}{C}}L^{-\frac{\Delta}{2}}\cosh{y}~.
\end{align}
\subsection*{$\mathcal{Q}=-\Gamma H (\rho_{de}+\rho_r)$:}
Similar to the closed universe case differential equations reads,
\begin{align}
    \frac{\ode '}{\ode(1-\ode)}&=-\frac{1}{1-\ode}\left[2\paren{\ode+\om+\ol-1}-\ol\paren{\frac{\Gamma\alpha\paren{1+r_2}}{r_2}+4}\right.\\& \nonumber\left.-\om\paren{\frac{\Gamma \paren{1+r_2}}{r_1}+3}\right]  
    +(\Delta-2)\Big(1-\sqrt{\frac{3M_{p}^2 \Omega_{de}}{C}}L^{-\frac{-\Delta}{2}}\cosh{y}\Big)~.
    \end{align} 
\begin{align}
    \frac{\om'}{\om}&=-\paren{\frac{\Gamma\paren{1+r_2}}{r_1}+3}
    -\left[2\paren{\ode+\om+\ol-1}-\ol\paren{\frac{\alpha\Gamma\paren{1+r_2}}{r_2}+4}\right.\\&\nonumber \left. -\om\paren{\frac{\Gamma\paren{1+r_2}}{r_1}+3}\right]-\ode\paren{\Delta-2}\paren{1-\sqrt{\frac{3M_P^2\ode}{C}}L^{-\frac{\Delta}{2}}\cosh{y}}
\end{align}

\begin{align}
    \frac{\ol'}{\ol}&=-\paren{\frac{\alpha\Gamma\paren{1+r_2}}{r_2}+4} -\left[2\paren{\ode+\om+\ol-1}-\ol\paren{\frac{\alpha\Gamma\paren{1+r_2}}{r_2}+4}\right.\\&\nonumber\left.-\om\paren{\frac{\Gamma\paren{1+r_2}}{r_1}+3}\right]-\ode\paren{\Delta-2}\paren{1-\sqrt{\frac{3M_P^2\ode}{C}}L^{-\frac{\Delta}{2}}\cosh{y}}
\end{align}
Eos for the dark energy reads,
\begin{align}
    \omega_{de}=\frac{\paren{1+\alpha}\paren{1+r_2}\Gamma}{3}-\Big(\frac{1+\Delta}{3}\Big)+\Big(\frac{\Delta+2}{3}\Big)\sqrt{\frac{3M_{p}^2 \Omega_{de}}{C}}L^{-\frac{\Delta}{2}}\cosh{y}~.
\end{align}
\subsection*{$\mathcal{Q}=-\Gamma H (\rho_{m}+\rho_r)$:}
For this type of interaction one have,
\begin{align}
    \frac{\ode '}{\ode(1-\ode)}&=-\frac{1}{1-\ode}\left[2\paren{\ode+\om+\ol-1}-\ol\paren{\frac{\Gamma\alpha\paren{r_1+r_2}}{r_2}+4}\right.\\& \nonumber\left.-\om\paren{\frac{\Gamma \paren{r_1+r_2}}{r_1}+3}\right]  
    +(\Delta-2)\Big(1-\sqrt{\frac{3M_{p}^2 \Omega_{de}}{C}}L^{-\frac{-\Delta}{2}}\cosh{y}\Big)~.
    \end{align} 
\begin{align}
    \frac{\om'}{\om}&=-\paren{\frac{\Gamma\paren{r_1+r_2}}{r_1}+3}
    -\left[2\paren{\ode+\om+\ol-1}-\ol\paren{\frac{\alpha\Gamma\paren{r_1+r_2}}{r_2}+4}\right.\\&\nonumber \left. -\om\paren{\frac{\Gamma\paren{r_1+r_2}}{r_1}+3}\right]-\ode\paren{\Delta-2}\paren{1-\sqrt{\frac{3M_P^2\ode}{C}}L^{-\frac{\Delta}{2}}\cosh{y}}
\end{align}
\begin{align}
    \frac{\ol'}{\ol}&=-\paren{\frac{\alpha\Gamma\paren{1+r_1}}{r_2}+4} -\left[2\paren{\ode+\om+\ol-1}-\ol\paren{\frac{\alpha\Gamma\paren{1+r_1}}{r_2}+4}\right.\\&\nonumber\left.-\om\paren{\frac{\Gamma\paren{1+r_1}}{r_1}+3}\right]-\ode\paren{\Delta-2}\paren{1-\sqrt{\frac{3M_P^2\ode}{C}}L^{-\frac{\Delta}{2}}\cosh{y}}
\end{align}
The EOS for de reads,
\begin{align}
    \omega_{de}=\frac{\paren{1+\alpha}\paren{r_1+r_2}\Gamma}{3}-\Big(\frac{1+\Delta}{3}\Big)+\Big(\frac{\Delta+2}{3}\Big)\sqrt{\frac{3M_{p}^2 \Omega_{de}}{C}}L^{-\frac{\Delta}{2}}\cosh{y}~.
\end{align}
\section*{Acknowledgments}
GG extends his heartfelt gratitude to CSIR, Govt. of India, for funding this research. SP would like to thank SNBNCBS for the Junior Research Fellowship. The authors also thank Shashank Sekhar Pandey for making them aware of different cosmological data sets and the required Python libraries to handle them.
\bibliographystyle{JHEP}
\bibliography{ref.bib} 
\end{document}